\documentclass[journal]{IEEEtran}

\usepackage[english]{babel}

\usepackage{ifpdf}

\usepackage{cite} 
\usepackage{url}
\usepackage{hyperref}

\ifCLASSINFOpdf
	\usepackage[pdftex]{graphicx}
	\graphicspath{{./figures/}}
\else
	\usepackage[dvips]{graphicx}
	\graphicspath{./figures/}
\fi
\usepackage{color}
\usepackage{amssymb}
\usepackage{pgf}
\usepackage{tikz-cd}
\usepackage{pgf, tikz, pgfplots}
\usepackage{tikz-3dplot}
\usepackage{tikz-qtree}
\usetikzlibrary{shapes, arrows, automata, plotmarks}
\usetikzlibrary{calc,hobby,decorations,math}
\usepackage{tcolorbox}

\usepackage[cmex10]{amsmath}
\usepackage{amsfonts, amssymb, amsthm}
\usepackage{mathrsfs}

\usepackage{algorithm,algpseudocode}
	\algnewcommand{\LeftComment}[1]{\Statex \(\triangleright\) #1}

\usepackage{enumerate}
\usepackage{multirow}
\usepackage{rotating}
\usepackage{subcaption}
\usepackage{needspace}

\input{./latex_resources/mySymbol.sty}
\input{./latex_resources/pennColors.sty}

\newcommand{\End}[1]{\textrm{End}\left(#1\right)}

\newtheorem{proposition}{\hspace{0pt}\bf Proposition}
\newtheorem{example}{\hspace{0pt}\bf Example}

\newtheorem{theorem}{\hspace{0pt}\bf Theorem}
\newtheorem{corollary}{\hspace{0pt}\bf Corollary}

\newtheorem{remark}{\hspace{0pt}\bf Remark}

\newtheorem{definition}{\hspace{0pt}\bf Definition}

\begin{document}
\title{Lie Group Algebra Convolutional Filters}
\author{Harshat Kumar\textsuperscript{\textdagger \S}, Alejandro~Parada-Mayorga\textsuperscript{\textdagger$\star$},
        and Alejandro Ribeiro\textsuperscript{$\star$}
\thanks{\textsuperscript{\textdagger}These authors contributed equally to the paper. \textsuperscript{\S}Johns Hopkins Applied Physics Laboratory \textsuperscript{$\star$}Department of Electrical and Systems Engineering, University of Pennsylvania. Email: harshat.kumar@jhuapl.edu, alejopm@seas.upenn.edu, aribeiro@seas.upenn.edu.}}

\markboth{IEEE Transactions on Signal Processing (submitted)}%
{Shell \MakeLowercase{\textit{et. al.}}: Bare Demo of IEEEtran.cls for Journals}
\maketitle




\begin{abstract}

In this paper we propose a framework to leverage Lie group symmetries on arbitrary spaces exploiting \textit{algebraic signal processing} (ASP). We show that traditional group convolutions are one particular instantiation of a more general Lie group algebra homomorphism associated to an algebraic signal model rooted in the Lie group algebra $L^{1}(G)$ for given Lie group $G$. Exploiting this fact, we decouple the discretization of the Lie group convolution elucidating two separate sampling instances: the filter and the signal. To discretize the filters, we exploit the exponential map that links a Lie group with its associated Lie algebra. We show that the discrete Lie group filter learned from the data determines a unique filter in $L^{1}(G)$, and we show how this uniqueness of representation is defined by the bandwidth of the filter given a spectral representation. We also derive error bounds for the approximations of the filters in $L^{1}(G)$ with respect to its learned discrete representations. The proposed framework allows the processing of signals on spaces of arbitrary dimension and where the actions of some elements of the group are not necessarily well defined. Finally, we show that multigraph convolutional signal models come as the natural discrete realization of Lie group signal processing models, and we use this connection to establish stability results for Lie group algebra filters. To evaluate numerically our results, we build neural networks with these filters and we apply them in multiple datasets, including a knot classification problem.
\end{abstract}


\IEEEpeerreviewmaketitle




\section{Introduction}

Deep neural network models built with prior knowledge of inherent symmetries learn faster than models built without. In the same way Euclidean convolutional neural networks (CNNs) capture translation symmetries \cite{bruna2013invariant} and graph convolutional neural networks (GNNs) capture label permutation symmetries \cite{gama2020stability}, group convolutions - and by their extension group convolutional neural networks (G-CNNs) - capture group symmetries in the signal \cite{kondor2018generalization,ravanbakhsh2017equivariance}. Empirically, G-CNNs have been applied to a wide range of problems from classification and segmentation on images~\cite{weiler2018learning,romero2020attentive} and point-cloud data \cite{esteves2018learning}, to predicting and controlling dynamical systems \cite{finzi2020generalizing}. In this work, introduce an algebraic generalization and propose a novel method which efficiently computes the group convolution by using Lie group algebra convolutional filters. 

The related work on group convolutions can be broadly categorized into three main approaches. The first approach involves augmenting the kernels of traditional CNNs with copies that have been transformed by a group action \cite{cohen2016group}. Such methods were generalized to continuous groups by utilizing steerable filters, which solve a constraint on the kernel using group representation theory \cite{worrall2017harmonic, weiler2018learning, weiler2019general}. Departing from the image domain, the second method uses the group harmonics in order to compute the group convolution in the harmonic domain. In particular, by working out the specific group harmonics which is analogous to frequency domain for time series signals \cite{esteves2020theoretical}, the convolution can be computed by point-wise multiplication of the filter and signal after the transformation \cite{esteves2018learning, esteves2020spin}, enabling the extension to homogeneous spaces on groups. The third approach directly estimates the group convolution with the Haar measure with Monte Carlo samples \cite{finzi2020generalizing,macdonald2022enabling}. Related but tangential to this work are approaches which consider group augmentation or learning group invariances \cite{chen2020group, benton2020learning, dehmamy2021automatic}. However, these works are tangential because we consider problems where the group symmetry is known. Additionally, Spatial Transformer Networks \cite{jaderberg2015spatial} - which learn a transformation to convert the signal into its canonical pose through resampling - have been extended to canonical Lie coordinates \cite{tai2019equivariant}. In the new basis, group convolution is computed through the familiar planar convolution \cite{esteves2017polar}, so the architecture becomes a preprocessing step before a CNN network, bearing similarity to the first approach.

In this work, we propose a novel group convolution implementation that is derived through the lens of \textit{algebraic signal processing} (ASP) \cite{algSP0}. ASP emerged as a fundamental tool that allows one to formally describe convolutional signal models across different domains relying on algebraic structure~\cite{algSP0,algSP1,algSP2,algSP3,algSP4,algSP5,algSP6,algSP7,algSP8,ParadaMayorga2020QuiverSP}. More specifically, ASP leverages the \textit{representation theory of algebras} to define each signal model as the representation of a specific algebra in the sense of~\cite{repthybigbook,repthysmbook}. This allows the consistent description of models that range from discrete time signal processing models~\cite{algSP1}, discrete space models with symmetric shift operators~\cite{algSP2}, signal models on 2D spatial hexagonal lattices~\cite{algSP6}, models on abstract lattices~\cite{puschel_asplattice}, signals on sets~\cite{puschel_aspsets} among others~\cite{algSP7}. Additionally, ASP has been central for deriving Cooley-Tukey algorithms to efficiently compute frequency representations~\cite{algSP4,algSP5,algSP8} and to understand the relationship via sampling between several ASP models~\cite{algSP3}.

Focusing on the case of Lie group symmetries, we extend the ASP signal model taking into account Banach $*$-algebras and signals in Hilbert spaces. From here, we use the Lie group algebra homomorphism to compute the group convolution in the representation space, of which the conventional group convolution is a particular case.  This brings us to our first major contribution. 
\begin{enumerate}
    \item [\textbf{(C1)}] We define group convolutions when a signal is \emph{not} presented in the canonical basis of the group or as a signal of the group itself.   
\end{enumerate}

Translation on pixels and rotation on angles are examples where the signal is on the canonical basis of the group. Our method allows for translations on angles and rotations on pixels. More specifically, we show that there exists a Lie group algebra homomorphism that maps an element from the group algebra (i.e. the filter) to the space of bounded operators acting on the Hilbert space. The main insight comes from the realization that the group homomorphism translates the action of the filters into a matrix multiplication acting on the signal space. This inherently removes the requirement to lift signals onto the group, which can be an expensive operation (see Section \ref{sec:experiments}). To the best of our knowledge, we are the only work to consider the disconnect between signal basis and symmetries outside the \emph{rotations on pixels} which have been studied~\cite{cohen2016group, finzi2020generalizing, worrall2017harmonic}.

Because a Lie group is both a group and a manifold and is by definition continuous, the classical ideal convolution performed between a filter and a signal that belongs to a dense space of functions on the Lie group. We sample both the filter and signal to make the convolution tractable, leading to our second main contribution. 
\begin{enumerate}
    \item [\textbf{(C2)}] We decouple the discretization of the group convolution elucidating two separate sampling instances: the Lie group algebra (filter) and the signal.
\end{enumerate}

Drawing on sampling theory from manifolds \cite{pesenson2000sampling}, we show that every bandlimited filter is \emph{uniquely} represented by its sampled values. To establish a principled sampling method, we propose the use of the Lie algebra basis and the exponential map. The Lie algebra, the tangent space of the Lie group at the identity, corresponds to the Lie group by the exponential map, which is bijective around the identity \cite{hall_liealg}.  For most problems of interest, the sampling of the input signal is intrinsic. To handle these potentially non-homogeneous signals, we use interpolation to realize induced group actions. Interpolation on structured and unstructured data is well studied \cite{lehmann1999survey}, and we show that the choice of interpolation is algebraically justified.

Given that conventional group convolutions \cite{kondor2018generalization} are a particular case of our generalization, it follows naturally that equivariance -the property that engendered much of the renewed interest in group convolutions- holds for that specification. In addition, we consider the notion of filter stability: that a deformation on the operator should be proportional to the size of the deformation. This leads us to our third major contribution.

\begin{enumerate}
    \item [\textbf{(C3)}]  We establish stability through the lens of multigraph signal processing \cite{parada2021convolutional,msp_journal,msp_conference}. Namely, the realization of the Lie group algebra homomorphism can be thought of as a shift operator, which enables the characterization of our filters as a traditional ASP model with a polynomial algebra. 
\end{enumerate}

The main insight is that the matrix representations of the group actions can be thought of as graph shift operators. Each basis element of the Lie algebra corresponds to a unique graph that disseminates the information across the nodes, which are the values on the sampled input signal. 

The remainder of the paper is organized as follows. In Section \ref{sec_asp_on_lie_groups} we introduce ASP and its extension to Banach $\ast$-algebras, and we also introduce Lie group algebra signal models. In Section \ref{sec_grpasm}, we discuss the first homomorphism approximation via the sampling on the Lie group algebra. In Section~\ref{sec:discrete_sample} we discuss the approximation of the homomorphisms using sampling on the signal space. In Section~\ref{sec:equivariance_and_stability}, we state the fundamental connection between discrete Lie algebraic signal models and multigraph signal processing. Additionally, we leverage this connection to state stability results. In section \ref{sec:experiments}, we evaluate our findings numerically on three classification datasets, including a novel knot classification problem. Finally, in Section~\ref{sec:discussion} we present some conclusions and discussions on future endeavors.

\vspace{3mm}
\noindent \textbf{Notation:} A general Lie group will be denoted by $G$ while its associated Lie algebra is denoted by $\mathfrak{g}$. If $g\in G$ the action of $g$ on a set $\ccalX$ is denoted by $T_{g}: \ccalX \to \ccalX$. If $\ccalH$ is a Hilbert space of functions on $\ccalX$, we denote its elements by italic bold symbols such as $\boldsymbol{f},~\boldsymbol{g},~\ldots$ The value of $\boldsymbol{f}\in\ccalH$ at $x\in\ccalH$ is denoted by $\boldsymbol{f}(x)$. We use the overline symbol to indicate the complex conjugate of a given value, i.e. $\overline{\boldsymbol{f}(x)}$ is the complex conjugate of $\boldsymbol{f}(x)$. If $T_g$ is the action of $g\in G$ on $\ccalX$, there is unique operator $\bbT_{g}: \ccalH \to \ccalH$ associated to $T_g$ and given by $\bbT_{g}(\boldsymbol{f}(x)) = \boldsymbol{f}(T_g x)$. A Haar measure on $G$ is typically denoted by $\mu$ and the set of absolutely integrable functions on $G$ under $\mu$ is denoted by $L^{1}(G)$. The symbol $\ccalA$ denotes a Banach $\ast$-algebra while $\ccalH$ denotes in general a Hilbert space and $\ccalB (\ccalH)$ is the set of bounded linear operators from $\ccalH$ onto itself. We also denote calligraphic letters to denote sets and spaces. The set of generators of a Lie algebra, $\mathfrak{g}$, is denoted by $\mathfrak{G}$, while arbitrary elements elements of $\mathfrak{g}$ are denoted by capital Fraktur letters such as $\mathfrak{X}_{1}, \mathfrak{X}_{1}, \ldots$. We use the hat math accent to represent discrete subsets and functions related to infinitely continuous counterparts. Additionally, if $\widehat{G}\subset G$ is a discrete subset of $G$, the elements of $\widehat{G}$ are in general endowed with the math-hat accent as well. We denote the Laplace-Beltrami operator on a Lie group by $\boldsymbol{\Delta}$.




\section{Algebraic Signal Processing for Lie Groups}~\label{sec_asp_on_lie_groups}

We consider the problem of processing signals defined on sets with Lie group symmetries. Formally, let $\ccalX\subset\mbR^{n}$ be a set, then our signals of interest live on the Hilbert space $\ccalH$ which consists on the set of functions on $f: \ccalX \to \mbC$. In this section, we focus our attention on homogenous transformations, where the Lie group $G$ acts on $\ccalX$ through the family of transformations $T_g:\ccalX \to \ccalX$ homomorphic to $G$, i.e. for any $g_1, g_2 \in G$ there is $T_{g_1 g_2}: \ccalX \to \ccalX$ such that $T_{g_1 g_2} = T_{g_1}T_{g_2}$. We discuss the case where $T_{g}\ccalX\notin\ccalX$ in Section~\ref{sec:discrete_sample}.

The group action $T_g$ on $\ccalX$ extends to $f$ by the induced transformation $\bbT_g: \ccalH \to \ccalH$ which maps $f\mapsto f'$ such that $f'(T_g(x)) = f(x)$~\cite{kondor2018generalization}. For example, consider the group of translations on the real line which shifts each element. In this case, $T_t(x) = x+t$ and $f'(x) = f(x-t)$. The collection of maps $\bbT_g$ for $g\in G$ define a \textit{representation} of $G$ on $\ccalH$, and we will denote this representation by $\bbT$ using a subindex explicitly when we refer to specific representations of an element $g\in G$. If for any $x\in\ccalX$ there exists $y\in\ccalX$, and there exists $g\in G$ such that $T_{g}y = x$, we call $\ccalX$ a homogeneous domain, and we will refer to the space of signals whose domain is $\ccalX$ as a homogeneous space. Homogeneous spaces are a particular case of more general cases of signals with Lie group symmetries which we consider.


\subsection{Extending ASP to Banach $*$-algebras}
\label{subsec_asp_BanachA}

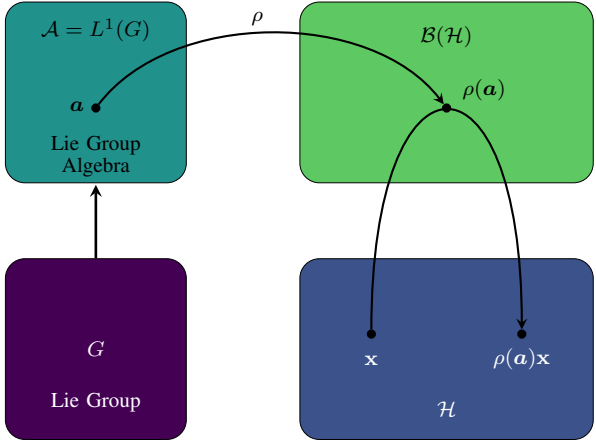
\begin{figure}
        \centering


\definecolor{my_cp5_col1}{RGB}{253, 231, 37}
\definecolor{my_cp5_col2}{RGB}{94, 201, 98}
\definecolor{my_cp5_col3}{RGB}{33, 145, 140}
\definecolor{my_cp5_col4}{RGB}{59, 82, 139}
\definecolor{my_cp5_col5}{RGB}{68, 1, 84}

\usetikzlibrary{positioning,decorations.pathreplacing,shapes}


\def \scale {1.3}
\def \unit  { \scale cm}
\def \layerinterdist {4}

\tikzstyle{set} = [rectangle,color=black,
                    rounded corners = 0.2*\unit,
                    fill=black,
                    inner sep=0pt,
                    draw,
                    anchor = center,
                    line width=0.1mm]
                    
\tikzstyle{shadow_squared} = [rectangle,color=black,
rounded corners = 0.2*\unit,
fill=black,
inner sep=0pt,
draw,
anchor = south east,
line width=0.1mm,
minimum width  = 4.1*\unit,
minimum height = 3.9*\unit]

\tikzstyle{vectorspace} = [ set, 
                             fill=my_cp5_col4,
                             minimum width  = 3*\unit,
                             minimum height = 1.85*\unit]
                             
\tikzstyle{endomorphisms} = [ vectorspace,
                              fill=my_cp5_col2]
                                
\tikzstyle{algebra} = [ endomorphisms,
                        fill=my_cp5_col3,
                        minimum width = 1.85*\unit]

\tikzstyle{group_math} = [endomorphisms,
minimum width = 1.85*\unit,
minimum height = 1.85*\unit]

\tikzstyle{dot} = [ circle,
                    minimum width  = 0.05*\unit,
                    fill=black,
                    color=black,
                    inner sep=0pt,
                    draw,
                    anchor = center ]

{\fontsize{8}{8}\selectfont

\begin{tikzpicture}[rounded corners,ultra thick]




   \path (0,0) node [vectorspace] (M0) {};
   \path (M0.south) ++ (0, 0.2) node [above, color=white] {$\mathcal{H}$};
   
   \path (M0) ++ (-1,0.2) node [dot] (x) {};
   \path (x.south)++(0,-0.1) node [below, color=white] {$\mathbf{x}$}; 
   
   \path (M0) ++ (1,0.2) node [dot] (ex) {};
   \path (ex.south) node [below, color=white] {$\rho(\boldsymbol{a})\mathbf{x}$};


   \path (M0.north) ++ (0, 1) 
         node [endomorphisms, anchor=south] (End0) {};
   \path (End0.north) ++ (0.0, -0.2) node [below, color=black] {$\mathcal{B}(\mathcal{H})$};

   \path (End0.south) ++ (0.0, 1) node [dot] (e) {};      
   \path (e.center) ++ (0.5,0.25) node [color=black] {$\rho(\boldsymbol{a})$}; 
   
   \path (e)+(-1,0.95) coordinate (c1);
   \path (e)+(1,0.95) coordinate (c2);   
   \path [draw, -stealth, line width=0.8,color=black] (x) .. controls (c1) and (c2) .. (ex);

%
%


   \path (End0.north west) ++ (-1.5, 0) 
         node [algebra, anchor = north east] (A) {};   
   \path (A.north) ++ (0,-0.1) node [below,color=black] {$\mathcal{A} = L^{1}(G)$};
   \path (A.center) ++ (-0.25,0) node [below,color=black] {$\boldsymbol{a}$};  
   \path (A.south) ++ (0,0.75) node [below,color=black, align=center] {Lie Group \\ Algebra};

    
    \path (A.south) ++ (0.0, 1) node [dot] (a1) {};

  
  \path (A.south) ++ (0, -1) 
  node [group_math, anchor = north, fill=my_cp5_col5] (G) {};   
  \path (G.center) ++ (0,0) node [color=white] {$G$}; 
  \path (G.south) ++ (0,0.75) node [below,color=white] {Lie Group};

   \path (A.north) + (1,-0.1) coordinate (c1);
   \path (End0.north) + (-1,-0.1) coordinate (c2);   
   \path [draw, -stealth,line width = 0.8,color=black] (a1) .. controls (c1) and (c2) .. (e) node[midway,above left,rotate=0,color=black] {$\rho$};


   \path (A.south) ++ (0.0, 0.1) node [] (a1) {};  
   \path [->,draw, -stealth,line width = 1,color=black] (G.north) -- (a1) node[midway,right,rotate=0,color=black] {};

\end{tikzpicture}
} 
        \caption{Schematic diagram of the Lie group algebra signal processing framework. The symmetries of a Lie group $G$ are embedded in the Banach-$\ast$ algebra $L^{1}(G)$. The algebraic filters in $L^{1}(G)$ are realized/implemented in the set of bounded operators acting on $\ccalH$, $\ccalB (\ccalH)$. The Hilbert space $\ccalH$ contains the signals transformed according to the symmetries in $G$, and its properties do not depend on the properties of $L^{1}(G)$.}
        \label{fig_asp_model}
\end{figure}

In algebraic signal processing (ASP), any signal model is defined by the triplet $(\ccalA, \ccalH, \rho)$, where $\ccalA$ is an associative algebra with unity, $\ccalH$ is a vector space, and $\rho:\ccalA \to \End{\ccalH}$ is a homomorphism between the algebra $\ccalA$ and the set of endomorphisms on $\ccalH$~\cite{algSP0} -- see Fig.~\ref{fig_asp_model}. The vector space $\ccalH$ and the homomorphism $\rho$ together make a representation $(\ccalH, \rho)$ of the associative algebra $\ccalA$ by preserving multiplication and unit. Signals are modeled as the elements in $\ccalH$, while filters are elements of $\ccalA$. Using $\rho$ we instantiate the \textit{abstract} filters in $\ccalA$ into concrete operators in $\text{End}(\ccalH)$ that transform the signals in $\ccalH$. Figure~\ref{fig_asp_model} depicts the basic components of an algebraic signal model (ASM).

Although ASP models encapsulate into algebraic principles the properties of a signal model, by themselves they do not attach \textit{topological properties} to any of the objects in the triplet $(\ccalA, \ccalH, \rho)$. Lie groups are not only algebraic objects but also manifolds that link the differentiability and continuity of the manifold to the algebraic properties of the group. Therefore, in order to capture Lie group symmetries in general algebraic convolutional signal models we need to extend ASP to include topological properties. Then, we need to extend the classical notion of algebraic signal model (ASM) introduced by P{\"u}schel to include topological properties. We achieve this by exploiting the notions of Banach $\ast$-algebra, Hilbert space, and $\ast$-homomorphism~\cite{folland2016course,deitmar2014principles}. We recall that a Banach $\ast$-algebra, $\ccalA$, is an algebra which is also a Banach space and is endowed with a closed operation $(\cdot)^\ast: \ccalA \rightarrow \ccalA$ called the adjoint. The adjoint must satisfy that $(ab)^\ast = b^\ast a^\ast$ for all $a,b\in\ccalA$, where $ab\in\ccalA$ indicates the product between $a$ and $b$. We also recall that $\ccalH$ is a Hilbert space if $\ccalH$ is endowed with an inner product and is complete as a metric space concerning the metric induced by the inner product. If we denote by $\ccalB (\ccalH)$ the set of bounded operators acting on $\ccalH$, we say that $\rho: \ccalA \rightarrow \ccalB (\ccalH)$ is a $\ast$-homomorphism if $\rho (ab) = \rho(a)\rho(b)$ and $\rho\left( a^\ast \right) = \rho(a)^\ast$. With these concepts at hand we now formalize the notion of algebraic signal model (ASM) used in this paper.


\begin{definition}\label{def_ASM}
 
 An algebraic signal model (ASM) is a triplet $(\ccalA, \ccalH, \rho)$, where $\ccalA$ is a Banach $\ast$-algebra, $\ccalH$ is a Hilbert space, and $\rho$ is a $\ast$-homomorphism between $\ccalA$ and $\ccalB (\ccalH)$.
 
\end{definition}


From Definition~\ref{def_ASM}, a rich variety of convolutional signal models can be derived. Fixing $\ccalH$ and choosing different algebras and/or homomorphisms, it is possible to leverage different types of symmetries of the data. Likewise, while fixing the algebra it is possible to leverage specific symmetries into different types of signals on different domains.

For our discussion, and for the rest of the paper, we choose $\ccalA = L^1 (G)$ for a given Lie group $G$, where $L^{1}(G)$ is the set of continuous and integrable functions defined on $G$ -- see Appendix~\ref{sec_background} in the supplementary material. To see that $L^{1}(G)$ is an algebra, we start verifying that $L^1 (G)$ is a vector space. Let $\boldsymbol{a},\boldsymbol{b} \in L^{1}(G)$ and $\alpha, \beta\in\mbC$. Then, it follows that

\begin{equation}
 \int_{G}\left\vert 
              \alpha\boldsymbol{a}
              +
              \beta\boldsymbol{b}
         \right\vert 
         d\mu
         \leq 
         \vert\alpha\vert
         \int_{G}\left\vert 
              \boldsymbol{a}
         \right\vert 
         d\mu
         +
         \vert\beta\vert
         \int_{G}\left\vert 
              \boldsymbol{b}
         \right\vert 
         d\mu
         ,
\end{equation}
where $d\mu$ is the Haar measure defined on $G$. Therefore, $\boldsymbol{a},\boldsymbol{b}\in L^{1}(G)$ implies $\left( \alpha\boldsymbol{a}+\beta\boldsymbol{b}\right)\in L^{1}(G)$. The vector space $L^{1}(G)$ becomes an algebra choosing a product $\ast$ between elements of $L^{1}(G)$ given according to

\begin{equation}\label{eq_conv_LieG_classic}
\left(
      \boldsymbol{a} \ast \boldsymbol{b}
\right)(x)
=
 \int_G \boldsymbol{a}(y)\boldsymbol{b}\left( y^{-1}x \right) d\mu(y)
 ,
\end{equation}
where 
$
\boldsymbol{a} \ast \boldsymbol{b}\in L^{1}(G)
$
\cite{hewitt1994abstract,folland2016course}.

Now, we define on $L^{1}(G)$ an involution operation given by $\boldsymbol{a}^{\ast}(x) = \Delta (x^{-1})\overline{\boldsymbol{a}(x^{-1})}$, where $\Delta$ is the modular function of $G$ -- see Appendix~\ref{sec_background} in the supplementary material -- and $\overline{\boldsymbol{a}(x^{-1})}$ is the complex conjugate of $\boldsymbol{a}(x^{-1})$. Putting all these attributes together, $L^{1}(G)$ is a Banach $\ast$-algebra.


\subsection{Lie Group Algebra Filters}

With the selection of $\ccalA = L^{1}(G)$ as the algebra in all the algebraic signal models considered from now on, we formally define the homomorphism that realizes the abstract filters in $L^{1}(G)$ into concrete operators on a space of signals $\ccalH$ whose symmetries are associated to the Lie group $G$. Before presenting such a definition we recall that for a group $G$ and element $g\in G$ the group transformations $\bbT_g$ constitute a representation of $G$ and $\bbT_g\in\ccalB(\ccalH)$.


\begin{definition}[~Lie group algebra homomorphism~\cite{lang2012sl2}]
\label{def:LieGroupAlgHomomorph}

Let $G$ be a Lie group and let $\bbT$ be a representation of $G$ on $\ccalH$. Let $\boldsymbol{a}\in L^1(G)$. The Lie group algebra homomorphism is the map 
\begin{equation*}
        \rho:L^1(G) \to \ccalB(\ccalH),
\end{equation*}
by letting
\begin{equation}\label{equ_rho_ideal}
        \rho(\boldsymbol{a})f = \int_G \boldsymbol{a}(g)\bbT_g fd\mu(g),
\end{equation}
    where $d\mu$ is the Haar measure on $G$.
\end{definition}


Indeed \eqref{equ_rho_ideal} describes the filter we want to compute to process our signals $f\in\ccalH$. The integral can be intractable in some scenarios, and approximating its value requires discretizing the Haar measure, which is a non-trivial source of error \cite{macdonald2022enabling}. In the next section, we describe our sampling-based approach to directly find unique approximations of the algebraic filters in $L^{1}(G)$ and their instantiations by $\rho$.

\begin{remark} \label{rem:conventional_grp_conv}\normalfont
Interestingly, by letting $\ccalH$ be $L^2(G)\subset L^{1}(G)$, we recover the conventional group convolution given by~\eqref{eq_conv_LieG_classic}. Specifically, by letting the representation $\bbT_g$ act on the signal $f$ by $\bbT_g f(x) = f(g^{-1}x)$, the Lie group algebra homomorphism in \ref{equ_rho_ideal} is equivalent to \eqref{eq_conv_LieG_classic}. Group convolutions, where the signal is defined on the group, is a particular case of our framework. Notice also that the group algebra on discrete groups as presented in~\cite{terrasFG, steinbergrepg} and whose product is given according to 
\begin{equation}
\left( 
      \boldsymbol{f}
      \ast
      \boldsymbol{g}
\right)(x)
      =
      \sum_{y\in G}
      \boldsymbol{f}(y)
      \boldsymbol{g}\left( y^{-1}x\right)
      ,
\end{equation}
is a particular case of~\eqref{eq_conv_LieG_classic}. This can be seen when selecting the Haar measure as $d\mu (H) = \vert H\vert/\vert G\vert$, where $H\subset G$ and $\vert H \vert$ and $\vert G\vert$ indicate the number of elements in $H$ and $G$, respectively~\cite{hewitt1994abstract}(page 198).
\end{remark}




\section{Sampling of the Lie Group Algebra}
\label{sec_grpasm}

In this section, we show that it is possible to build a tractable realization of the homomorphism in~\eqref{equ_rho_ideal} by sampling the filters in $L^{1}(G)$. Let $\widehat{G}\subset G$ be a discrete subset of the Lie group $G$. We will refer to $\widehat{G}$ as a \textit{sampling set} for the filters in $L^{1}(G)$. We will show that for bandlimited algebraic filters in $L^{1}(G)$, we can build unique representations using their samples on $\widehat{G}$.


\begin{figure*}
    \centering
    	\begin{subfigure}{.3\linewidth}
		\centering
       \resizebox{150pt}{150pt}{
  \includegraphics[trim={2cm 0 2cm 0},clip]{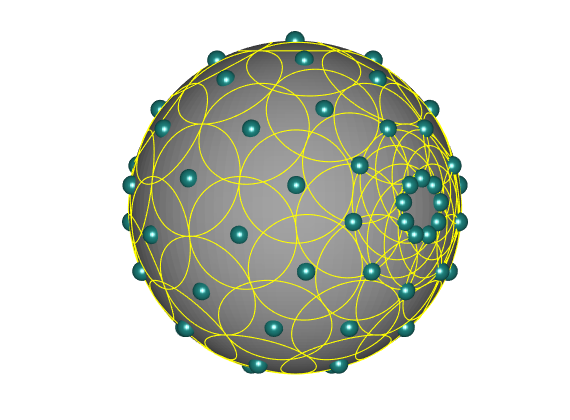} }
\end{subfigure}
    	\begin{subfigure}{.3\linewidth}
		\centering
       \resizebox{150pt}{150pt}{
  \includegraphics[trim={2cm 0 2cm 0},clip]{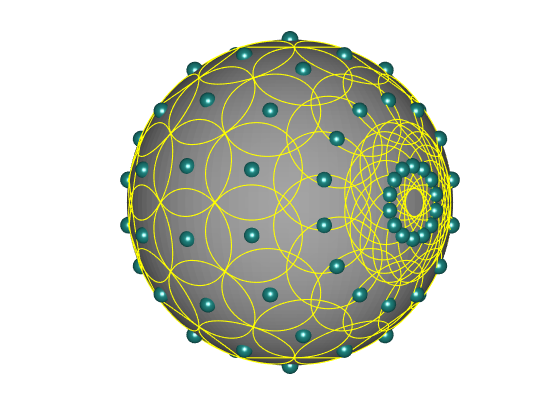} }
\end{subfigure}
    	\begin{subfigure}{.3\linewidth}
		\centering
              \resizebox{150pt}{150pt}{
  \includegraphics[trim={2cm 0 2cm 0},clip]{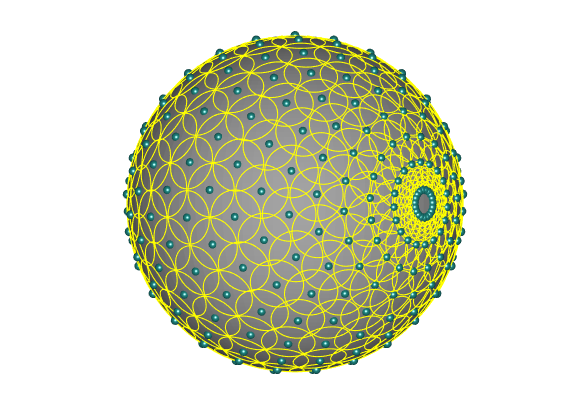} }
\end{subfigure}
    \caption{Consider the Lie group $G = \mathsf{SO}(3)$, which can be visualized on the sphere. Further, consider the sampling set $\widehat{G}_{\delta,N}^{k}$ from Definition \ref{def_Ghat_monomials}. Increasing the number of samples on $G$, i.e. increasing the size of $\widehat{G}_{\delta,N}^{k}$, allows coverings $Y_{\widehat{G}}(r^*)$ with smaller radius ($r^*$) -- shown by the yellow circles. The smaller radius allows a larger bandwidth $\omega$ of the class of bandlimited signals that are uniquely determined by their samples on $\widehat{G}_{\delta,N}^{k}$.}
    \label{fig:Lie_group_sampling}
\end{figure*}



\subsection{Bandlimited filters}\label{sec_bandlimitted}

In this subsection, we introduce the notion of bandlimited filters in $L^{1}(G)$. To do so, we follow the characterization proposed in~\cite{pesenson2000sampling} relying on the spectral decomposition of an operator defined on the Lie group~\footnote{Notice that this is a standard approach for defining bandlimited signals on graphs~\cite{ortega_gsp}, graphons~\cite{parada2022graphon,diao2016model,alejopm_gpooling_c} and other domains~\cite{algSP0,parada2021convolutional,parada2021algebraic}.}. In particular, since any Lie group is a manifold, we consider the Laplace-Beltrami operator, $\boldsymbol{\Delta}: L^{2}(G) \to L^{2}(G)$, acting on filters with finite energy (square integrable). We then leverage the spectral decomposition of the auxiliary operator $\boldsymbol{D} = \boldsymbol{\Delta}^{\frac{1}{2}}$ which is self-adjoint and positive definite~\cite{pesenson2000sampling}, guaranteeing the spectral decompositions have support (frequencies) contained in $[0,\infty)$.

As stated in~\cite{pesenson2000sampling} one can leverage the spectral decomposition of $\boldsymbol{D}$ to define projections of elements in $L^{2}(G)$ on eigenspaces associated with the spectrum of $\boldsymbol{D}$. Additionally, such projections define a unique unitary operator $F$ that maps elements from $L^{2}(G)$ into a direct integral of Hilbert spaces indexed by the spectrum of $\boldsymbol{D}$. The reader unfamiliar with the technical details in~\cite{pesenson2000sampling}, can consider a particular instantiation of these ideas on the $n$-dimensional sphere $\mbS^{n-1}$, where a spectral decomposition of $\boldsymbol{D}$ allows the decomposition of functions in $L^{2}(\mbS^{n-1})$ in terms of countable sums of the so-called spherical harmonics~\cite{hall_quantum,atkinson2012spherical,rosenberg1997laplacian}. We can think about the operator $F$ as a Fourier transform. We leverage these ideas to formally introduce the notion of bandwidth for signals defined on a Lie group.


\begin{definition}[$E_\omega(\boldsymbol{D})$-Bandlimitted filters]
\label{def_band_limited_filters_L1G}

Let us represent by $\int H(\lambda)dm(\lambda)$ the decomposition of $L^2(G)$ as a direct integral indexed by the spectrum of $\boldsymbol{D}$. Let $F:L^2(G)\to \int H(\lambda)dm(\lambda)$ be a projection operator. Then, if the action of $F$ on $\boldsymbol{a}\in L^2(G)$ has support on $[0, \omega]$, we say that it is $\omega$-bandlimitted. The set of all $\omega$-bandlimitted filters is called $E_\omega(\boldsymbol{D})$.

\end{definition}


Definition~\ref{def_band_limited_filters_L1G} allows the specification of families of filters in $L^{2}(G)$ on the frequency domain. Now, we introduce some definitions and notation that will enable the characterization of filters in $L^{2}(G)$ using their samples on $G$, and we link this characterization to the aforementioned frequency representations.

Let us denote by $B(g, r)$ the open ball of radius (geodesic) $r$ around the group element $g\in G$, and let $\widehat{G}$ be a finite subset of $G$. Then, we denote by $Y_{\widehat{G}}(r)$ the union of all $B(g, r)$ for all $g\in \widehat{G}$, i.e. $Y_{\widehat{G}}(r):= \bigcup_{g\in G}B(g,r)$. Let $r_\textrm{max}$ be the maximum distance (geodesic) between any two points on the Lie group $G$, formally
$r_\textrm{max}(G):= \max_{g,g' \in G} d_{geo}(g,g^{'}),$ where $d_{geo}(\cdot, \cdot)$ refers to the geodesic distance. In this work, we assume that $r_\textrm{max}(G)$ exists. Given that $r_\textrm{max}$ exists, it follows that there is a $r$ such that $Y_{\widehat{G}}(r)$ covers the entire manifold $G$ for any finite subset $\widehat{G}\subset G$. We will denote by $r^*$ the smallest $r\in (0, r_\textrm{max}(G)]$ such that $Y_{\widehat{G}}(r^*)$ covers $G$. Formally, 
\begin{equation}\label{equ:r_star}
\begin{split}
    r^*_{\widehat{G}} & := \min~ r\\
   \textrm{s.t.}& ~G\subset Y_{\widehat{G}}(r).
\end{split}
\end{equation}

We can now state a particular case of the main result from~\cite{pesenson2000sampling}, which we adapt for convenience here.


\begin{theorem}\cite[Adapted from Theorem 1.6]{pesenson2000sampling}\label{thm:pesenson} 
Let $G$ be a Lie group and $\widehat{G}$ a finite subset of $G$. If $G\subset Y_{\widehat{G}}(r)$, then there exists $C_{0}>0$ such that every bandlimited algebraic filter in $E_{\omega}(\boldsymbol{D})$ with $\omega>0$ is uniquely determined by its values on $\widehat{G}$ as long as
\begin{equation} \label{equ:thm1_ineqality}
r^{*}_{\widehat{G}} < (C_0\omega)^{-1} \leq r_\textrm{max}(G) 
.
\end{equation}
\end{theorem}


Intuitively, this means that the smaller the radius $r^*$ to cover the Lie group, the larger $\omega$ is permitted to uniquely reconstruct the signal. We let $\omega(r^{*})$ denote the maximum bandwidth such that the inequality on the left-hand side of \eqref{equ:thm1_ineqality} holds. An example of the sampling set $\widehat{G}$ and and the covering $Y_{\widehat{G}}(r)$ is shown in Figure \ref{fig:Lie_group_sampling}. The closer the samples (by having more samples), the smaller the radius $r$ and the larger the bandwidth $\omega$. 

Finding the optimal sampling set for functions on an arbitrary manifold is an open problem. In fact, finding the optimal sampling pattern for functions on $\mathsf{SO}(n+1)$ is equivalent to finding the solution of the spherical code problem which is still open for general dimensions~\cite{conway2013sphere}. Therefore, in this work, we propose a sampling method that exploits the exponential map and the basis of the associated Lie algebra of a given Lie group $G$.

\subsection{The Exponential Map} 
\label{sec:generators}


\begin{figure}
        \centering
        \includegraphics{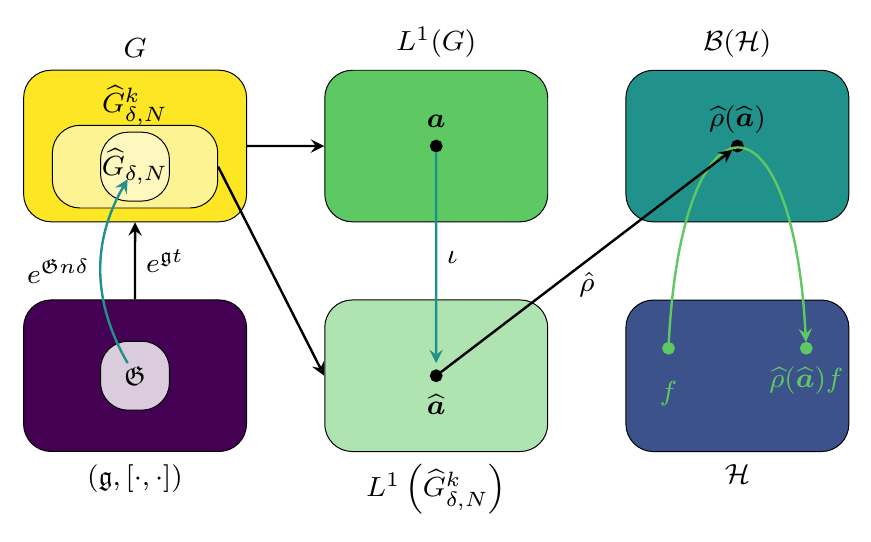} 
        \caption{Algebraic Signal Processing model of the Lie group convolutions. The elements in $\ccalB(\ccalH)$ are filters that act on elements in the Hilbert space $\ccalH$ which are the signals. We search over elements in the Lie group algebra $L^1(G)$ which comes from the Lie group $G$. To make the filter tractable, the exponential map from the Lie algebra $(\mathfrak{g},[\cdot, \cdot])$ is sampled to create the subset $\widehat{G}_{\delta, N}$. These elements are used to create monomials which make the subset $\widehat{G}_{\delta, N}^k$. It is the signals defined on that subset $\widehat{\bba}$ which are realized through the homomorphism $\widehat{\rho}$ in $\ccalB(\ccalH)$. }
\end{figure}


For any Lie group $G$ and its associated Lie algebra $\mathfrak{g}$, we have that $e^{t\mathfrak{X}}\in G$ for any $\mathfrak{X}\in\mathfrak{g}$ and $t\in\mbR$ -- see Appendix~\ref{sec_background} in the supplementary material. By the discretization of the parameter $t$, we obtain a sampling set on the Lie group. We formalize this idea with the following definition.


\begin{definition}\label{def_Ghat} 

Let $G$ be a Lie group and let $\mathfrak{g}$ be its associated Lie algebra. Let $\mathfrak{G}\subset\mathfrak{g}$ be a basis of $\mathfrak{g}$ as a vector space, and let $\widehat{G}_{\delta, N}$ be the subset of $G$ given by

\begin{equation}
\widehat{G}_{\delta, N}
         =
         \left\lbrace
              \left.
              e^{\delta n \mathfrak{X}} 
              \right\vert \mathfrak{X}\in\mathfrak{G}, n\in[-N,N], \delta\in\mbR^{+}
         \right\rbrace
         ,
\end{equation}
where $\delta$ and $N$ are fixed. We say that $\widehat{G}_{\delta, N}$ is a discrete approximation of $G$ with resolution $\delta$ and range $N\delta$.
\end{definition}


In general, the basis elements of Lie algebras corresponding to common Lie groups are well known. As such, we leverage the basis elements of the Lie Algebra equipped with the exponential map to obtain a sampling set on $G$ for the filters in $L^1(G)\cap L^{2}(G)$. We illustrate this with the special orthogonal group $\mathsf{SO}(2)$. It is known that the Lie algebra of $\mathsf{SO}(2)$ is the skew-symmetric matrices $\mathsf{so}(2)$, which is generated by the matrix 
$\mathfrak{X} =     \begin{bmatrix}
           0 & -1 \\
           1 & 0 
         \end{bmatrix}.$
By applying the exponential map on the generators of the Lie Algebra, we are able to recover the Lie group

\begin{equation}
\mathrm{exp}(\mathfrak{X}t) =    \begin{bmatrix}
           \cos(t) & -\sin(t) \\
           \sin(t) & \cos(t) 
         \end{bmatrix} \in \mathsf{SO}(2).
\end{equation}

By taking the sampling as described in Definition \ref{def_Ghat}, we obtain 

\begin{equation}
\mathrm{exp}(\delta n\mathfrak{X}) =    \begin{bmatrix}
           \cos(n\delta) & -\sin(n\delta) \\
           \sin(n\delta) & \cos(n\delta) 
         \end{bmatrix} \in \mathsf{SO}(2).
\end{equation}

In short, by applying the exponential map on the generators of the Lie algebra, we obtain a method to sample the filters.

The exponential map is both injective and surjective in a neighborhood around the identity for \emph{any} Lie group. In the particular case of $\mathsf{SO}(3)$, the exponential map is bijective \cite[Theorem 14.2.2]{gallier2001basics}. In general, however, the map is not bijective, so we invoke the following corollary.


\begin{corollary}[Corollary 3.47~\cite{hall_liealg}]\label{corll_group_decomp_exp} 

If $G$ is a connected matrix Lie group and $\mathfrak{g}$ its associated Lie algebra. Then, every element $g\in G$ can be written in the form
\begin{equation}
g =
   e^{\mathfrak{X}_1}e^{\mathfrak{X}_2}\cdots e^{\mathfrak{X}_{M(g)}}
   ,
\end{equation}
for some $\mathfrak{X}_1, \ldots, \mathfrak{X}_{M(g)} \in\mathfrak{g}$, and where $M(g)\in\mbN$.

\end{corollary}


Corollary \ref{corll_group_decomp_exp} shows that not only do the discrete samples of the exponential map need to be included in the sampling pattern but also their products. We, therefore, extend the exponential map of Definition \ref{def_Ghat} to include their monomials up to some order $k$. We formally define this below.


\begin{definition}\label{def_Ghat_monomials} 

Let $G$ be a Lie group and let $\hat G_{\delta, N}$ the finite subset of $G$ given according to Definition~\ref{def_Ghat} with resolution $\delta$ and range $\delta N$. The extended approximation of order $k$ of $\hat G_{\delta,N}$ is the set of monomials up to order $k$ on $\hat G_{\delta, N}$, i.e.
\begin{equation}
 \widehat{G}^k_{\delta, N} 
         := 
           \left\{
               \prod_i \left(
                          \hat{g}_i
                       \right)^{n_i}
                       \Bigg\vert \sum_i n_i \leq k, \hat{g}_i\in \widehat{G}_{\delta,N}
           \right\}
           .   
\end{equation}
\end{definition}


For different values of $k$, the set $ \widehat{G}^k_{\delta, N} $ defines sampling sets of increasing complexity and size, and as one might expect intuitively the higher the value of $k$ the larger the class of bandlimited signals one can uniquely represent on $\widehat{G}^k_{\delta, N}$. We formalize this in the following theorem.


\begin{theorem}\label{corr:add_degree} 

Let $G$ be a Lie group and let $\hat G^k_{\delta, N}$ be a sampling set on $G$ given according to Definition~\ref{def_Ghat_monomials}. It follows that for any $k'>k$,
\begin{equation}
 \omega\left(
           r^*_{\hat G^k_{\delta, N}}
        \right) 
           \leq 
\omega\left(
            r^*_{\hat G^{k'}_{\delta, N}}
        \right)  
        .            
\end{equation}
Similarly, if $N' > N$ it follows that for any $k'>k$, 
\begin{equation}
\omega\left(
          r^*_{\hat G^k_{\delta, N}}
       \right) 
       \leq 
\omega\left(
         r^*_{\hat G^{k^{'}}_{\delta, N'}}
       \right)  
       .
\end{equation}
\end{theorem}

\begin{proof}
    See Appendix \ref{sec:proof_thm_add_degree}
\end{proof}


Theorem~\ref{corr:add_degree} states that increasing the degree of the monomials considered or the range of the sampling set allows for more complexity -- a larger bandwidth -- in the sampled filter. Now suppose that the range $\delta N$ is fixed. Next, consider the sampling set where $\delta' = \delta /2$. To achieve the same range, it must be that $N' = 2N$ so that $N\delta = N'\delta' = 2N/(\delta/2)$. This can be generalized to any $\delta/h$ where $h \in \mbZ^{+}$. We formalize this with the following theorem.


\begin{theorem} \label{corr:decrease_resolution} 

Let $\widehat{G}^k_{\delta, N}$ be a sampling set on $G$. Consider the sampling set $\widehat{G}^k_{\delta^{'}, N^{'}}$ with $\delta' = \delta/ h$, $N' = hN$ and $h\in\mbZ^{+}$. Then,
\begin{equation}
\omega\left(
         r^*_{\hat G_{\delta, N}^k}
       \right) 
       \leq 
\omega\left(
         r^*_{\hat G_{\delta', N'}^k}
       \right). 
\end{equation}
\end{theorem}


\begin{proof}
    See Appendix~\ref{sec:proof_them_decrease_resolution}
\end{proof}


\begin{figure}
        \centering
\begin{tikzpicture}

\def\axisdefaultwidth{9cm}
\def\axisdefaultheight{5.562cm}

\definecolor{darkgray176}{RGB}{176,176,176}
\definecolor{steelblue31119180}{RGB}{31,119,180}
\definecolor{my_cp5_col1}{RGB}{253, 231, 37}
\definecolor{my_cp5_col2}{RGB}{180, 222,44}
\definecolor{my_cp5_col3}{RGB}{94, 201, 98}
\definecolor{my_cp5_col4}{RGB}{33, 145, 140}
\definecolor{my_cp5_col5}{RGB}{59, 82, 139}
\definecolor{my_cp5_col6}{RGB}{68, 1, 84}

\begin{axis}[
log basis x={10},
tick align=outside,
tick pos=left,
x grid style={darkgray176},
xlabel={\(\displaystyle \delta\)},
xmin=0.170932203389347, xmax=3.60874231324937,
xmode=log,
xtick style={color=black},
y grid style={darkgray176},
ylabel={\(\displaystyle 1/ r^*\)},
y label style={at={(axis description cs:0.1,.5)},rotate=0,anchor=south},
ymin=0.225765890800892, ymax=10.8213793672985,
ytick style={color=black}
]
\addplot [draw=black, fill=my_cp5_col5, mark=*, only marks]
table{%
x  y
3.14159265358979 0.707384685187147
2.0943951023932 1.0272756173643
1.5707963267949 1.10452482447043
1.25663706143592 1.66355207307081
1.0471975511966 1.52186293154756
0.897597901025656 2.32985386090474
0.785398163397448 1.95294679257257
0.698131700797732 3.01004295084412
0.628318530717959 2.40070545400498
0.571198664289053 3.67780731768262
0.523598775598299 2.8473461135398
0.483321946706122 4.35051164145169
0.448798950512828 3.30152813702414
0.418879020478639 4.9732835747762
0.392699081698724 3.77499566088433
0.369599135716446 5.63234504818631
0.349065850398866 4.2936688550935
0.330693963535768 6.37167045975337
0.314159265358979 4.77394980951399
0.299199300341885 6.94875669115158
0.285599332144527 5.21698054829118
0.273181969877373 7.67592910050012
0.26179938779915 5.65254375850435
0.251327412287183 8.24990311995419
0.241660973353061 6.12504046928963
0.232710566932577 9.02167576006639
0.224399475256414 6.59419449690012
0.216661562316538 9.62509329914296
0.209439510239319 7.02324629301415
0.202683397005793 10.3397605729123
0.196349540849362 7.54737833487019
};
\addplot [draw=black, fill=my_cp5_col2, mark=square*, only marks]
table{%
x  y
3.14159265358979 0.707384685187147
1.5707963267949 1.10452482447043
0.785398163397448 1.95294679257257
0.392699081698724 3.77499566088433
0.196349540849362 7.54737833487019
};
\end{axis}

\end{tikzpicture} 
        \caption{In general, as $\delta$ in $\widehat{G}^k_{\delta, N}$ increases, the bandwidth (represented here by the inverse of $r^*$) decreases. The green squares show the progression of the bandwidth when the $\delta$ is adjusted by a factor of two. As suggested by Theorem~\ref{corr:decrease_resolution}, the bandwidth is non-decreasing as $\delta$ decreases.}
        \label{fig:pesenson_sample}
\end{figure}
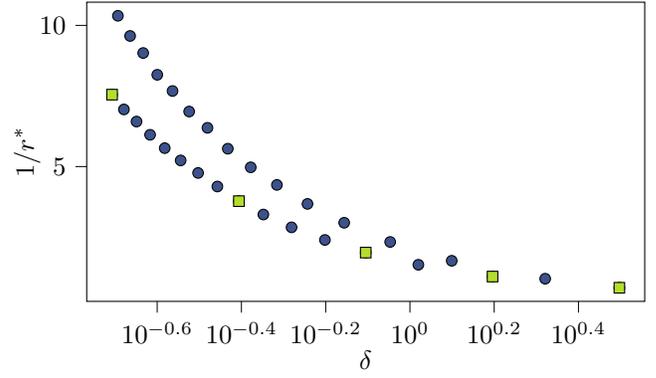


Theorem \ref{corr:decrease_resolution} shows that decreasing the value of $\delta$ by an integer factor while keeping the range the same allows for a larger bandwidth of those filters that are uniquely represented by their samples on $\widehat{G}$. Note that the value of $\delta$ does not need to be reduced by an integer factor, however, this does not necessarily result in an increased bandwidth as shown in Figure \ref{fig:pesenson_sample}.

By using the sampling set $\widehat{G}_{\delta,N}^{k}$, we have a bijective relationship between bandlimited filters in $ L^{2}(G)$ and their samples. We can then use the samples of the filter $\boldsymbol{a}\in L^{2}(G)$ on $\widehat{G}_{\delta,N}^{k}$ to define an approximate version of $\boldsymbol{a}$ that is uniquely associated to $\boldsymbol{a}$. To do this we take into account that the set of sampled filters on $\widehat{G}_{\delta,N}^{k}$ is bijective to the vector space 
\begin{equation}
L^{1}\left(
        \widehat{G}_{\delta,N}^{k}
     \right)
     =
 \left\lbrace 
   \left.
     \sum_{\hat{g}}
             \boldsymbol{a}(\hat{g})\hat{g}
   \right\vert 
          \boldsymbol{a}(\hat{g})\in\mbC,~\hat{g}\in\widehat{G}_{\delta,N}^{k}
 \right\rbrace
 .
\end{equation}
In the light of the Theorem~\ref{thm:pesenson} we approximate $\boldsymbol{a}$ by its unique image in 
$
L^{1}\left(
        \widehat{G}_{\delta,N}^{k}
     \right)
     ,
$
which we represent by $\widehat{\boldsymbol{a}}$ and which is given by
\begin{equation}\label{eq_afilter_hat}
 \widehat{\boldsymbol{a}}
    =
     \sum_{\hat{g}\in\widehat{G}_{\delta,N}^k}
             \boldsymbol{a}(\hat{g})\hat{g}
             .
\end{equation}

Then, we approximate the action of $\rho(\boldsymbol{a})$ on $f\in\ccalH$ by means of the realization of $\widehat{\boldsymbol{a}}$ via the linear map $\widehat{\rho}:L^{1}\left(\widehat{G}_{\delta,N}^{k}\right) \to \ccalB (\ccalH)$ as
\begin{equation} \label{equ:lieGAH_discrete}
\widehat{\rho}\left( 
         \widehat{\boldsymbol{a}} 
    \right) f
         = 
          \sum_{\hat{g}\in \hat G_{\delta,N}^{k}} \boldsymbol{a}(\hat{g}) \bbT_{\hat{g}} f.
\end{equation}

For homogeneous signals, the filter in \eqref{equ:lieGAH_discrete} can be directly implemented. In the next section, we explore the realization on non-homogeneous spaces of signals that are sampled from homogeneous spaces.


\subsection{Non-Expanding Approximation Error}

In this subsection, we discuss the approximation attributes associated to the homomorphism $\rho$ in any Lie group convolutional model $(L^{1}(G),\ccalH,\rho)$. We will show that the implementation of the learned filters in $\ccalB (\ccalH)$ carried out by $\rho$ has an approximation error that does not exceed the approximation error of the filters in $L^{1}(G)$ on a given sampling set.

As stated in~\cite{pesenson2000sampling}, given $\boldsymbol{a}\in L^{2}(G)$ and its samples on $\widehat{G}_{\delta,N}^{k}\subset G$, there is an interpolator function $\mathtt{I}_{\widehat{G}_{\delta,N}^{k}}^{\ell}\left( \boldsymbol{a} \right)$ of order $\ell$ such that
\begin{equation}\label{eq_interpolator_filter}
 \left\Vert 
        \boldsymbol{a}
        -
        \mathtt{I}_{\widehat{G}_{\delta,N}^{k}}^{(\ell)}(\boldsymbol{a})
 \right\Vert
        \leq 
        \left(
            C_{0}r^{*}_{\widehat{G}}\omega
        \right)^{\ell}
           \Vert
               \boldsymbol{a}
           \Vert 
           ,
\end{equation}
as long as $r^{*}_{\widehat{G}}$ and $\omega$ are selected according to Theorem~\ref{thm:pesenson}. Additionally, by Theorem~\ref{thm:pesenson} we have $C_{0}r^{*}_{\widehat{G}}\omega<1$, and therefore the accuracy of interpolation by $\mathtt{I}_{\widehat{G}_{\delta,N}^{k}}^{(\ell)}(\boldsymbol{a})$ increases with $\ell$.

We now show that the images of $\mathtt{I}_{\widehat{G}_{\delta,N}^{k}}^{(\ell)}(\boldsymbol{a})$ under a homomorphism $\rho$ in the algebraic signal model $(L^{1}(G),\ccalH,\rho)$,  define interpolators for $\rho(\boldsymbol{a})$ on $\ccalB\left( \ccalH\right)$ whose approximation error is bounded by the approximation error associated to $\mathtt{I}_{\widehat{G}_{\delta,N}^{k}}^{(\ell)}(\boldsymbol{a})$ in $L^{1}(G)$.


\begin{theorem}\label{thm_imageinterpolator}
 Let $\boldsymbol{a}\in L^{2}(G)$ be a Lie algebraic filter with bandwidth $\omega$ and let $\widehat{G}_{\delta,N}^{k}\subset G$ be a sampling set on $G$. If $\omega$ and $\widehat{G}_{\delta,N}^{k}\subset G$ are selected according to Theorem~\ref{thm:pesenson} and $\mathtt{I}_{\widehat{G}_{\delta,N}^{k}}^{(\ell)}(\boldsymbol{a})$ is an interpolator of $\boldsymbol{a}$ of order $\ell$ it follows that
\begin{equation}
\left\Vert 
      \rho\left( 
             \boldsymbol{a}
          \right)
      -    
      \rho\left( \mathtt{I}_{\hat{g}}^{(\ell)}
                \left(
                            \boldsymbol{a}
                         \right)   
                \right)
\right\Vert
        \leq
        \left(
            C_{0}r^{*}_{\widehat{G}}\omega
        \right)^{\ell}
        \Vert 
            \boldsymbol{a}
        \Vert
        .
\end{equation}

\end{theorem}

\begin{proof}
See Appendix~\ref{sec_proof_of_thm_imageinterpolator}.
\end{proof}

Theorem~\ref{thm_imageinterpolator} has two fundamental consequences. First, it shows that the terms $\rho\left( \mathtt{I}_{\hat{g}}^{(\ell)}\left(\boldsymbol{a}\right)   \right)$ are interpolators of $\rho(\boldsymbol{a})$ in $\ccalB (\ccalH)$. Second, the error associated to the interpolator $\rho\left(\mathtt{I}_{\hat{g}}^{(\ell)}\left(\boldsymbol{a}\right)\right)$ is bounded by the error of the interpolator of $\boldsymbol{a}$ in $L^{2}(G)$. This is, the error associated to the sampling set $\widehat{G}_{\delta,N}^{k}$ for the reconstruction of $\boldsymbol{a}$ from its samples on $\widehat{G}_{\delta,N}^{k}$ is preserved by $\rho$. This emphasizes the usefulness and importance of modeling convolutions in an ASM.

Notice that while the uniqueness results and the error bounds derived are for filters in $L^{2}(G)$, it is possible to obtain a set of step functions built from the sampling points in $\widehat{G}^{k}_{\delta,N}$. Since step (simple) functions constitute a dense set in $L^{1}(G)$~\cite[p.~5]{stein2011functional}, this implies that in the limit, the family of step functions built from $\widehat{G}^{k}_{\delta,N}$ converges to an element in $L^{1}(G)$ as well. This is, a step function built from $\widehat{G}^{k}_{\delta,N}$ approximates elements in $L^{1}(G)$ although the dependency of the error estimate in terms of the bandwidth is unknown. Additionally, it is worth pointing out that a different choice of basis for $\mathfrak{g}$ leads to a different sampling pattern on $G$ since the sampling points are determined by the exponential map $e^{\delta n \mathfrak{B}} $ where $\mathfrak{B}$ is the basis selected for $\mathfrak{g}$. However, it is important to remark that given a fixed number of samples, the factor that determines the bandwidth of the signals that can be represented uniquely is the minimum distance between sampling points and not the location of the sampling points themselves. This is indeed something that can be derived from Theorem~\ref{thm:pesenson}.


\section{Discretization of the Signals Space} 
\label{sec:discrete_sample}


\begin{figure*}[t]
	\centering
	\begin{subfigure}{.48\textwidth}
		\centering


\definecolor{my_cp5_col1}{RGB}{253, 231, 37}
\definecolor{my_cp5_col2}{RGB}{94, 201, 98}
\definecolor{my_cp5_col3}{RGB}{33, 145, 140}
\definecolor{my_cp5_col4}{RGB}{59, 82, 139}
\definecolor{my_cp5_col5}{RGB}{68, 1, 84}

\usetikzlibrary{positioning,decorations.pathreplacing,shapes}
\usetikzlibrary{arrows}


\tikzstyle{dot} = [ circle,
                    minimum width  = 0.05*\unit,
                    fill=black,
                    color=black,
                    inner sep=0pt,
                    draw,
                    anchor = center ]

\tdplotsetmaincoords{0}{0} 
\tdplotsetrotatedcoords{0}{0}{0}

\def\gridnumber{6}
\def\imagesize{4}
\def\imagedotsize{5}

\def\scale{0.45}

{\fontsize{6}{6}\selectfont

\begin{tikzpicture}[scale=\scale, tdplot_rotated_coords, grid/.style={very thin,gray}]



\foreach \x in {0,1,...,\gridnumber}
\foreach \y in {0,1,...,\gridnumber}
{
	\draw[grid,rotate=0,my_cp5_col3,line width=1*\scale] (\x,0) -- (\x,\gridnumber);
	\draw[grid,rotate=0,my_cp5_col3,line width=1*\scale] (0,\y) -- (\gridnumber,\y);
};

\foreach \x in {1,...,\imagesize}
\foreach \y in {1,...,\imagesize}
{
\draw[fill=my_cp5_col2, opacity=1]  (\x,\y,0) -- (\x+1,\y,0) -- (\x+1,\y+1,0) -- (\x,\y+1,0) -- cycle;
}


\begin{scope}[shift={(\gridnumber+1.6,-0.5)},rotate=10]
\foreach \x in {1,...,\imagesize}
\foreach \y in {1,...,\imagesize}
{
	\draw[fill=my_cp5_col2, opacity=1]  (\x,\y,0) -- (\x+1,\y,0) -- (\x+1,\y+1,0) -- (\x,\y+1,0) -- cycle;
}
\end{scope}

\begin{scope}[shift={(\gridnumber+1,0)},rotate=0]
\foreach \x in {0,1,...,\gridnumber}
\foreach \y in {0,1,...,\gridnumber}
{
	\draw[grid,rotate=0,my_cp5_col3,line width=1*\scale] (\x,0) -- (\x,\gridnumber);
	\draw[grid,rotate=0,my_cp5_col3,line width=1*\scale] (0,\y) -- (\gridnumber,\y);
};
\end{scope}



\begin{scope}[shift={(2*\gridnumber+2,0)},rotate=0]
\foreach \x in {0,1,...,\gridnumber}
\foreach \y in {0,1,...,\gridnumber}
{
	\draw[grid,rotate=0,my_cp5_col3,line width=1*\scale] (\x,0) -- (\x,\gridnumber);
	\draw[grid,rotate=0,my_cp5_col3,line width=1*\scale] (0,\y) -- (\gridnumber,\y);
};
\end{scope}

\begin{scope}[shift={(2*\gridnumber+2,0)},rotate=0]
\foreach \x in {1,...,\imagesize}
\foreach \y in {1,...,\imagesize}
{
	\draw[fill=my_cp5_col3, opacity=1]  (\x,\y,0) -- (\x+1,\y,0) -- (\x+1,\y+1,0) -- (\x,\y+1,0) -- cycle;
};
\end{scope}



\begin{scope}[shift={(0,-\gridnumber-1)},rotate=0]
\foreach \x in {0,1,...,\imagedotsize}
\foreach \y in {0,1,...,\imagedotsize}
{
	\pgfmathtruncatemacro{\label}{\x + 5 * (4 - \y)+6};
	\node[dot,draw=black,fill=white,minimum size=15*\scale, opacity=0.5]
	(\label) at (\x+0.5,\y+0.5) {\label};
}
\end{scope}

\begin{scope}[shift={(0,-\gridnumber-1)},rotate=0]
\foreach \x in {1,...,\imagesize}
\foreach \y in {1,...,\imagesize}
{
	\pgfmathtruncatemacro{\label}{\x + 5 * (4 - \y)+6};
	\node[dot,draw=black,fill=my_cp5_col2,minimum size=15*\scale, opacity=0.7]
	(\label) at (\x+0.5,\y+0.5) {\label};
}
\end{scope}


\begin{scope}[shift={(\gridnumber+1,-\gridnumber-1)},rotate=0]
\foreach \x in {0,1,...,\imagedotsize}
\foreach \y in {0,1,...,\imagedotsize}
{
	\pgfmathtruncatemacro{\label}{\x + 5 * (4 - \y)+6};
	\node[dot,draw=black,fill=white,minimum size=15*\scale, opacity=0.5]
	(u\label) at (\x+0.5,\y+0.5) {\label};
}
\end{scope}

\begin{scope}[shift={(\gridnumber+1.6,-\gridnumber-1-0.5)},rotate=10]
\foreach \x in {1,...,\imagesize}
\foreach \y in {1,...,\imagesize}
{
	\pgfmathtruncatemacro{\label}{\x + 5 * (4 - \y)+6};
	\node[dot,draw=black,fill=my_cp5_col2,minimum size=15*\scale, opacity=0.7]
	(v\label) at (\x+0.5,\y+0.5) {\label};
}
\end{scope}


\node[circle,fill=my_cp5_col4,draw,opacity=0.4,inner sep=20*\scale] () at (u7){};

\node[dot,draw=black,fill=my_cp5_col4,minimum size=15*\scale, opacity=1,text=white](u7) at (u7) {7};


\begin{scope}[shift={(2*\gridnumber+2,-\gridnumber-1)},rotate=0]
\foreach \x in {0,1,...,\imagedotsize}
\foreach \y in {0,1,...,\imagedotsize}
{
	\pgfmathtruncatemacro{\label}{\x + 5 * (4 - \y)+6};
	\node[dot,draw=black,fill=white,minimum size=15*\scale, opacity=0.5]
	() at (\x+0.5,\y+0.5) {\label};
}
\end{scope}

\begin{scope}[shift={(2*\gridnumber+2,-\gridnumber-1)},rotate=0]
\foreach \x in {1,...,\imagesize}
\foreach \y in {1,...,\imagesize}
{
	\pgfmathtruncatemacro{\label}{\x + 5 * (4 - \y)+6};
	\node[dot,draw=black,fill=my_cp5_col3,minimum size=15*\scale, opacity=0.7]
	() at (\x+0.5,\y+0.5) {\label};
}
\end{scope}

%
%
%
%
%
%
%

\end{tikzpicture}

} 
		\caption{}
		\label{fig_rotation_image_example}
	\end{subfigure}
 \hfill
	\begin{subfigure}{.48\textwidth}
		\centering

\definecolor{my_cp4_col1}{RGB}{255, 86, 87}
\definecolor{my_cp4_col2}{RGB}{55, 108, 138}
\definecolor{my_cp4_col3}{RGB}{242, 217, 187}
\definecolor{my_cp4_col4}{RGB}{99, 143, 169}

\definecolor{my_cp5_col1}{RGB}{253, 231, 37}
\definecolor{my_cp5_col2}{RGB}{94, 201, 98}
\definecolor{my_cp5_col3}{RGB}{33, 145, 140}
\definecolor{my_cp5_col4}{RGB}{59, 82, 139}
\definecolor{my_cp5_col5}{RGB}{68, 1, 84}


\usetikzlibrary{positioning,decorations.pathreplacing,shapes}
\usetikzlibrary{arrows}


\def \scale {1.4}
\def \unit { \scale cm}

\def \vertsep {0.5*\scale}

\def \horzsep {0.5*\scale}

\def \mylinewidth {0.5*\scale}


\tikzstyle{set} = [rectangle,color=black,
rounded corners = 0.2*\unit,
fill=black,
inner sep=0pt,
draw,
anchor = center,
line width=0.1mm]

\tikzstyle{vectorspace} = [ set, 
fill=my_cp5_col1,
minimum width  = 1.61*\unit,
minimum height = 1*\unit]

\tikzstyle{endomorphisms} = [ vectorspace,
fill=black!15,
minimum width  = 1.61*\unit,
minimum height = 1*\unit]

\tikzstyle{subendomor2in1} = [endomorphisms, rounded corners=0.2*\unit,
fill=black!30,
minimum width  = 1*\unit,
minimum height = 0.62*\unit]      

\tikzstyle{algebra} = [ endomorphisms,
fill=my_cp5_col1,
minimum width = 1.5*\unit]

\tikzstyle{group_math} = [endomorphisms,
minimum width = 1.5*\unit,
minimum height = 0.5*\unit]

\tikzstyle{dot} = [ circle,
minimum width  = 0.05*\unit,
fill=black,
color=black,
inner sep=0pt,
draw,
anchor = center ]



{\fontsize{8}{8}\selectfont
	
\begin{tikzpicture}[scale=\scale,rounded corners,ultra thick]

	
	\path (0,0) node [algebra, anchor = north east, fill=my_cp5_col2] (LG) {};   
	\path (LG.north) ++ (0,0) node [above,color=black] {$L^{1}\left( \widehat{G}_{\delta,N}^{k}\right)$}; 
	
	\path (LG.south) ++ (0,0.3) node [dot] (a) {};
    \path (a.north) ++ (0,0) node [above,color=black] {$\hat{g}$};


		
	\path (LG.east) ++ (\horzsep, 0) 
	node [endomorphisms, anchor=west, fill=my_cp5_col3] (End1) {};
	\path (End1.north) ++ (0.0, 0) node [above, color=black] {$\mathcal{B}(\ccalH)$}; 
		
	 \path (End1.south) ++ (0.0, 0.3) node [dot] (e1) {};      
	\path (e1) node [above] {$\rho(\hat{g})=\bbT_{\hat{g}}$};

	
	\path (End1.south)++(0,-\vertsep) node [vectorspace, anchor=north, fill=my_cp5_col4] (M1) {};
	\path (M1.south) ++ (0, 0.0) node [below, color=black] {$\ccalH$};

		 
		 \path (M1.center) ++ (-0.2, 0.0) 
		 node [subendomor2in1, anchor = center, fill=my_cp5_col4!50] (V) {}; 
		 \path (V.north) ++ (0,-0.1) node [above,color=black,my_cp5_col1] {$\overline{\mathcal{H}}$};

	   \path (M1) ++ (-0.5,0.2) node [dot] (x) {};
	   \path (x.south)++(0,0) node [below, color=black] {$f$}; 
	   
	   	   \path (x)++(0.7,0)  node [dot] (xint) {};
	   	   \path (xint) node [below, color=black] {$\tau\left(\mathbf{z}\right)$}; 
	   
	   \path (M1) ++ (0.6,0.2) node [dot,color=my_cp5_col2] (ex) {};
	   \path (ex.south) node [below, color=my_cp5_col2] {$\mathbf{z}$};      
	   
		\path (e1)+(-0.5,0.45) coordinate (c1);
		\path (e1)+(0.5,0.45) coordinate (c2);   
		\path [draw, -stealth, line width=\mylinewidth,color=my_cp5_col2] (x) .. controls (c1) and (c2) .. (ex) node [very near end, opacity=1,color=black]{$\mathbf{z}=\bbT_{\hat{g}}f$};

	    \path [draw, -stealth, line width=\mylinewidth,color=my_cp5_col1] (ex) edge[bend right,above, midway] node [above,midway] {$\tau$} (xint) ;



\path (LG.west) ++ (-\horzsep, 0) 
node [endomorphisms, anchor=east,fill=my_cp5_col3] (End2) {};
\path (End2.north) ++ (0.0, 0) node [above, color=black] {$\mathcal{B}\left( \widehat{\ccalH}\right)$}; 

 Adding an endomorphism element 
 \path (End2.south) ++ (0.0, 0.3) node [dot] (e2) {};      
\path (e2) node [above] {$\widehat{\rho}(\hat{g})=\widehat{\bbT}_{\hat{g}}$};    



\path (End2.south)++(0,-\vertsep) node [vectorspace, anchor=north,fill=my_cp5_col4] (M2) {};
\path (M2.south) ++ (0, 0) node [below, color=black] {$\widehat{\ccalH}$};

   \path (M2) ++ (-0.5,0.2) node [dot,color=my_cp5_col2] (y) {};
   \path (y.south)++(0,-0.1) node [below, color=my_cp5_col2] {$\widehat{f}$}; 
   
   \path (M2) ++ (0.5,0.2) node [dot,color=my_cp5_col2] (ey) {};
   \path (ey.south) node [below, color=my_cp5_col2] {$\widehat{\bbT}_{\hat{g}} \widehat{f}$};      
   
		\path (e2)+(-0.5,0.45) coordinate (c1);
		\path (e2)+(0.5,0.45) coordinate (c2);   
		\path [draw, -stealth, line width=\mylinewidth,color=my_cp5_col2] (y) .. controls (c1) and (c2) .. (ey);

	
    	
	\path [draw, -stealth, line width = \mylinewidth, color=my_cp5_col5] 
	(a) edge [bend right] node [below] {$\rho~~~$} (e1);

	    
   \path [draw, -stealth, line width = \mylinewidth, color=my_cp5_col5] 
	    (a) edge [bend left] node [below] {$\widehat{\rho}~~~$} (e2);

	
	  \path [draw, -stealth, line width = \mylinewidth, color=black] 
	   (y) edge [bend left] node [below] {$\theta~~~$} (x);
	
     \path [draw, -stealth, line width = \mylinewidth, black] 
		  (xint) edge [bend left] node [below] {$\theta^{-1}~~~$} (ey);

	\end{tikzpicture}

} 
		\caption{}
		\label{fig_asp_interpolation}
	\end{subfigure}
 %
 %
 \caption{The action of Lie group on a discrete space of functions and its representation in terms of an algebraic signal model. Left (top row): Let $\ccalH$ the space of scalar functions whose domain is $\mbR^2$, i.e. $\ccalH = \{ f:\mbR^2 \to \mbR  \} $. The image $f$ is modeled as an element in a vector space, $\overline{\ccalH}\subset\ccalH$, of piecewise constant functions on $\mbR^{2}$ with support on a specific grid. Then, $f$ is rotated by the action of an element in $g\in\mathsf{SO}(2)$ and using interpolation the element $\bbT_{g}f\notin\overline{\ccalH}$ is mapped back into $\overline{\ccalH}$. Left (low row): A depiction of how interpolation is carried out to ensure $\bbT_{g}f\notin\overline{\ccalH}$ is mapped to $\overline{\ccalH}$. Right: Diagram representing how interpolation is carried out considering algebraic signal models. To realize or implement the action of a monomial $\hat{g}\in L^{1}\left( \widehat{G}_{\delta,N}^{k}\right)$ in $\ccalB( \widehat{\ccalH} )$ we use the realization of $\hat{g}$ on $\ccalB (\ccalH)$ and the isomorphism, $\theta$, between $\widehat{\ccalH}$ and $\overline{\ccalH}\subset\ccalH$. Then, the action $\bbT_{g}( \theta (\widehat{f}))$ is well defined and the interpolation map $\tau$ puts 
 $\bbT_{g}( \theta (\widehat{f}))$ in $\overline{\ccalH}$ and then is mapped back in $\overline{\ccalH}$ by means of $\theta^{-1}$.
 }
\end{figure*}
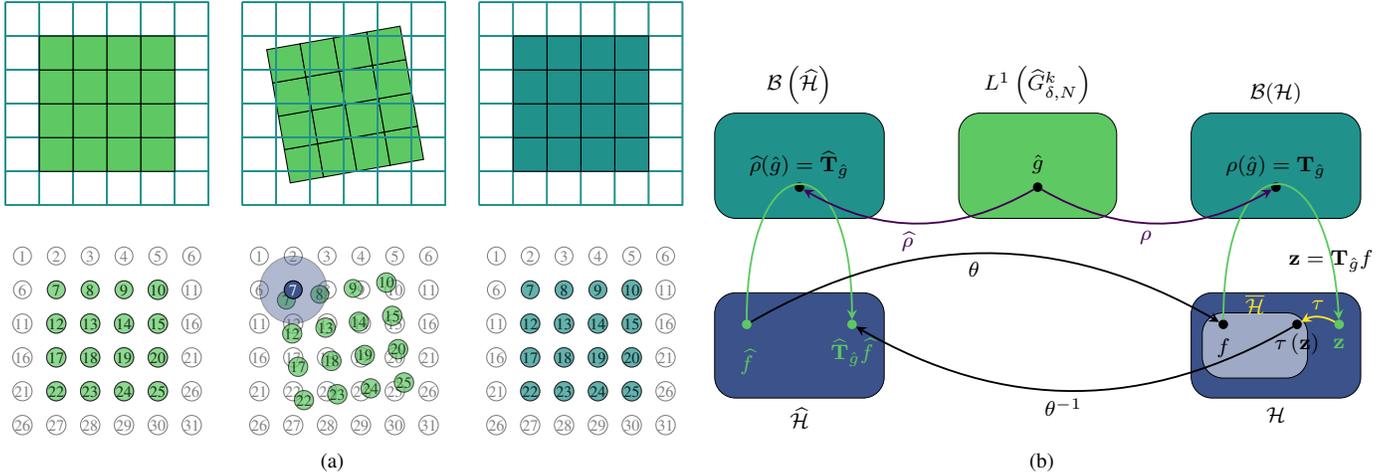


Up until this point, we have considered spaces of signals, $\ccalH$, where the action of $\bbT_g f$ is well defined, i.e. for any $f\in\ccalH$ and any $g\in G$ we have $\bbT_{g}f\in\ccalH$. In several real-life applications, this is an unlikely scenario. For instance, images are intended to describe patterns that are ideally functions on $\mbR^2$, however, their realization comes as a set of functions on a discrete grid --see Figure~\ref{fig_rotation_image_example}. Then, the action of a rotation group on the set of functions on the discrete grid is not well defined for all the rotations.

To see this formally, let $\widehat{\ccalX}\subset\mbR^2$ be the subset of points located on the grid in $\mbR^2$ whose lattice points are defined by integer coordinates. If $\ccalH$ is the set of signals defined on $\mbR^2$ and $\ccalM = \{ f: \mbR^2 \to \mbC,~\text{with}~f(\mbR^2\setminus\widehat{\ccalX})=0 \}$, then we have $\ccalM\subset\ccalH$. Images can be seen as particular elements of $\ccalM$. Now, let $G = \mathsf{SO}(2)$ be the group of rotations on the circle acting on $\mbR^2$. Then, for any $g\in G$ and any $f\in\ccalH$ we have $\bbT_{g}f \in\ccalH$. As indicated in Fig.~\ref{fig_rotation_image_example} if $f\in\ccalM$ there are rotations for which we have $\bbT_{g}f\notin\ccalM$. To circumvent this, we propose the use of interpolation to estimate the induced group transformation. In this section, we describe interpolation through the ASP lens. Interpolation on images to approximate affine transformations is a well-studied problem with many applications~\cite{lehmann1999survey}.

We will now formally describe the process of interpolation from the ASP lens. To do this, we will consider a set of ideal signals, $\ccalH$, and a set of discrete signals, $\widehat{\ccalH}$. The elements in $\widehat{\ccalH}$ are discretized versions of the signals in $\ccalH$. The elements in $\ccalH$ are functions defined on a subset $\ccalX\subset\mbR^n$ such that $\bbT_g f\in\ccalH$ for any $g\in G$. The elements in $\widehat{\ccalH}$ are obtained as the functions defined on $\widehat{\ccalX} = \{  x_i \in \ccalX  \}_{i=1}^{N}$. Then, for each $f\in\ccalH$ there is $\widehat{f} = f(\widehat{\ccalX}) \in \widehat{\ccalH}$. Naturally, we will refer to $\widehat{f}$ as the discrete version of $f$. 

There is a space of piecewise constant signals with domain on $\ccalX$ denoted by $\overline{\ccalH}\subset\ccalH$, which is isomorphic to $\widehat{\ccalH}$. To state the piecewise constant characterization of the elements in $\overline{\ccalH}$ we consider a Voronoi cell representation of $\ccalX$ using the elements in $\widehat{\ccalX}$. In particular, we build the Voronoi cells on $\ccalX$ around each point $x_i \in\widehat{\ccalX}$ using the inner product metric $d(a, b) := \sqrt{\langle a-b, a - b\rangle}$ to obtain
\begin{equation} \label{equ:voronoi_cells}
    \ccalC_i 
        := 
           \left\lbrace
                x\in \ccalX \vert d(x, x_i) < d(x, x_j) ~\textrm{for all}~ i \neq j
            \right\rbrace
            .
\end{equation}
Then, given the Voronoi cells $\ccalC_i$, for $i = 0, \dots N-1$, we specify the elements in $\overline{\ccalH}$ as
\begin{equation}\label{eq_step_functions_H}
\overline{\ccalH}
        := 
           \left\{
               f\in \ccalH | f(x)= c_i \in \mbC ~\textrm{for all}~ x \in \ccalC_i
            \right\}.
\end{equation}
We emphasize that an element belongs to $\overline{\ccalH}$ with the overline symbol. Then, if $f$ is an arbitrary signal in $\ccalH$, we denote its discrete version in $\widehat{\ccalH}$ by $\widehat{f}$ and its piecewise constant representation in $\overline{\ccalH}$ by $\overline{f}$. If we denote the isomorphism between $\overline{\ccalH}$ and $\widehat{\ccalH}$ by $\theta$ we have $\theta\left( \widehat{f}\right) = \overline{f}$.

As we mentioned before, for some group elements $g\in G$ and some $\widehat{f}\in\widehat{\ccalH}$ we have $\bbT_g \widehat{f}\notin\widehat{\ccalH}$. However, we always have that $\bbT_g \overline{f}\in\ccalH$. We leverage this fact to actually carry out the action of $g\in G$ on $\widehat{f}\in\widehat{\ccalH}$. For a given group element $g\in G$ let us denote by $\widehat{\bbT}_{g}$ the induced action of $g$ on $\widehat{\ccalH}$ by means of $\theta$ and $\overline{\ccalH}$. Then, we have that
\begin{equation}\label{eq_approx_Tg}
\widehat{\bbT}_{g}
    \left( \widehat{f} \right)
         \equiv 
              \theta^{-1}\left( 
                              \bbT_{g} \left(
                                           \theta \left(
                                                         \widehat{f}
                                                   \right)
                                       \right) 
                         \right)
                        .    
\end{equation}
Notice that~\eqref{eq_approx_Tg} is well defined as long as $\bbT_{g} \left(\theta \left( \widehat{f}\right)\right)\in\overline{\ccalH} $. Since this is not guaranteed in general, interpolation is necessary. In algebraic terms, the interpolation can be represented by a map $\tau:\ccalH \to \overline{\ccalH}$, where $\tau\vert_{\overline{\ccalH}} = \text{Id}$ and as expected, the map $\tau$ is not injective. Taking into account $\tau$ we can re-write~\eqref{eq_approx_Tg} as
\begin{equation}\label{eq_approx_Tg_interp}
\widehat{\bbT}_{g}
    \left( \widehat{f} \right)
         \equiv 
              \theta^{-1}\left( 
                     \tau\left(
                              \bbT_{g} \left(
                                           \theta \left(
                                                         \widehat{f}
                                                   \right)
                                       \right) 
                          \right)             
                         \right)
                        .    
\end{equation}

Now, let us recall~\eqref{equ:lieGAH_discrete} where we approximate the implementation of $\boldsymbol{a}\in L^{1}(G)\cap L^{2}(G)$ by means of $\widehat{\boldsymbol{a}}$ -- see~\eqref{eq_afilter_hat} -- which is uniquely defined from the samples on $\widehat{G}_{\delta,N}^{k}\subset G$. We can then obtain an approximate implementation of $\boldsymbol{a}$ as
\begin{equation} \label{equ:lieGAH_discrete_final}
\widehat{\rho}\left( 
                 \widehat{\boldsymbol{a}}
              \right)
           \widehat{f}
           = 
          \sum_{\hat{g}\in \hat G_{\delta,N}^{k}} \boldsymbol{a}(\hat{g}) \widehat{\bbT}_{\hat{g}} \left(
                                      \widehat{f}
                                    \right)  
          ,
\end{equation}
where $\widehat{\bbT}_{\hat{g}}\left(\widehat{f}\right)$ is given according to~\eqref{eq_approx_Tg_interp}. Due to the interpolation, there is a loss of information in $\widehat{\rho}$. We note, however, that this error is minimized by the choice of interpolation schemes, which are well studied both for scattered points as well as structured grids~\cite{davis1975interpolation,mitas1999spatial,han2013comparison,chen2019interpolation}.

We describe the formal procedure to compute the transformation which we denote by $\widehat{\bbT}_{\hat{g}}$ in Algorithm \ref{alg:Transformation}. It is important to emphasize that the transformation should be computed offline, as it depends entirely on the geometry of the sampled Hilbert space. 

Furthermore, it is true that in general, the number of neighbors considered for the interpolation is much smaller than the number of sampled points (see Figure \ref{fig_rotation_image_example}). As such, the matrix $\widehat{\bbT}_{\hat{g}}$ is sparse, and does not take up much memory even though it belongs to $\mbC^{M \times M}$, where $M = |\widehat{G}^k_{\delta, N}|$. The choice of the interpolation scheme is up to the designer. In Algorithm \ref{alg:BarycentricInterpolator}, we show the generalized Barycentric interpolation used in our numerical experiments with $\mathsf{SO}(3)$ (See section \ref{sec:experiments}). Generalized Barycentric interpolation is equivalent to natural neighbor interpolation \cite{bobach2009natural}.


\begin{remark}\normalfont
\label{rem:comparison_finzi}

Although the computation of Lie group convolutions is most often performed through spectral representations, other contributions estimate \eqref{equ_rho_ideal} directly as we do here; see e.g., \cite{finzi2020generalizing}. These contributions require that input and output signals be represented as group signals. Namely, that we have a collection of values $f(g)$ associated with each element of the group. Since signals are not in general given in this format a computationally expensive lifting process is required for each input. This is different from what we do here where we use ASP concepts to decouple the filter algebra from the vector space of signals. This results in approximations of Lie group convolutions with scalable computational cost.

\end{remark}



\subsection{Constructing the induced operator}

In this subsection, we describe the process of building each of the operators $\widehat{\bbT}_{\hat g}$ in~\eqref{equ:lieGAH_discrete_final} for each element in the sampling set $\widehat{G}^k_{\delta, N}$. The central idea is to estimate $\widehat{\bbT}_{\hat g}$ considering the effect of the transformations $T_g$ on $\ccalX$ when using $\ccalX$ as a reference. First, we initialize a sparse $M\times M$ matrix $\widehat{\bbT}_{\hat g}$ with $M = |\widehat{G}^k_{\delta, N}|$. The transformation $T_g$ is then applied to each $x_i\in \widehat{\ccalX}$, and we denote this set by $T_g \widehat{\ccalX}$.
Then, for each element $x_i \in \widehat{\ccalX}$, we use an interpolation scheme to find its neighbors in the group-actioned set $T_g\widehat{\ccalX}$. We let $\mathsf{N}$ denote the set of the indices of $k$-nearest neighbors. An interpolation scheme is used to find a collection of weights $\mathsf{W}:\mathsf{N}\to \mbR$. The sparse matrix $\widehat{\bbT}_{\hat g}$ is populated by $\widehat{\bbT}_{i, n_j} \leftarrow \mathsf{W} (n_j)$, where $i$ denotes the element of $x_i\in \widehat{\ccalX}$. Note that in the process described, the obtained matrix $\widehat{\bbT}_{\hat g}$ acts directly on the elements of $\widehat{\ccalH}$ and such values are obtained exploiting the geometry of $\widehat{\ccalX}$ and $T_{g}\widehat{\ccalX}$. This procedure is summarized in Algorithm \ref{alg:Transformation}. We also provide an example of the interpolation scheme for Barycentric interpolation in Algorithm \ref{alg:BarycentricInterpolator}, which is achieved by solving the system of equations shown in step three. It is important to remark that there is no dependence on the signal $\widehat{f}$ in either algorithm; this permits the operators to be constructed offline.


\begin{algorithm}[t]
\caption{Offline transformation operator generation}
\begin{algorithmic}[1]
\Require Group action $\hat g\in \widehat{G}^k_{\delta, N}$, \textbf{InterpScheme}, sampling points $\widehat{\ccalX}$
\State Initialize sparse matrix $\widehat{\bbT}_{\hat g} \in \mbR^{M \times M}$, where $M = |\widehat{G}^k_{\delta, N}|$
\State Apply the transformation $T_g$ to all $x_i \in \widehat{\ccalX}$, $i = 0, \dots, M-1$.
\For{$i = 0, \dots, M -1$}
\State $\mathsf{N}, \mathsf{W} \leftarrow$ \textbf{InterpScheme}$(x_i, T_g \widehat{\ccalX}, k)$
\For{$n_j\in \mathsf{N}$ }
\State Assign $\widehat{\bbT}_{i,n_j} \leftarrow \mathsf{W}(n_j)$
\EndFor
\EndFor
\State\Return $\widehat{\bbT}_{\hat g}$
\end{algorithmic}
\label{alg:Transformation}
\end{algorithm}



\begin{algorithm}[t]
\caption{InterpScheme: Barycentric Interpolation}
\begin{algorithmic}[1]
\Require Query $x_i$, Transformed points $T_g \widehat{\ccalX}$, number of neighbors $k$
\State Compute the distance of $x_i$ to each point in $T_g \widehat{\ccalX}$
\State Find indices of the $k$ smallest distances $n_0, \dots, n_{k-1}$ 
\State Solve the system of equations for $\pmb{\lambda}$
$$\begin{bmatrix}
1 & 1 & 1 & \dots & 1\\
x_{n_0}^0 & x_{n_1}^0 & x_{n_2}^0 & \dots & x_{n_{k-1}}^0 \\
\vdots & \vdots & \vdots & \ddots & \vdots\\
x_{n_0}^p & x_{n_1}^p & x_{n_2}^p & \dots & x_{n_{k-1}}^p \\
\end{bmatrix}
\begin{bmatrix}
\lambda_0 \\ \lambda_1 \\ \vdots \\ \lambda_{k-1} 
\end{bmatrix}
=
\begin{bmatrix}
x_i^0 \\ x_i^1 \\ \vdots \\ x_i^p
\end{bmatrix}
$$
\State \Return $\mathsf{N}=n_{0, \dots, k-1}$ and weights $\mathsf{W}$ given by $n_i \mapsto \lambda_i$
\end{algorithmic}
\label{alg:BarycentricInterpolator}
\end{algorithm}





\section{Equivariance and Stability} \label{sec:equivariance_and_stability}

Much of the recent interest in group convolutional neural networks stems from the desire to exploit the property of equivariance - that applying a group action on a signal then applying the convolution is equal to applying the convolution first then applying the group action\cite{cohen2016group}. Although it is well known that group convolutions on homogeneous spaces are equivariant, the class of networks which are truly mathematically equivariant is actually relatively small. The main restriction in almost all equivariant proofs is that the domain of both the filter and the signal is the group \cite{kondor2018generalization,finzi2020generalizing}. Our proposed framework is a generalization of conventional group convolutions. Naturally, it follows that we recover equivariance when we apply the ASM to group signals, discussed in Remark \ref{rem:conventional_grp_conv}. This is elaborated on in Section \ref{sec:equivariance}.

We additionally consider the notion of filter stability - when the deformation of the operator is proportional to the size of the deformation. Stability has been explored through the lens of graph signal processing \cite{gama2020stability, bruna2013invariant} and recently generalized to ASMs \cite{parada2021convolutional,msp_journal,msp_conference}. In section \ref{sec:multigraphs}. we draw connections of our proposed \emph{discretized} filter \eqref{equ:lieGAH_discrete_final} to multigraphs. Unlike equivariance, the established stability encapsulates the sampling and interpolation errors incurred.


\subsection{Conditions which allow for equivariance} \label{sec:equivariance}

Recall the definition of equivariance. Let us consider the map $\varphi:\ccalH \to \ccalH$ from a space of signals $\ccalH$ onto itself. Let $G$ be a Lie group and let $\bbT_g$ the action of $g\in G$ on $\ccalH$ for a given representation of $G$ on $\ccalB (\ccalH)$. We say that $\varphi$ is equivariant to the actions of the group $G$ if 
\begin{equation}\label{eq_equivariance_condition}
\varphi (\bbT_g f) = \widetilde{\bbT}_{g}\left( \varphi (f)  \right)
 ,
\end{equation}
where $\widetilde{\bbT}_{g}$ is the action of $g$ on $\ccalH$ for possibly another representation of $G$ on $\ccalH$ \cite{cohen2016group,hall_liealg}. Taking into account this concept we say that the Lie algebraic filter $\boldsymbol{a}\in L^{1}(G)$ acting on the space of signals $\ccalH$ is equivariant to the action of $g\in G$ if $\widetilde{\bbT}_{g} \rho(\boldsymbol{a})f = \rho(\boldsymbol{a})\bbT_{g}f$,
%
%
%
%
%
%
where $\rho$ is the Lie group algebra homomorphism that realizes the elements in $L^{1}(G)$ as concrete operators acting on a space of signals $\ccalH$.

From~\eqref{eq_equivariance_condition} we see that when $G$ is commutative equivariance is guaranteed with $\bbT_{g} = \widetilde{\bbT}_{g}$. When $G$ is noncommutative and the ASM is the one associated to traditional group signal processing -- see \eqref{eq_conv_LieG_classic} --, we have equivariance associated to left and right actions of the group~\cite{folland2016course}. As pointed out in~\cite{folland2016course} (page 56), if 
%
%
%
$
    \bbT_{g}\left( \boldsymbol{a}(x) \right) 
    =
    \boldsymbol{a}\left( T_{g}(x) \right) 
    = 
    \boldsymbol{a}\left( g^{-1}x\right)
    ,
$
%
%
%
we have
%
%
%
$
\bbT_{g}\left( 
            \boldsymbol{a}
             \ast 
             \boldsymbol{b}
        \right)
        =
        \left( 
         \bbT_g \boldsymbol{a}
        \right)
        \ast 
        \boldsymbol{b},
$
%
%
%
and if
%
%
%
$
    \bbT_{g}\left( \boldsymbol{a}(x) \right) 
    =
    \boldsymbol{a}\left( T_{g}(x) \right) 
    = 
    \boldsymbol{a}\left( xg\right)
    ,
$    
%
%
%
we have
%
%
%
$
\bbT_{g}\left( 
            \boldsymbol{a}
             \ast 
             \boldsymbol{b}
        \right)
        =
        \boldsymbol{a}
        \ast 
                \left( 
         \bbT_g \boldsymbol{b}
        \right)
        .
$        
%
%
%
This is, the traditional group convolution operation is endowed with two equivariant transformations, which are generally referred to as right and left equivariance~\cite{folland2016course}. 

Notice that a similar notion of equivariance -- left and right equivariance -- can be considered in scenarios where the group actions are defined by matrix transformations on the domain of the signals~\cite{taylor1986noncommutative}. In what follows we formalize this notion. Let us consider the space, $\ccalH$, of functions from $\mbR^n$ into $\mbC$. Let $\bbT$ be a representation of the Lie group $G$ in $\ccalB (\ccalH)$ and let $L_h$ and $R_h$ the operators on $L^{1}(G)$ given by $L_{h}\boldsymbol{a}(g)=\boldsymbol{a}(hg)$ and $R_{h}\boldsymbol{a}(g)=\boldsymbol{a}(gh)$. Then, the following result takes place.


\begin{theorem}\label{thm:equivariance_LR}
Let $\rho$ be a Lie group homomorphism -- Definition~\ref{def:LieGroupAlgHomomorph} eqn.~\eqref{equ_rho_ideal} -- from $L^{1}(G)$ on $\ccalB (\ccalH)$ and let $\bbT$ be the representation of $G$ in $\ccalB(\ccalH)$. Then, if the measure $\mu$ in~\eqref{equ_rho_ideal} is right-invariant to the actions of the group we have
\begin{equation}\label{equ:R_transform_L_rho}
\rho\left( 
           \boldsymbol{a}
       \right)
       \bbT_{h}
       f 
       =
\rho\left(
         R_{h^{-1}}\boldsymbol{a}
    \right)
    f   
    .
\end{equation}
Additionally, if the measure $\mu$ is left-invariant to the actions of the group it follows that
\begin{equation}\label{equ:L_transform_L_rho}
\bbT_{h}\rho\left( 
           \boldsymbol{a}
       \right)
       f
       =
\rho\left(
         L_{h^{-1}}\boldsymbol{a}
    \right)
    f  
    .
\end{equation}
%
%
%
\end{theorem}

\begin{proof}
    See Appendix \ref{sec:proof_equivariance_LR}
\end{proof}


It is important to note in Theorem~\ref{thm:equivariance_LR} that the results hold for the ideal filter~\eqref{equ_rho_ideal} and not the discrete approximation~\eqref{equ:lieGAH_discrete_final}. This is because interpolation is needed to consider
signals in nonhomogeneous spaces. Indeed, this is a common occurrence in GCNNs. Apart from the image classification domain, where the filters are specifically designed to satisfy the constraints \cite{cohen2016steerable, cohen2018spherical, worrall2017harmonic, cohen2019gauge}, equivariance is not guaranteed after considering the practical applications to estimate the Haar integral~\cite{esteves2018learning, macdonald2022enabling}. The closest form of equivariance comes from the work of \cite{finzi2020generalizing}, which establishes equivariance in distribution only for homogeneous spaces. Instead, equivariance is generally shown empirically. Therefore, in this work, we do not restrict our filters to be equivariant, but instead, we emphasize stability, which is discussed in the following subsection. We emphasize that the stability conditions hold for the discretized filter \eqref{equ:lieGAH_discrete_final}.


\subsection{Connection to Multigraphs and Stability} \label{sec:multigraphs}


\begin{figure}
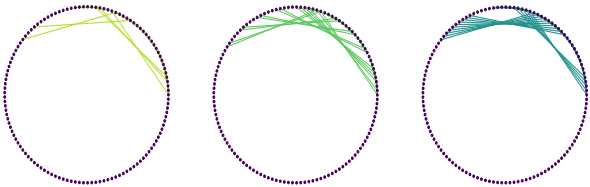

    \centering
    	\begin{subfigure}{.3\linewidth}
		\centering
  \resizebox{75pt}{75pt}{
     \input{./figures/1005_figures/multi1.tex}
     }
\end{subfigure}
    	\begin{subfigure}{.3\linewidth}
		\centering
       \resizebox{75pt}{75pt}{
     \input{./figures/1005_figures/multi3.tex}
     }
\end{subfigure}
    	\begin{subfigure}{.3\linewidth}
		\centering
       \resizebox{75pt}{75pt}{
     \input{./figures/1005_figures/multi2.tex}
     }
\end{subfigure}
    \caption{Consider the group $\mathsf{SO}(3)$, whose Lie algebra has three generators which can be thought of as a rotation around the $x, y,$ and $z$ axes. Let each node in the above graph represent a point on the sampled signal (here we see a signal sampled with 125 points). An edge is drawn between two nodes when the weight of the $\hat \bbT_{i, n_j}$ element of the induced transformation of Algorithm \ref{alg:Transformation} is non-negative. Notice how each generator admits a unique graph between the points, allowing us to draw connections to multigraph signal processing.}
    \label{fig:Multigraph}
\end{figure}


In the light of~\eqref{equ:lieGAH_discrete_final}, we can see that the discrete implementation of Lie group algebra filters on spaces of signals of finite dimension is determined by a collection of operators realizing the actions of the group. This is, the discrete implementation of the filters is given by multinomial operators whose independent variables are $\left\lbrace \widehat{\bbT}_{\hat{g}} \right\rbrace_{\widehat{G}_{\delta,N}^{k}}$. Such realization of the filters is equivalent to the realization of noncommutative polynomial filters on a \textit{multigraph}~\cite{parada2021convolutional,msp_journal,msp_conference}. To see this, let us consider Fig.~\ref{fig:Multigraph} where we depict the realization of three group actions $\widehat{\bbT}_{\hat{g}}$ acting on the space of signals $\ccalH = \mbR^{125}$. Since each
$\widehat{\bbT}_{\hat{g}}$ is acting on $\mbR^{125}$ we associate each $\widehat{\bbT}_{\hat{g}}$ to an adjacency matrix of a multigraph with 125 nodes. Then, the signals to be processed live on the nodes of this multigraph and are associated with vectors in $\mbR^{125}$. Recall that a convolutional multigraph signal model is defined by the triplet $(\ccalA,\ccalH,\rho)$, where $\ccalA$ is a polynomial algebra with multiple generators, $\ccalH$ is a finite-dimensional vector space and $\rho$ is the homomorphism that translates the abstract polynomials in $\ccalA$ into concrete operators in $\ccalB (\ccalH)$. Additionally, for $a\in\ccalA$ we have that the independent variables of $\rho (a)$ are the matrix representations of the multigraph~\cite{parada2021convolutional,msp_journal,msp_conference}. 

This observation comes as a fundamental difference between the discrete realizations associated to convolutional signal models on manifolds and the discrete realizations of convolutional signal models on Lie groups. The essential attributes of a manifold can be fairly well approximated by a graph~\cite{apm_thesis}, while in the case of the Lie group only a multigraph can capture both the topological and the algebraic properties. Given the importance of this observation, we state it in the form of the following proposition.


\begin{proposition}\label{prop_liegalg_as_msp}

Let $G$ be a Lie group and $L^{1}(G)$ its Lie group algebra. Let $(L^{1}(G),L^{2}(G),\rho)$ be the algebraic convolutional signal model representing the classical group signal processing model on $G$. Let $\widehat{\ccalH}$ be a finite dimensional vector space and $\widehat{\rho}: L^{1}(G) \to \ccalB (\widehat{\ccalH})$ the approximate homomorphism given according to~\eqref{equ:lieGAH_discrete_final}. Then, the discrete finite dimensional realization of $(L^{1}(G),L^{2}(G),\rho)$ given by $(L^{1}(G),\widehat{\ccalH},\widehat{\rho})$ is a convolutional multigraph signal model.

\end{proposition}


Then, we can describe the properties of Lie group algebraic filters using tools and results from multigraph signal processing and multigraph neural networks. In particular, in virtue of the stability results in~\cite{parada2021convolutional} for multigraph convolutional filters and multigraph neural networks, we can obtain stability results for the Lie group algebraic filters given by~\eqref{equ:lieGAH_discrete_final}. In what follows we specify the details of the multigraph model associated to $(L^{1}(G),\widehat{\ccalH},\widehat{\rho})$.

Let us consider that $(L^{1}(G),\widehat{\ccalH},\widehat{\rho})$ is obtained using a sampling set $\widehat{G}_{\delta,N}^{k}$. Then, we introduce the non commutative algebra $\ccalA$ generated by the elements in $\widehat{G}_{\delta, N}$ -- notice that $\ccalA$ has $2N+1$ generators. Let $\widehat{\ccalH}$ be a space of discrete signals and let $\widehat{\rho}$ be given by~\eqref{equ:lieGAH_discrete_final}. Then, $(L^{1} (G),\widehat{\ccalH},\widehat{\rho})$ is equivalent to the multigraph signal model $(\ccalA,\widehat{\ccalH},\zeta)$ with shift operators given by $\{ \widehat{\bbT}_{\hat{g}} \}_{\hat{g}\in\widehat{G}_{\delta,N}}$, and where $\zeta: \ccalA \to \ccalB (\widehat{\ccalH})$ is given by $\zeta (\hat{g}) = \widehat{\rho} (\hat{g})$.

To state our stability results we consider perturbations on the shift operators according to 
\begin{equation}\label{eq_perturb_model}
\bbQ \left(
         \widehat{\bbT}_{\hat{g}}
     \right)
           =
           \bbQ_{0,\hat{g}} 
           +
           \bbQ_{1,\hat{g}} \widehat{\bbT}_{\hat{g}}
           ,
\end{equation}
where $\bbQ_{i,\hat{g}} \in\mbR^{\text{dim}(\widehat{\ccalH})\times\text{dim}(\widehat{\ccalH})}$ and $\Vert \bbQ_{i,\hat{g}} \Vert_{F} \leq c \Vert \bbQ_{i,\hat{g}} \Vert$ for $i=0,1$, and $c\in\mbR^{+}$. Then, $\bbQ \left(\widehat{\bbT}_{\hat{g}}\right)$ is a perturbed version of $\widehat{\bbT}_{\hat{g}}$ and it models a realization of the actions of the group affected by small and undesired changes. This is, $\widehat{\bbT}_{\hat{g}}$ are the shift operators we think we are implementing while $\bbQ (\widehat{\bbT}_{\hat{g}})$ are the shift operators we get in the final realization of the Lie group algebraic signal processing model. Notice that the perturbation of shift operators indicated in~\eqref{eq_perturb_model} corresponds indeed to a perturbation of the homomorphism in the algebraic signal model. If $(\ccalA,\widehat{\ccalH},\zeta)$ is the multigraph signal model of $(L^{1} (G),\widehat{\ccalH},\widehat{\rho})$, we denote by $(\ccalA,\widehat{\ccalH},\widetilde{\zeta})$ its perturbed version under the perturbation model in~\eqref{eq_perturb_model}.

We now define operator stability for algebraic filters.


\begin{definition}\label{def_operator_stab} \emph{(Filter Stability)}
    Let $(\ccalA, \widehat{\ccalH}, \zeta)$ be an algebraic signal model and let $(\ccalA,\widehat{\ccalH}, \widetilde{\zeta})$ be its perturbed version according to~\eqref{eq_perturb_model}. We denote by  $p\left( \widehat{\bbT} \right)$ and $p\left( \bbQ\left(\widehat{\bbT}\right) \right)$ the realizations of the filter $p\in\ccalA$ on $\ccalB(\widehat{\ccalH})$ for $(\ccalA, \widehat{\ccalH}, \zeta)$ and $(\ccalA,\widehat{\ccalH}, \zeta)$, respectively. Then, we say that $p$ is Lipschitz stable if there exist constants $C_0, C_1 > 0$ such that
\begin{multline} \label{equ:filter_stability}
\left\Vert 
        p\left(\widehat{\bbT}\right)\bbf 
        -
        p\left(\bbQ\left(\widehat{\bbT}\right)\right)\bbf
\right\Vert 
\leq
\left[
   C_{0}\sup_{\widehat{\bbT}}\left\Vert 
                        \bbQ\left( \widehat{\bbT}\right)
                   \right\Vert
                      + 
    \right. 
    \\
    \left.
   C_1 \sup_{\widehat{\bbT}}\left\Vert
                          D_{\bbQ}\left( 
                                            \bbQ\left(\widehat{\bbT}\right)
                                            \right)
                    \right\Vert 
   + 
   \ccalO\left(
           \left\Vert
                \bbQ\left(\widehat{\bbT}\right)
           \right\Vert^2
         \right)
\right]
\|\bbf\|,
\end{multline}
    where $D_{\bbQ}$ is the Fr\'echet derivative of the perturbation operator $\bbQ$ with respect to $\widehat{\bbT}$ and $\bbf \in \widehat{\ccalH}$.
    
\end{definition}


Definition \ref{def_operator_stab} states that a filter is stable when the deformation of the operator is proportional to the size of the deformation. Given that the realized form of the algebraic Lie group convolution can be viewed as a multigraph signal model, the stability follows directly.


\begin{corollary} (Corollary 1~\cite{parada2021convolutional})
\label{corollary_stability}
   Let $(\ccalA, \widehat{\ccalH}, \zeta)$ be a multigraph signal model and let $(\ccalA,\widehat{\ccalH}, \widetilde{\zeta})$ be its perturbed version according to~\eqref{eq_perturb_model}. If $p\in\ccalA$ is $L_0$-Lipschitz and $L_1$-integral Lipschitz, and $\ccalA$ has $2N+1$ generators, then $p$ is stable in the sense of Definition \ref{def_operator_stab} with $C_0 = (2N+1)c L_0$ and $C_1 = (2N+1)c L_1$. 
\end{corollary}


Notice that the conditions of being $L_{0}$-Lipschitz and $L_{1}$-integral Lipschitz are imposed on the spectral representation of the filters in $\ccalA$ -- see~\cite{parada2021convolutional,msp_journal} for more details. We remark that these restrictions are mentioned here only to emphasize that there are subsets of filters in $\ccalA$ that can be implemented to achieve stability.


The stability result stated in Corollary~\ref{corollary_stability} provides the guarantee that the implementation or realization of the Lie algebraic group filters is consistent when such implementation is affected by small changes. We can think about $p\left(\widehat{\bbT}\right)$ as the filter we aim to implement or use in the signal model and $ p\left(\bbQ\left(\widehat{\bbT}\right)\right)$ as the filter that we get. Then, if $p$ is stable, the size of the change in $p$ is proportional to the size of the change associated with the deformation $\bbQ\left(\widehat{\bbT}\right)$ -- when seen as a diffeomorphism acting in $\ccalB(\ccalH)$. Therefore, stability aims to ensure the consistency of the implementation of a Lie group algebra filter. It is important to emphasize that the changes considered in Corollary~\ref{corollary_stability} are associated with the actions of the generators of the group when implemented in $\ccalB(\ccalH)$ by a homomorphism.

One important attribute of the constants $C_i = (2N+1)cL_0$ is that they unveil a trade-off between accuracy and stability. To see this, notice that when $N$ is increased, the number of sampling points on the Lie group increases and therefore the bandwidth of the filters that can be learned on $G$. At the same time, increasing $N$ translates into an increment in the size of $C_i$ that can only be compensated when reducing the value $L_1$ and $L_0$. Then, in the case of Lie group algebra filters, one has to keep in check the number of sampling on $G$ to guarantee that small changes will not build up into significant changes on the filters implemented.

In Section~\ref{sec:experiments} we provide a set of numerical experiments that will help the reader to validate, visualize, and understand the stability results derived above.





\section{Numerical Results}  \label{sec:experiments}


\begin{figure*}
        \centering
        \input{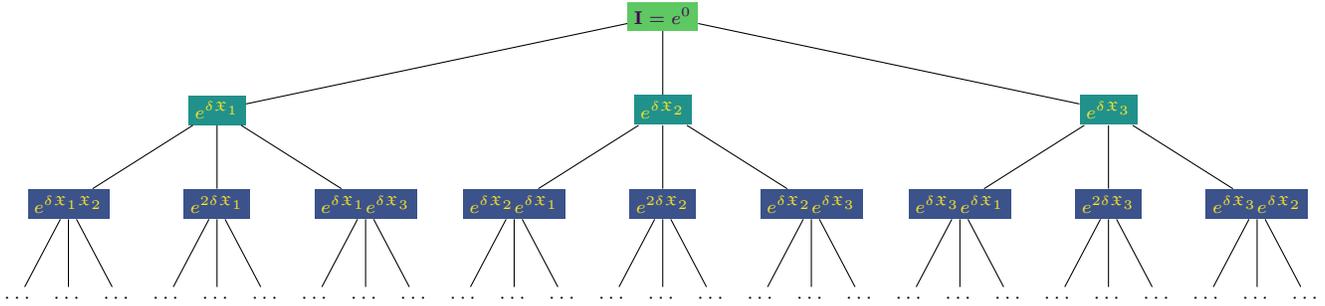} 
        \caption{Diffusion tree to depict the generation of monomials associated to a sampling set $\hat G_{\delta,N}^{k}\subset G$ in Definition \ref{def_Ghat_monomials}. The depth of the tree indicates the degree of the monomials generated by taking products of the terms $e^{\delta\mathfrak{X}_i}$. The tree representation of the diffusion operators also allows us to visualize the connection to Multigraph signal processing, where $\bbS_{i} = e^{\delta\mathfrak{X}_i}$ represent the shift operators corresponding to different graphs connecting the edges of the sampled points $\ccalX\subset\mbR^n$.}
        \label{fig:diffusion_tree}
\end{figure*}



\begin{figure}
\centering


\colorlet{my_alejocolg1}{black!30}
\colorlet{my_alejocolg2}{black!35}
\colorlet{my_alejocolg3}{black!40}
\colorlet{my_alejocolg4}{black!45}
\colorlet{my_alejocolg5}{black!50}
\colorlet{my_alejocolg6}{black!55}
\colorlet{my_alejocolg7}{black!60}

\definecolor{my_cp4_col1}{RGB}{255, 86, 87}
\definecolor{my_cp4_col2}{RGB}{55, 108, 138}
\definecolor{my_cp4_col3}{RGB}{242, 217, 187}
\definecolor{my_cp4_col4}{RGB}{99, 143, 169}

\definecolor{my_cp5_col1}{RGB}{253, 231, 37}
\definecolor{my_cp5_col2}{RGB}{94, 201, 98}
\definecolor{my_cp5_col3}{RGB}{33, 145, 140}
\definecolor{my_cp5_col4}{RGB}{59, 82, 139}
\definecolor{my_cp5_col5}{RGB}{68, 1, 84}


\def \myfactor {0.65}
\def \unit  {\myfactor cm}


\tikzstyle{block} = [ rectangle,
                      minimum width = \unit,
                      minimum height = \unit,
                      fill = gray,
                      draw = black,
                      text = black]

\tikzstyle{filter} = [block,
                      minimum width  = 5.5*\unit,
                      minimum height = 1.8*\unit,
                      fill=my_cp5_col4]

\tikzstyle{nonlinearity} = [ filter,
                             minimum width  = 3.9*\unit,
                             fill = my_cp5_col3]

\tikzstyle{pooling} = [ filter,
                             minimum width  = 3.9*\unit,
                             fill = my_cp5_col2]

\def \deltainput     {( 0.0,-1.7)}
\def \deltaoutput    {( 0.0,-1.2)}
\def \deltalayer     {4}
\def \deltaconnector {1.35}
\def \deltasigma     {( 7.5, 0.0)}

\def \one   {$\textcolor{white}{\displaystyle{\mathbf{y}_{1}  = \sum_{\hat{g}\in \widehat{G}_{\delta,N}^{k}} \boldsymbol{a}_{1}(\hat{g}) \widehat{\bbT}_{\hat{g}}\mathbf{x}}}$}
\def \two   {$\textcolor{white}{\displaystyle{\mathbf{y}_2  =  \sum_{\hat{g}\in \widehat{G}_{\delta,N}^{k}} \boldsymbol{a}_{2}(\hat{g}) \widehat{\bbT}_{\hat{g}}\mathbf{x}_{1}}}$}

\def \sigmaone   {$\textcolor{white}{\displaystyle{\mathbf{x}_{1} = {\eta_{1}} \Big[\, \mathbf{y}_1 \, \Big]}}$}
\def \sigmatwo   {$\textcolor{white}{\displaystyle{\mathbf{x}_{2} = {\eta_{2}} \Big[\, \mathbf{y}_2 \, \Big]}}$}
\def \sigmathree {$\textcolor{white}{\displaystyle{\mathbf{z}_{3} = {\eta_{3}} \Big[\, \mathbf{y}_3 \, \Big]}}$}
\def \proyone   {$\displaystyle{\mathbf{x}_{1} = {P_{1}} \Big[\, \mathbf{z}_1 \, \Big]}$}
\def \proytwo   {$\displaystyle{\mathbf{x}_{2} = {P_{2}} \Big[\, \mathbf{z}_2 \, \Big]}$}
\def \proythree   {$\displaystyle{\mathbf{x}_{3} = {P_{3}} \Big[\, \mathbf{z}_3 \, \Big]}$}


{\fontsize{7}{7}\selectfont\begin{tikzpicture}[scale = \myfactor]

  \pgfdeclarelayer{bg}     
  \pgfsetlayers{bg,main}   

  \node (input) [rectangle, minimum width = 0.1*\unit] {$\mathbf{x}$};
  
  
  \path (input.east)      ++ \deltainput node [filter]       (L1 Filter1) {\one};
  \path (L1 Filter1) ++ \deltasigma node [nonlinearity] (L1 F1)      {\sigmaone};
  \path[draw, -stealth] (L1 Filter1.east) -- node [above] {$\mathbf{y}_1$} (L1 F1.west);

  
  \path (L1 Filter1) ++ (0,-\deltalayer) node [filter]       (L2 Filter1) {\two};
  \path (L2 Filter1) ++ \deltasigma      node [nonlinearity] (L2 F1)      {\sigmatwo};
  \path[draw, -stealth] (L2 Filter1.east) --  node [above] {$\mathbf{y}_2$} (L2 F1.west);

  
  \path[draw, -stealth] (input.east) -- (L1 Filter1.north);
  \path (L1 F1.south) ++ (0,-\deltaconnector) node [] (aux1) {};
  \path[draw, -stealth] (L1 F1.south) -- node [below right] {$\mathbf{x}_1$} (aux1.north) 
                                      --                         (aux1.north -| L2 Filter1.north) 
                                      -- node [above left]  {$\mathbf{x}_1$} (L2 Filter1.north);
  \path[draw, -stealth](L2 F1.south) -- ++ \deltaoutput -- ++ (0.5, 0) 
  node [right]{$\mathbf{x}_2$};

  
  \begin{pgfonlayer}{bg} 
      \path (L1 Filter1.west |- L1 F1.south) ++ (-0.4,-0.7)
           node [filter, anchor = south west,
                 fill = black!5, 
                 minimum width  = 13*\unit,
                 minimum height = 3*\unit,] 
        (layer)
        {}; 
       \path (layer.south west) ++ (0.0,0.0) node [above right] {Layer 1};
      \path (L1 Filter1.west |- L2 F1.south) ++ (-0.4,-0.7)
           node [filter, anchor = south west,
                 fill = black!5, 
                 minimum width  = 13*\unit,
                 minimum height = 3*\unit,] 
        (layer)
        {}; 
       \path (layer.south west) ++ (0.0,0.0) node [above right] {Layer 2};
    \end{pgfonlayer}

\end{tikzpicture}} 
\caption{Two layer neural network design used during numerical experiments. The input is processed by a filter then passed through a swish non-linearity $\eta$\cite{ramachandran2017searching}.}
\label{fig:GNNbasicfig}
\end{figure}
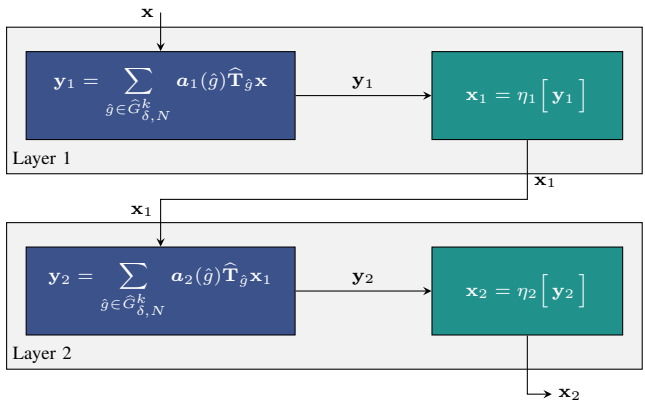


In this section, we evaluate the Lie group algebra filters on two problems with $\mathsf{SO}(3)$ group symmetries and one problem with $\mathsf{SO}(2)$ group symmetries. In particular, for the $\mathsf{SO}(3)$ group, we consider a novel binary classification for three-dimensional knots and the ModelPoint10 dataset. For the $\mathsf{SO}(2)$ group, we consider the RotatedMNIST dataset. The classification on knots is specifically chosen so that the group agnostic graph signal processing methods for pointcloud classification \cite{shi2020point} would not be applicable, as all of the graphs would be the same (a cyclic graph). 

Consistent with our claim in \textbf{(C2)}, we indeed decouple the discretization of the convolution through the sampling of the Lie Group Algebra and the sample space. We describe how that is done for the three-dimensional data in the following subsections.

\subsection{Sampling of $L^1(\mathsf{SO}(3))$}
Recall the sampling of the Lie group given by definition \ref{def_Ghat_monomials}. For the $\mathsf{SO}(3)$ group, the resolution $\delta$ and range $\delta N$ can be expressed in terms of angles. In all of our experiments, we consider a resolution $\delta = \pi/18$ radians (or 10 degrees). The range is set to $\delta N = \pi/6$ radians (or 60 degrees) with $N = 6$. The multiplicity is equal to $k = 3$.


\subsection{Sampling in the signal space}~\label{sec:num_signal_space}
The second part of the discretization of the group convolution comes from the sampling of the signal. An advantage to our approach is the ability to process data with an arbitrary domain (i.e. the sampling of the signal space). To highlight this advantage, we consider projecting the pointcloud data onto four distinct samplings of the domain in $\mathbb{R}^3$. 

First, we describe the projection operation. from the pointcloud  Consider a pointcloud signal represented by $\ccalS \subset \mbR^3$. Each pointcloud point $s\in \ccalS$ has a corresponding value $f(s)$. For our experiments, the value of $f(s) = 1$ for all points. The size of the pointcloud is finite $|\ccalS| < \infty$ and can vary for each signal. Let $S$ denote a sampling of the domain. For each point $\bbs$ in the set $S$, we find the $k$ nearest neighbors from the pointcloud points $s\in \ccalS$. We compute the mean of these distances and define the new signal on the sampled domain by
\begin{equation} \label{equ_point_cloud_proj}
    \hat f(\bbs) = \left[\xi - \frac{1}{|\textrm{KNN}(\bbs)|}\sum_{x \in \textrm{KNN}(s)} f(x)^{-1} \|x - s\|_2\right]_+,
\end{equation}
where $\left[\cdot\right]_+$ clips negative values, $\textrm{KNN}(s)$ denotes the set of the $k$ nearest neighbors around $s$, and $\xi$ is some threshold. For all of our experiments, the value of $f(s) = 1$ and $\xi = 1$. We emphasize that the projection operation we consider is a design choice, as any projection that maps the pointcloud to the sampled domain could be considered. 

Next, we will describe the four distinct sample schemes for the domain $S$. For all sampling, we assume $a_1, a_2, a_3 < b_1, b_2, b_3$
\begin{itemize}
    \item Grid: The set of points in $[a_1, a_2, a_3]\times [b_1, b_2, b_3]$ such that $$s = \left[a_1 + i_1\frac{b_1 -a_1}{n_1-1}, a_2 + i_2\frac{b_2 -a_2}{n_2-1},a_3 + i_3\frac{b_3 -a_3}{n_3-1}\right]$$
    \item Uniform: Samples drawn uniformly on the cube $[a_1, a_2, a_3]\times [b_1, b_2, b_3]$.  
    \item Sphere: Samples at random on a unit sphere with radius $R>0$. In terms of polar coordinates $(r, \theta, \varphi)$ points are sampled randomly with $r\sim (0,1]$, $\theta\sim (0, \pi]$, and $\varphi \sim (0, 2\pi]$.
    \item Gaussian: Samples drawn from a multivariate Gaussian distribution $\ccalN(\mathbf{0}, R\cdot I_3)$ where $R > 0$.
\end{itemize}
We emphasize that the random sampling on the sphere is not uniform.


\subsection{Knot Dataset}


\begin{figure}[t]
	\centering
	\begin{subfigure}{.21\textwidth}
		\centering
		\includegraphics[width=1.00\linewidth]{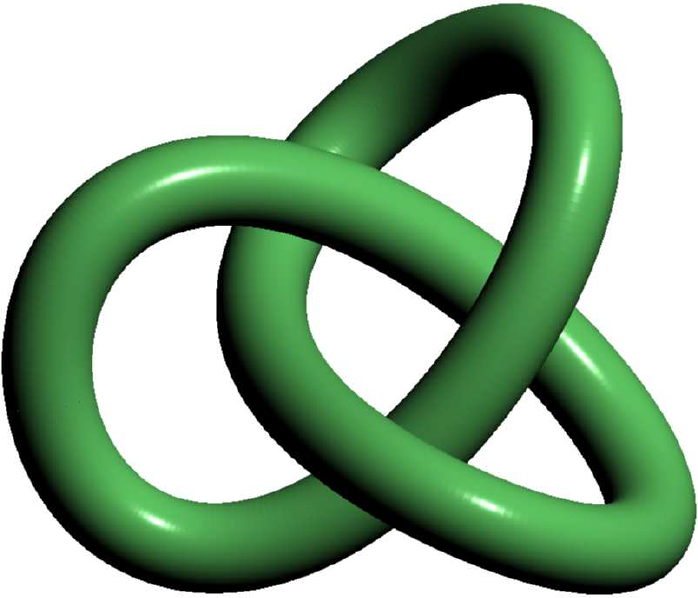} 
		\caption{}
		\label{fig:multiData-one}
	\end{subfigure}
	\begin{subfigure}{.2\textwidth}
		\centering
		\includegraphics[width=1.00\linewidth]{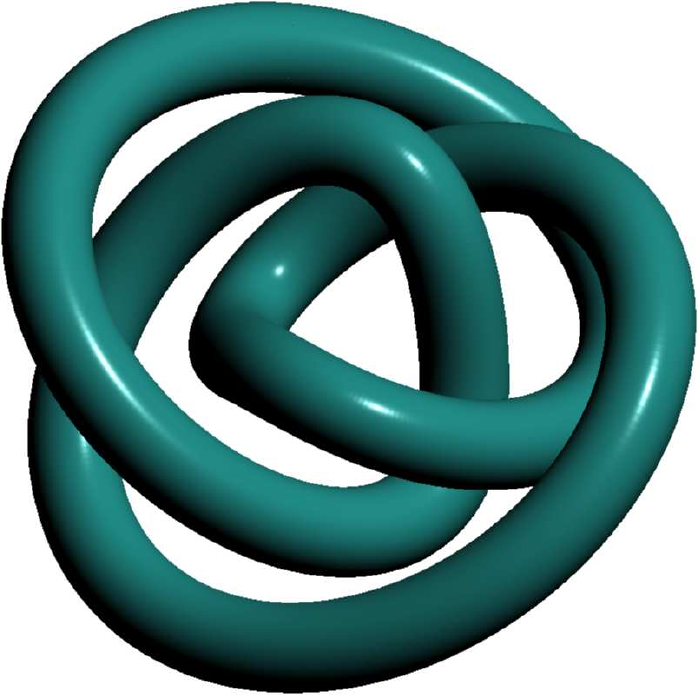} 
		\caption{}
		\label{fig:multiData-two}
	\end{subfigure}
	\caption{(a) Trefoil knot and (b) Listing's knot used in binary classification experiments.}
	\label{fig_knot_numsim}
\end{figure}


We introduce the problem of binary classification on knots. In particular, we would like to distinguish between a Trefoil knot, characterized by 
$
 x = \cos(t) + 2\cos(2t),  y = \sin(t) - 2\sin(2t),  z = -\sin(3t),
$
and the Listing's knot (also known as the figure-eight knot), characterized by 
$
x = (2+ \cos(2t))\cos(3t), y = (2+ \cos(2t))\sin(3t), z = \sin(4t),
$
shown in Figure \ref{fig_knot_numsim}.

Parameterized by $t$ as described in the aforementioned paragraph, the knots are sampled $N$ times evenly with $t = \left\{2\pi i /N \right\}{i = 0, \dots, N}$. The sampled points on the knot are then rotated by selecting a random point $g\in \mathsf{SO}(3)$. The positions of the points are then perturbed by a Gaussian noise with zero mean and $\sigma$ standard deviation. As such, each knot signal is a collection of $N$ points in $\mathbb{R}^3$. These points are then projected onto each of the sampled domain sets (Grid, Uniform, Sphere, and Gaussian) as described by \eqref{equ_point_cloud_proj} in Section \ref{sec:num_signal_space}. In particular, we let $a_1 = a_2 - a_3 = -2$, $b_1= b_2 = b_3 = 2$, and $R = 2$.

For our experiments, we set $N = 200$, and the Gaussian jitter position noise drawn from $\ccalN(0,0.01)$ during training and $\ccalN(0,0.1)$ for testing. There are 120 knots used during training and 140 knots used during test with equal contributions from each knot class.

\def\pathresultsrev{./figures/1113_figures}

\def\pathresults{./figures/1005_figures}
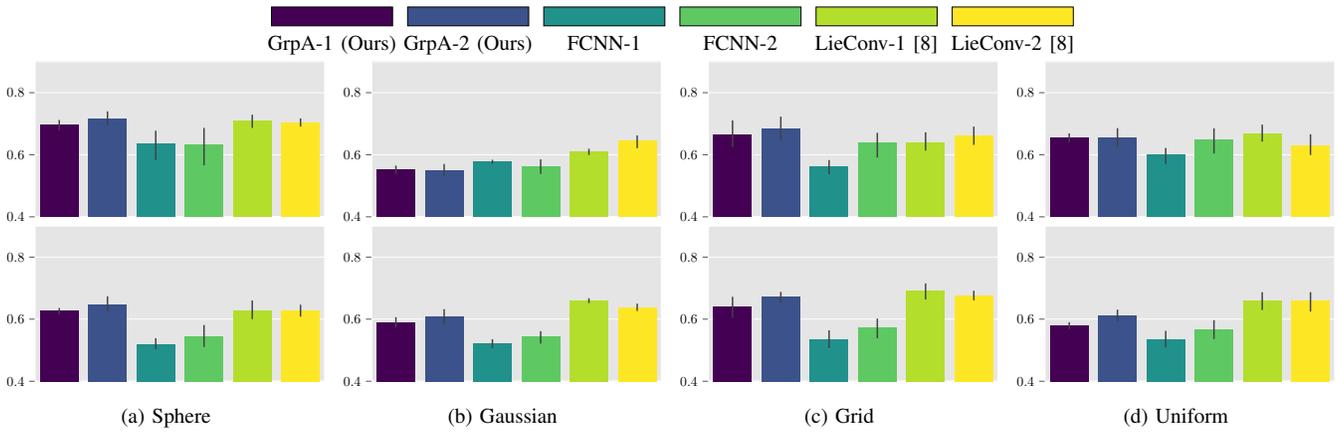
\begin{figure*}[t]
\centering


\definecolor{my_cp5_col1}{RGB}{253, 231, 37}
\definecolor{my_cp5_col2}{RGB}{180, 222,44}
\definecolor{my_cp5_col3}{RGB}{94, 201, 98}
\definecolor{my_cp5_col4}{RGB}{33, 145, 140}
\definecolor{my_cp5_col5}{RGB}{59, 82, 139}
\definecolor{my_cp5_col6}{RGB}{68, 1, 84}

\usetikzlibrary{positioning,decorations.pathreplacing,shapes}


\def \scale {1}
\def \unit { \scale cm}

\def \vertsep {0.5*\scale}

\def \horzsep {1*\scale}


\tikzstyle{set} = [rectangle,color=black,
                    rounded corners = 0*\unit,
                    fill=black,
                    inner sep=0pt,
                    draw,
                    anchor = center,
                    line width=0.1mm]
                    
 
\tikzstyle{myboxlabel} = [set,
fill=my_cp5_col3,
minimum width  = 1.61*\unit,
minimum height = 0.25*\unit]

\tikzstyle{dot} = [ circle,
                    minimum width  = 0.05*\unit,
                    fill=black,
                    color=black,
                    inner sep=0pt,
                    draw,
                    anchor = center ]


{\fontsize{8}{8}\selectfont

\begin{tikzpicture}[scale=\scale,rounded corners,ultra thick]


   
   \path (0,0) node [myboxlabel,fill=my_cp5_col6] (L1) {};
   \path (L1.south) ++ (0, 0) node [below, color=black] {GrpA-1 (Ours)};

   
    \path (L1.east)++(\horzsep,0) node [myboxlabel,fill=my_cp5_col5] (L2) {};
    \path (L2.south) ++ (0, 0) node [below, color=black] {GrpA-2 (Ours)};

     
     \path (L2.east)++(\horzsep,0) node [myboxlabel,fill=my_cp5_col4] (L3) {};
     \path (L3.south) ++ (0, 0) node [below, color=black] {FCNN-1};

         
     \path (L3.east)++(\horzsep,0) node [myboxlabel,fill=my_cp5_col3] (L4) {};
     \path (L4.south) ++ (0, 0) node [below, color=black] {FCNN-2};

      
      \path (L4.east)++(\horzsep,0) node [myboxlabel,fill=my_cp5_col2] (L5) {};
      \path (L5.south) ++ (0, 0) node [below, color=black] {LieConv-1\cite{finzi2020generalizing}};

      
      \path (L5.east)++(\horzsep,0) node [myboxlabel,fill=my_cp5_col1] (L6) {};
      \path (L6.south) ++ (0, 0) node [below, color=black] {LieConv-2\cite{finzi2020generalizing}};

%
%

\end{tikzpicture}

}
	\centering
	\begin{subfigure}{.24\textwidth}
		\centering
		\resizebox{\linewidth}{!}{
\begin{tikzpicture}

\def\axisdefaultwidth{9cm}
\def\axisdefaultheight{5.56cm}

\definecolor{my_cp5_col1}{RGB}{253, 231, 37}
\definecolor{my_cp5_col2}{RGB}{180, 222,44}
\definecolor{my_cp5_col3}{RGB}{94, 201, 98}
\definecolor{my_cp5_col4}{RGB}{33, 145, 140}
\definecolor{my_cp5_col5}{RGB}{59, 82, 139}
\definecolor{my_cp5_col6}{RGB}{68, 1, 84}

\definecolor{burlywood231187113}{RGB}{231,187,113}
\definecolor{darkgray158150204}{RGB}{158,150,204}
\definecolor{darkslategray66}{RGB}{66,66,66}
\definecolor{dimgray85}{RGB}{85,85,85}
\definecolor{gainsboro229}{RGB}{229,229,229}
\definecolor{gray119}{RGB}{119,119,119}
\definecolor{indianred2049072}{RGB}{204,90,72}
\definecolor{steelblue69133171}{RGB}{69,133,171}
\definecolor{yellowgreen13817181}{RGB}{138,171,81}

\begin{axis}[
axis background/.style={fill=gainsboro229},
axis line style={white},
tick align=outside,
x grid style={white},
xmajorticks=false,
xmin=-0.5, xmax=5.5,
xtick style={color=dimgray85},
xtick={0,1,2,3,4,5},
xticklabels={GrpA-1,GrpA-2,FCNN-1,FCNN-2,LieConv-1,LieConv-2},
y grid style={white},
ymajorgrids,
ymin=0.4, ymax=0.9,
ytick pos=left,
ytick style={color=dimgray85}
]
\draw[draw=none,fill=my_cp5_col6,very thin] (axis cs:-0.4,0) rectangle (axis cs:0.4,0.695744680851064);
\draw[draw=none,fill=my_cp5_col5,very thin] (axis cs:0.6,0) rectangle (axis cs:1.4,0.715602836879433);
\draw[draw=none,fill=my_cp5_col4,very thin] (axis cs:1.6,0) rectangle (axis cs:2.4,0.636170212765958);
\draw[draw=none,fill=my_cp5_col3,very thin] (axis cs:2.6,0) rectangle (axis cs:3.4,0.631914893617021);
\draw[draw=none,fill=my_cp5_col2,very thin] (axis cs:3.6,0) rectangle (axis cs:4.4,0.708510638297872);
\draw[draw=none,fill=my_cp5_col1,very thin] (axis cs:4.6,0) rectangle (axis cs:5.4,0.702836879432624);
\addplot [line width=1.08pt, darkslategray66]
table {%
0 0.678723404255319
0 0.712056737588652
};
\addplot [line width=1.08pt, darkslategray66]
table {%
1 0.693617021276596
1 0.739716312056738
};
\addplot [line width=1.08pt, darkslategray66]
table {%
2 0.582978723404255
2 0.677304964539007
};
\addplot [line width=1.08pt, darkslategray66]
table {%
3 0.565957446808511
3 0.686524822695036
};
\addplot [line width=1.08pt, darkslategray66]
table {%
4 0.686524822695035
4 0.729078014184397
};
\addplot [line width=1.08pt, darkslategray66]
table {%
5 0.690780141843972
5 0.717021276595745
};
\end{axis}

\end{tikzpicture}}
        \resizebox{\linewidth}{!}{
\begin{tikzpicture}

\def\axisdefaultwidth{9cm}
\def\axisdefaultheight{5.56cm}

\definecolor{my_cp5_col1}{RGB}{253, 231, 37}
\definecolor{my_cp5_col2}{RGB}{180, 222,44}
\definecolor{my_cp5_col3}{RGB}{94, 201, 98}
\definecolor{my_cp5_col4}{RGB}{33, 145, 140}
\definecolor{my_cp5_col5}{RGB}{59, 82, 139}
\definecolor{my_cp5_col6}{RGB}{68, 1, 84}

\definecolor{burlywood231187113}{RGB}{231,187,113}
\definecolor{darkgray158150204}{RGB}{158,150,204}
\definecolor{darkslategray66}{RGB}{66,66,66}
\definecolor{dimgray85}{RGB}{85,85,85}
\definecolor{gainsboro229}{RGB}{229,229,229}
\definecolor{gray119}{RGB}{119,119,119}
\definecolor{indianred2049072}{RGB}{204,90,72}
\definecolor{steelblue69133171}{RGB}{69,133,171}
\definecolor{yellowgreen13817181}{RGB}{138,171,81}

\begin{axis}[
axis background/.style={fill=gainsboro229},
axis line style={white},
tick align=outside,
x grid style={white},
xmajorticks=false,
xmin=-0.5, xmax=5.5,
xtick style={color=dimgray85},
xtick={0,1,2,3,4,5},
xticklabels={GrpA-1,GrpA-2,FCNN-1,FCNN-2,LieConv-1,LieConv-2},
y grid style={white},
ymajorgrids,
ymin=0.4, ymax=0.9,
ytick pos=left,
ytick style={color=dimgray85}
]
\draw[draw=none,fill=my_cp5_col6,very thin] (axis cs:-0.4,0) rectangle (axis cs:0.4,0.626028368794326);
\draw[draw=none,fill=my_cp5_col5,very thin] (axis cs:0.6,0) rectangle (axis cs:1.4,0.647375886524823);
\draw[draw=none,fill=my_cp5_col4,very thin] (axis cs:1.6,0) rectangle (axis cs:2.4,0.519290780141844);
\draw[draw=none,fill=my_cp5_col3,very thin] (axis cs:2.6,0) rectangle (axis cs:3.4,0.544042553191489);
\draw[draw=none,fill=my_cp5_col2,very thin] (axis cs:3.6,0) rectangle (axis cs:4.4,0.627517730496454);
\draw[draw=none,fill=my_cp5_col1,very thin] (axis cs:4.6,0) rectangle (axis cs:5.4,0.626666666666667);
\addplot [line width=1.08pt, darkslategray66]
table {%
0 0.615742907801418
0 0.635817375886525
};
\addplot [line width=1.08pt, darkslategray66]
table {%
1 0.625242907801418
1 0.673491134751773
};
\addplot [line width=1.08pt, darkslategray66]
table {%
2 0.503306737588652
2 0.53859219858156
};
\addplot [line width=1.08pt, darkslategray66]
table {%
3 0.510283687943262
3 0.581
};
\addplot [line width=1.08pt, darkslategray66]
table {%
4 0.600060283687943
4 0.660214539007092
};
\addplot [line width=1.08pt, darkslategray66]
table {%
5 0.608223404255319
5 0.646455673758865
};
\end{axis}

\end{tikzpicture}}
		\caption{Sphere}
		\label{a}
	\end{subfigure}
	\begin{subfigure}{.24\textwidth}
		\centering
        \resizebox{\linewidth}{!}{
\begin{tikzpicture}

\def\axisdefaultwidth{9cm}
\def\axisdefaultheight{5.56cm}

\definecolor{my_cp5_col1}{RGB}{253, 231, 37}
\definecolor{my_cp5_col2}{RGB}{180, 222,44}
\definecolor{my_cp5_col3}{RGB}{94, 201, 98}
\definecolor{my_cp5_col4}{RGB}{33, 145, 140}
\definecolor{my_cp5_col5}{RGB}{59, 82, 139}
\definecolor{my_cp5_col6}{RGB}{68, 1, 84}

\definecolor{burlywood231187113}{RGB}{231,187,113}
\definecolor{darkgray158150204}{RGB}{158,150,204}
\definecolor{darkslategray66}{RGB}{66,66,66}
\definecolor{dimgray85}{RGB}{85,85,85}
\definecolor{gainsboro229}{RGB}{229,229,229}
\definecolor{gray119}{RGB}{119,119,119}
\definecolor{indianred2049072}{RGB}{204,90,72}
\definecolor{steelblue69133171}{RGB}{69,133,171}
\definecolor{yellowgreen13817181}{RGB}{138,171,81}

\begin{axis}[
axis background/.style={fill=gainsboro229},
axis line style={white},
tick align=outside,
x grid style={white},
xmajorticks=false,
xmin=-0.5, xmax=5.5,
xtick style={color=dimgray85},
xtick={0,1,2,3,4,5},
xticklabels={GrpA-1,GrpA-2,FCNN-1,FCNN-2,LieConv-1,LieConv-2},
y grid style={white},
ymajorgrids,
ymin=0.4, ymax=0.9,
ytick pos=left,
ytick style={color=dimgray85}
]
\draw[draw=none,fill=my_cp5_col6,very thin] (axis cs:-0.4,0) rectangle (axis cs:0.4,0.552482269503546);
\draw[draw=none,fill=my_cp5_col5,very thin] (axis cs:0.6,0) rectangle (axis cs:1.4,0.549645390070922);
\draw[draw=none,fill=my_cp5_col4,very thin] (axis cs:1.6,0) rectangle (axis cs:2.4,0.577304964539007);
\draw[draw=none,fill=my_cp5_col3,very thin] (axis cs:2.6,0) rectangle (axis cs:3.4,0.563829787234042);
\draw[draw=none,fill=my_cp5_col2,very thin] (axis cs:3.6,0) rectangle (axis cs:4.4,0.609929078014184);
\draw[draw=none,fill=my_cp5_col1,very thin] (axis cs:4.6,0) rectangle (axis cs:5.4,0.64468085106383);
\addplot [line width=1.08pt, darkslategray66]
table {%
0 0.538297872340425
0 0.565248226950355
};
\addplot [line width=1.08pt, darkslategray66]
table {%
1 0.531914893617021
1 0.570212765957447
};
\addplot [line width=1.08pt, darkslategray66]
table {%
2 0.570212765957447
2 0.583687943262411
};
\addplot [line width=1.08pt, darkslategray66]
table {%
3 0.538280141843972
3 0.585106382978723
};
\addplot [line width=1.08pt, darkslategray66]
table {%
4 0.599290780141844
4 0.619166666666667
};
\addplot [line width=1.08pt, darkslategray66]
table {%
5 0.621276595744681
5 0.66241134751773
};
\end{axis}

\end{tikzpicture}}
		\label{fig:multiData-two}
		\resizebox{\linewidth}{!}{
\begin{tikzpicture}

\def\axisdefaultwidth{9cm}
\def\axisdefaultheight{5.56cm}

\definecolor{my_cp5_col1}{RGB}{253, 231, 37}
\definecolor{my_cp5_col2}{RGB}{180, 222,44}
\definecolor{my_cp5_col3}{RGB}{94, 201, 98}
\definecolor{my_cp5_col4}{RGB}{33, 145, 140}
\definecolor{my_cp5_col5}{RGB}{59, 82, 139}
\definecolor{my_cp5_col6}{RGB}{68, 1, 84}

\definecolor{burlywood231187113}{RGB}{231,187,113}
\definecolor{darkgray158150204}{RGB}{158,150,204}
\definecolor{darkslategray66}{RGB}{66,66,66}
\definecolor{dimgray85}{RGB}{85,85,85}
\definecolor{gainsboro229}{RGB}{229,229,229}
\definecolor{gray119}{RGB}{119,119,119}
\definecolor{indianred2049072}{RGB}{204,90,72}
\definecolor{steelblue69133171}{RGB}{69,133,171}
\definecolor{yellowgreen13817181}{RGB}{138,171,81}

\begin{axis}[
axis background/.style={fill=gainsboro229},
axis line style={white},
tick align=outside,
x grid style={white},
xmajorticks=false,
xmin=-0.5, xmax=5.5,
xtick style={color=dimgray85},
xtick={0,1,2,3,4,5},
xticklabels={GrpA-1,GrpA-2,FCNN-1,FCNN-2,LieConv-1,LieConv-2},
y grid style={white},
ymajorgrids,
ymin=0.4, ymax=0.9,
ytick pos=left,
ytick style={color=dimgray85}
]
\draw[draw=none,fill=my_cp5_col6,very thin] (axis cs:-0.4,0) rectangle (axis cs:0.4,0.588652482269504);
\draw[draw=none,fill=my_cp5_col5,very thin] (axis cs:0.6,0) rectangle (axis cs:1.4,0.607659574468085);
\draw[draw=none,fill=my_cp5_col4,very thin] (axis cs:1.6,0) rectangle (axis cs:2.4,0.520070921985816);
\draw[draw=none,fill=my_cp5_col3,very thin] (axis cs:2.6,0) rectangle (axis cs:3.4,0.542056737588652);
\draw[draw=none,fill=my_cp5_col2,very thin] (axis cs:3.6,0) rectangle (axis cs:4.4,0.66);
\draw[draw=none,fill=my_cp5_col1,very thin] (axis cs:4.6,0) rectangle (axis cs:5.4,0.637801418439716);
\addplot [line width=1.08pt, darkslategray66]
table {%
0 0.572471631205674
0 0.606248226950355
};
\addplot [line width=1.08pt, darkslategray66]
table {%
1 0.58361170212766
1 0.632132978723404
};
\addplot [line width=1.08pt, darkslategray66]
table {%
2 0.506666666666667
2 0.535460992907801
};
\addplot [line width=1.08pt, darkslategray66]
table {%
3 0.521271276595745
3 0.561209219858156
};
\addplot [line width=1.08pt, darkslategray66]
table {%
4 0.651771276595745
4 0.667379432624113
};
\addplot [line width=1.08pt, darkslategray66]
table {%
5 0.626310283687943
5 0.649934397163121
};
\end{axis}

\end{tikzpicture}}
		\caption{Gaussian}
	\end{subfigure}
	\begin{subfigure}{.24\textwidth}
		\centering
    \resizebox{\linewidth}{!}{
\begin{tikzpicture}

\def\axisdefaultwidth{9cm}
\def\axisdefaultheight{5.56cm}

\definecolor{my_cp5_col1}{RGB}{253, 231, 37}
\definecolor{my_cp5_col2}{RGB}{180, 222,44}
\definecolor{my_cp5_col3}{RGB}{94, 201, 98}
\definecolor{my_cp5_col4}{RGB}{33, 145, 140}
\definecolor{my_cp5_col5}{RGB}{59, 82, 139}
\definecolor{my_cp5_col6}{RGB}{68, 1, 84}
\definecolor{burlywood231187113}{RGB}{231,187,113}
\definecolor{darkgray158150204}{RGB}{158,150,204}
\definecolor{darkslategray66}{RGB}{66,66,66}
\definecolor{dimgray85}{RGB}{85,85,85}
\definecolor{gainsboro229}{RGB}{229,229,229}
\definecolor{gray119}{RGB}{119,119,119}
\definecolor{indianred2049072}{RGB}{204,90,72}
\definecolor{steelblue69133171}{RGB}{69,133,171}
\definecolor{yellowgreen13817181}{RGB}{138,171,81}

\begin{axis}[
axis background/.style={fill=gainsboro229},
axis line style={white},
tick align=outside,
x grid style={white},
xmajorticks=false,
xmin=-0.5, xmax=5.5,
xmajorticks=false,
xtick style={color=dimgray85},
xtick={0,1,2,3,4,5},
xticklabels={GrpA-1,GrpA-2,FCNN-1,FCNN-2,LieConv-1,LieConv-2},
y grid style={white},
ymajorgrids,
ymin=0.4, ymax=0.9,
ytick pos=left,
ytick style={color=dimgray85}
]
\draw[draw=none,fill=my_cp5_col6,very thin] (axis cs:-0.4,0) rectangle (axis cs:0.4,0.665957446808511);
\draw[draw=none,fill=my_cp5_col5,very thin] (axis cs:0.6,0) rectangle (axis cs:1.4,0.683687943262411);
\draw[draw=none,fill=my_cp5_col4,very thin] (axis cs:1.6,0) rectangle (axis cs:2.4,0.56241134751773);
\draw[draw=none,fill=my_cp5_col3,very thin] (axis cs:2.6,0) rectangle (axis cs:3.4,0.638297872340426);
\draw[draw=none,fill=my_cp5_col2,very thin] (axis cs:3.6,0) rectangle (axis cs:4.4,0.640425531914894);
\draw[draw=none,fill=my_cp5_col1,very thin] (axis cs:4.6,0) rectangle (axis cs:5.4,0.661702127659575);
\addplot [line width=1.08pt, darkslategray66]
table {%
0 0.624822695035461
0 0.71063829787234
};
\addplot [line width=1.08pt, darkslategray66]
table {%
1 0.646099290780142
1 0.722695035460993
};
\addplot [line width=1.08pt, darkslategray66]
table {%
2 0.537588652482269
2 0.582978723404255
};
\addplot [line width=1.08pt, darkslategray66]
table {%
3 0.590780141843972
3 0.670921985815603
};
\addplot [line width=1.08pt, darkslategray66]
table {%
4 0.613475177304964
4 0.672340425531915
};
\addplot [line width=1.08pt, darkslategray66]
table {%
5 0.631914893617021
5 0.690780141843972
};
\end{axis}

\end{tikzpicture}}
		\label{fig:multiData-two}
    \resizebox{\linewidth}{!}{
\begin{tikzpicture}

\def\axisdefaultwidth{9cm}
\def\axisdefaultheight{5.56cm}

\definecolor{my_cp5_col1}{RGB}{253, 231, 37}
\definecolor{my_cp5_col2}{RGB}{180, 222,44}
\definecolor{my_cp5_col3}{RGB}{94, 201, 98}
\definecolor{my_cp5_col4}{RGB}{33, 145, 140}
\definecolor{my_cp5_col5}{RGB}{59, 82, 139}
\definecolor{my_cp5_col6}{RGB}{68, 1, 84}

\definecolor{burlywood231187113}{RGB}{231,187,113}
\definecolor{darkgray158150204}{RGB}{158,150,204}
\definecolor{darkslategray66}{RGB}{66,66,66}
\definecolor{dimgray85}{RGB}{85,85,85}
\definecolor{gainsboro229}{RGB}{229,229,229}
\definecolor{gray119}{RGB}{119,119,119}
\definecolor{indianred2049072}{RGB}{204,90,72}
\definecolor{steelblue69133171}{RGB}{69,133,171}
\definecolor{yellowgreen13817181}{RGB}{138,171,81}

\begin{axis}[
axis background/.style={fill=gainsboro229},
axis line style={white},
tick align=outside,
x grid style={white},
xmajorticks=false,
xmin=-0.5, xmax=5.5,
xtick style={color=dimgray85},
xtick={0,1,2,3,4,5},
xticklabels={GrpA-1,GrpA-2,FCNN-1,FCNN-2,LieConv-1,LieConv-2},
y grid style={white},
ymajorgrids,
ymin=0.4, ymax=0.9,
ytick pos=left,
ytick style={color=dimgray85}
]
\draw[draw=none,fill=my_cp5_col6,very thin] (axis cs:-0.4,0) rectangle (axis cs:0.4,0.639574468085106);
\draw[draw=none,fill=my_cp5_col5,very thin] (axis cs:0.6,0) rectangle (axis cs:1.4,0.671631205673759);
\draw[draw=none,fill=my_cp5_col4,very thin] (axis cs:1.6,0) rectangle (axis cs:2.4,0.533262411347518);
\draw[draw=none,fill=my_cp5_col3,very thin] (axis cs:2.6,0) rectangle (axis cs:3.4,0.571489361702128);
\draw[draw=none,fill=my_cp5_col2,very thin] (axis cs:3.6,0) rectangle (axis cs:4.4,0.690780141843972);
\draw[draw=none,fill=my_cp5_col1,very thin] (axis cs:4.6,0) rectangle (axis cs:5.4,0.676241134751773);
\addplot [line width=1.08pt, darkslategray66]
table {%
0 0.603957446808511
0 0.672001773049645
};
\addplot [line width=1.08pt, darkslategray66]
table {%
1 0.654184397163121
1 0.687946808510638
};
\addplot [line width=1.08pt, darkslategray66]
table {%
2 0.506666666666667
2 0.563911347517731
};
\addplot [line width=1.08pt, darkslategray66]
table {%
3 0.538363475177305
3 0.601845744680851
};
\addplot [line width=1.08pt, darkslategray66]
table {%
4 0.663326241134752
4 0.715253546099291
};
\addplot [line width=1.08pt, darkslategray66]
table {%
5 0.66063829787234
5 0.691631205673759
};
\end{axis}

\end{tikzpicture}}
		\caption{Grid}
		\label{fig:multiData-two}
	\end{subfigure}
		\begin{subfigure}{.24\textwidth}
		\centering
\resizebox{\linewidth}{!}{
\begin{tikzpicture}

\def\axisdefaultwidth{9cm}
\def\axisdefaultheight{5.56cm}

\definecolor{my_cp5_col1}{RGB}{253, 231, 37}
\definecolor{my_cp5_col2}{RGB}{180, 222,44}
\definecolor{my_cp5_col3}{RGB}{94, 201, 98}
\definecolor{my_cp5_col4}{RGB}{33, 145, 140}
\definecolor{my_cp5_col5}{RGB}{59, 82, 139}
\definecolor{my_cp5_col6}{RGB}{68, 1, 84}
\definecolor{burlywood231187113}{RGB}{231,187,113}
\definecolor{darkgray158150204}{RGB}{158,150,204}
\definecolor{darkslategray66}{RGB}{66,66,66}
\definecolor{dimgray85}{RGB}{85,85,85}
\definecolor{gainsboro229}{RGB}{229,229,229}
\definecolor{gray119}{RGB}{119,119,119}
\definecolor{indianred2049072}{RGB}{204,90,72}
\definecolor{steelblue69133171}{RGB}{69,133,171}
\definecolor{yellowgreen13817181}{RGB}{138,171,81}

\begin{axis}[
axis background/.style={fill=gainsboro229},
axis line style={white},
tick align=outside,
x grid style={white},
xmajorticks=false,
xmin=-0.5, xmax=5.5,
xtick style={color=dimgray85},
xtick={0,1,2,3,4,5},
xticklabels={GrpA-1,GrpA-2,FCNN-1,FCNN-2,LieConv-1,LieConv-2},
y grid style={white},
ymajorgrids,
ymin=0.4, ymax=0.9,
ytick pos=left,
ytick style={color=dimgray85}
]
\draw[draw=none,fill=my_cp5_col6,very thin] (axis cs:-0.4,0) rectangle (axis cs:0.4,0.653900709219858);
\draw[draw=none,fill=my_cp5_col5,very thin] (axis cs:0.6,0) rectangle (axis cs:1.4,0.65531914893617);
\draw[draw=none,fill=my_cp5_col4,very thin] (axis cs:1.6,0) rectangle (axis cs:2.4,0.599290780141844);
\draw[draw=none,fill=my_cp5_col3,very thin] (axis cs:2.6,0) rectangle (axis cs:3.4,0.648936170212766);
\draw[draw=none,fill=my_cp5_col2,very thin] (axis cs:3.6,0) rectangle (axis cs:4.4,0.669503546099291);
\draw[draw=none,fill=my_cp5_col1,very thin] (axis cs:4.6,0) rectangle (axis cs:5.4,0.631205673758865);
\addplot [line width=1.08pt, darkslategray66]
table {%
0 0.640425531914894
0 0.668794326241135
};
\addplot [line width=1.08pt, darkslategray66]
table {%
1 0.625514184397163
1 0.685833333333333
};
\addplot [line width=1.08pt, darkslategray66]
table {%
2 0.570212765957447
2 0.621985815602837
};
\addplot [line width=1.08pt, darkslategray66]
table {%
3 0.604255319148936
3 0.685124113475177
};
\addplot [line width=1.08pt, darkslategray66]
table {%
4 0.642553191489362
4 0.697163120567376
};
\addplot [line width=1.08pt, darkslategray66]
table {%
5 0.599290780141844
5 0.665975177304964
};
\end{axis}

\end{tikzpicture}}
		\label{fig:multiData-two}
\resizebox{\linewidth}{!}{
\begin{tikzpicture}

\def\axisdefaultwidth{9cm}
\def\axisdefaultheight{5.56cm}

\definecolor{my_cp5_col1}{RGB}{253, 231, 37}
\definecolor{my_cp5_col2}{RGB}{180, 222,44}
\definecolor{my_cp5_col3}{RGB}{94, 201, 98}
\definecolor{my_cp5_col4}{RGB}{33, 145, 140}
\definecolor{my_cp5_col5}{RGB}{59, 82, 139}
\definecolor{my_cp5_col6}{RGB}{68, 1, 84}

\definecolor{burlywood231187113}{RGB}{231,187,113}
\definecolor{darkgray158150204}{RGB}{158,150,204}
\definecolor{darkslategray66}{RGB}{66,66,66}
\definecolor{dimgray85}{RGB}{85,85,85}
\definecolor{gainsboro229}{RGB}{229,229,229}
\definecolor{gray119}{RGB}{119,119,119}
\definecolor{indianred2049072}{RGB}{204,90,72}
\definecolor{steelblue69133171}{RGB}{69,133,171}
\definecolor{yellowgreen13817181}{RGB}{138,171,81}

\begin{axis}[
axis background/.style={fill=gainsboro229},
axis line style={white},
tick align=outside,
x grid style={white},
xmajorticks=false,
xmin=-0.5, xmax=5.5,
xtick style={color=dimgray85},
xtick={0,1,2,3,4,5},
xticklabels={GrpA-1,GrpA-2,FCNN-1,FCNN-2,LieConv-1,LieConv-2},
y grid style={white},
ymajorgrids,
ymin=0.4, ymax=0.9,
ytick pos=left,
ytick style={color=dimgray85}
]
\draw[draw=none,fill=my_cp5_col6,very thin] (axis cs:-0.4,0) rectangle (axis cs:0.4,0.57886524822695);
\draw[draw=none,fill=my_cp5_col5,very thin] (axis cs:0.6,0) rectangle (axis cs:1.4,0.609148936170213);
\draw[draw=none,fill=my_cp5_col4,very thin] (axis cs:1.6,0) rectangle (axis cs:2.4,0.535106382978723);
\draw[draw=none,fill=my_cp5_col3,very thin] (axis cs:2.6,0) rectangle (axis cs:3.4,0.566737588652482);
\draw[draw=none,fill=my_cp5_col2,very thin] (axis cs:3.6,0) rectangle (axis cs:4.4,0.658936170212766);
\draw[draw=none,fill=my_cp5_col1,very thin] (axis cs:4.6,0) rectangle (axis cs:5.4,0.657943262411347);
\addplot [line width=1.08pt, darkslategray66]
table {%
0 0.568297872340426
0 0.590141843971631
};
\addplot [line width=1.08pt, darkslategray66]
table {%
1 0.593182624113475
1 0.63063829787234
};
\addplot [line width=1.08pt, darkslategray66]
table {%
2 0.509290780141844
2 0.56220390070922
};
\addplot [line width=1.08pt, darkslategray66]
table {%
3 0.535886524822695
3 0.596813829787234
};
\addplot [line width=1.08pt, darkslategray66]
table {%
4 0.6295
4 0.68688475177305
};
\addplot [line width=1.08pt, darkslategray66]
table {%
5 0.624397163120567
5 0.687166666666667
};
\end{axis}

\end{tikzpicture}}
		\caption{Uniform}
		\label{fig:multiData-two}
	\end{subfigure}
	\caption{Test accuracy on binary knot classification: Signals are defined on Sphere, Gaussian, Grid, and Uniform grids with few samples (Top row: $|\widehat{\ccalX}| = 125$) and many samples (Bottom row: $|\widehat{\ccalX}| = 1000$). Each filter is evaluated with one and two layers. }
	\label{fig_knot_numsim_small}
\end{figure*}


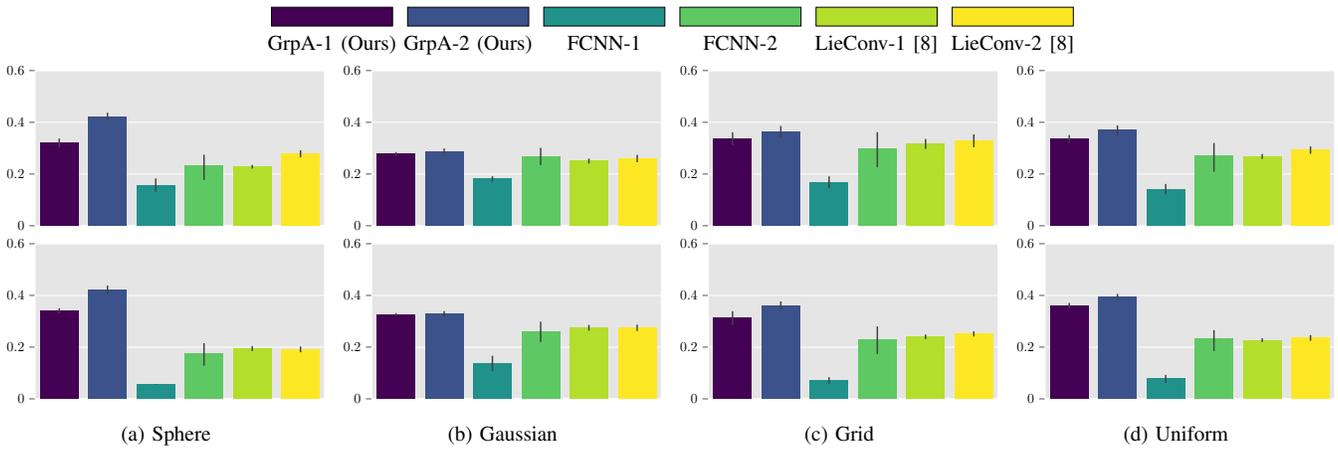
\begin{figure*}[t]
\centering


\definecolor{my_cp5_col1}{RGB}{253, 231, 37}
\definecolor{my_cp5_col2}{RGB}{180, 222,44}
\definecolor{my_cp5_col3}{RGB}{94, 201, 98}
\definecolor{my_cp5_col4}{RGB}{33, 145, 140}
\definecolor{my_cp5_col5}{RGB}{59, 82, 139}
\definecolor{my_cp5_col6}{RGB}{68, 1, 84}

\usetikzlibrary{positioning,decorations.pathreplacing,shapes}


\def \scale {1}
\def \unit { \scale cm}

\def \vertsep {0.5*\scale}

\def \horzsep {1*\scale}


\tikzstyle{set} = [rectangle,color=black,
                    rounded corners = 0*\unit,
                    fill=black,
                    inner sep=0pt,
                    draw,
                    anchor = center,
                    line width=0.1mm]
                    
 
\tikzstyle{myboxlabel} = [set,
fill=my_cp5_col3,
minimum width  = 1.61*\unit,
minimum height = 0.25*\unit]

\tikzstyle{dot} = [ circle,
                    minimum width  = 0.05*\unit,
                    fill=black,
                    color=black,
                    inner sep=0pt,
                    draw,
                    anchor = center ]


{\fontsize{8}{8}\selectfont

\begin{tikzpicture}[scale=\scale,rounded corners,ultra thick]


   
   \path (0,0) node [myboxlabel,fill=my_cp5_col6] (L1) {};
   \path (L1.south) ++ (0, 0) node [below, color=black] {GrpA-1 (Ours)};

   
    \path (L1.east)++(\horzsep,0) node [myboxlabel,fill=my_cp5_col5] (L2) {};
    \path (L2.south) ++ (0, 0) node [below, color=black] {GrpA-2 (Ours)};

     
     \path (L2.east)++(\horzsep,0) node [myboxlabel,fill=my_cp5_col4] (L3) {};
     \path (L3.south) ++ (0, 0) node [below, color=black] {FCNN-1};

         
     \path (L3.east)++(\horzsep,0) node [myboxlabel,fill=my_cp5_col3] (L4) {};
     \path (L4.south) ++ (0, 0) node [below, color=black] {FCNN-2};

      
      \path (L4.east)++(\horzsep,0) node [myboxlabel,fill=my_cp5_col2] (L5) {};
      \path (L5.south) ++ (0, 0) node [below, color=black] {LieConv-1\cite{finzi2020generalizing}};

      
      \path (L5.east)++(\horzsep,0) node [myboxlabel,fill=my_cp5_col1] (L6) {};
      \path (L6.south) ++ (0, 0) node [below, color=black] {LieConv-2\cite{finzi2020generalizing}};

%
%

\end{tikzpicture}

}
	\centering
	\begin{subfigure}{.24\textwidth}
		\centering
    \resizebox{\linewidth}{!}{
\begin{tikzpicture}

\def\axisdefaultwidth{9cm}
\def\axisdefaultheight{5.56cm}

\definecolor{my_cp5_col1}{RGB}{253, 231, 37}
\definecolor{my_cp5_col2}{RGB}{180, 222,44}
\definecolor{my_cp5_col3}{RGB}{94, 201, 98}
\definecolor{my_cp5_col4}{RGB}{33, 145, 140}
\definecolor{my_cp5_col5}{RGB}{59, 82, 139}
\definecolor{my_cp5_col6}{RGB}{68, 1, 84}

\definecolor{burlywood231187113}{RGB}{231,187,113}
\definecolor{darkgray158150204}{RGB}{158,150,204}
\definecolor{darkslategray66}{RGB}{66,66,66}
\definecolor{dimgray85}{RGB}{85,85,85}
\definecolor{gainsboro229}{RGB}{229,229,229}
\definecolor{gray119}{RGB}{119,119,119}
\definecolor{indianred2049072}{RGB}{204,90,72}
\definecolor{steelblue69133171}{RGB}{69,133,171}
\definecolor{yellowgreen13817181}{RGB}{138,171,81}

\begin{axis}[
axis background/.style={fill=gainsboro229},
axis line style={white},
tick align=outside,
x grid style={white},
xmajorticks=false,
xmin=-0.5, xmax=5.5,
xtick style={color=dimgray85},
xtick={0,1,2,3,4,5},
xticklabels={GrpA-1,GrpA-2,FCNN-1,FCNN-2,LieConv-1,LieConv-2},
y grid style={white},
ymajorgrids,
ymin=0, ymax=0.6,
ytick pos=left,
ytick style={color=dimgray85}
]

\draw[draw=none,fill=my_cp5_col6,very thin] (axis cs:-0.4,0) rectangle (axis cs:0.4,0.320925110132159);

\draw[draw=none,fill=my_cp5_col5,very thin] (axis cs:0.6,0) rectangle (axis cs:1.4,0.422356828193833);

\draw[draw=none,fill=my_cp5_col4,very thin] (axis cs:1.6,0) rectangle (axis cs:2.4,0.155947136563877);
\draw[draw=none,fill=my_cp5_col3,very thin] (axis cs:2.6,0) rectangle (axis cs:3.4,0.231057268722467);
\draw[draw=none,fill=my_cp5_col2,very thin] (axis cs:3.6,0) rectangle (axis cs:4.4,0.227973568281938);
\draw[draw=none,fill=my_cp5_col1,very thin] (axis cs:4.6,0) rectangle (axis cs:5.4,0.278854625550661);

\addplot [line width=1.08pt, darkslategray66]
table {%
0 0.30352422907489
0 0.337555066079295
};
\addplot [line width=1.08pt, darkslategray66]
table {%
1 0.408480176211454
1 0.437007158590308
};
\addplot [line width=1.08pt, darkslategray66]
table {%
2 0.13182268722467
2 0.182714757709251
};
\addplot [line width=1.08pt, darkslategray66]
table {%
3 0.176960352422907
3 0.274672356828194
};
\addplot [line width=1.08pt, darkslategray66]
table {%
4 0.221137114537445
4 0.235134911894273
};
\addplot [line width=1.08pt, darkslategray66]
table {%
5 0.264424559471366
5 0.291646475770925
};
\end{axis}

\end{tikzpicture}}
    \resizebox{\linewidth}{!}{
\begin{tikzpicture}

\def\axisdefaultwidth{9cm}
\def\axisdefaultheight{5.56cm}

\definecolor{my_cp5_col1}{RGB}{253, 231, 37}
\definecolor{my_cp5_col2}{RGB}{180, 222,44}
\definecolor{my_cp5_col3}{RGB}{94, 201, 98}
\definecolor{my_cp5_col4}{RGB}{33, 145, 140}
\definecolor{my_cp5_col5}{RGB}{59, 82, 139}
\definecolor{my_cp5_col6}{RGB}{68, 1, 84}

\definecolor{burlywood231187113}{RGB}{231,187,113}
\definecolor{darkgray158150204}{RGB}{158,150,204}
\definecolor{darkslategray66}{RGB}{66,66,66}
\definecolor{dimgray85}{RGB}{85,85,85}
\definecolor{gainsboro229}{RGB}{229,229,229}
\definecolor{gray119}{RGB}{119,119,119}
\definecolor{indianred2049072}{RGB}{204,90,72}
\definecolor{steelblue69133171}{RGB}{69,133,171}
\definecolor{yellowgreen13817181}{RGB}{138,171,81}

\begin{axis}[
axis background/.style={fill=gainsboro229},
axis line style={white},
tick align=outside,
x grid style={white},
xmajorticks=false,
xmin=-0.5, xmax=5.5,
xtick style={color=dimgray85},
xtick={0,1,2,3,4,5},
xticklabels={GrpA-1,GrpA-2,FCNN-1,FCNN-2,LieConv-1,LieConv-2},
y grid style={white},
ymajorgrids,
ymin=0, ymax=0.6,
ytick pos=left,
ytick style={color=dimgray85}
]
\draw[draw=none,fill=my_cp5_col6,very thin] (axis cs:-0.4,0) rectangle (axis cs:0.4,0.340264317180617);
\draw[draw=none,fill=my_cp5_col5,very thin] (axis cs:0.6,0) rectangle (axis cs:1.4,0.421607929515419);
\draw[draw=none,fill=my_cp5_col4,very thin] (axis cs:1.6,0) rectangle (axis cs:2.4,0.055726872246696);
\draw[draw=none,fill=my_cp5_col3,very thin] (axis cs:2.6,0) rectangle (axis cs:3.4,0.1759140969163);
\draw[draw=none,fill=my_cp5_col2,very thin] (axis cs:3.6,0) rectangle (axis cs:4.4,0.19557268722467);
\draw[draw=none,fill=my_cp5_col1,very thin] (axis cs:4.6,0) rectangle (axis cs:5.4,0.190903083700441);
\addplot [line width=1.08pt, darkslategray66]
table {%
0 0.330406662995595
0 0.350661343612335
};
\addplot [line width=1.08pt, darkslategray66]
table {%
1 0.403525330396476
1 0.438338931718062
};
\addplot [line width=1.08pt, darkslategray66]
table {%
2 0.0550660792951542
2 0.0570484581497797
};
\addplot [line width=1.08pt, darkslategray66]
table {%
3 0.12847577092511
3 0.21555781938326
};
\addplot [line width=1.08pt, darkslategray66]
table {%
4 0.184900605726872
4 0.204009911894273
};
\addplot [line width=1.08pt, darkslategray66]
table {%
5 0.180272852422907
5 0.202469713656388
};
\end{axis}

\end{tikzpicture}}

		\caption{Sphere}
	\end{subfigure}
	\begin{subfigure}{.24\textwidth}
		\centering
		    \resizebox{\linewidth}{!}{
\begin{tikzpicture}

\def\axisdefaultwidth{9cm}
\def\axisdefaultheight{5.56cm}

\definecolor{my_cp5_col1}{RGB}{253, 231, 37}
\definecolor{my_cp5_col2}{RGB}{180, 222,44}
\definecolor{my_cp5_col3}{RGB}{94, 201, 98}
\definecolor{my_cp5_col4}{RGB}{33, 145, 140}
\definecolor{my_cp5_col5}{RGB}{59, 82, 139}
\definecolor{my_cp5_col6}{RGB}{68, 1, 84}

\definecolor{burlywood231187113}{RGB}{231,187,113}
\definecolor{darkgray158150204}{RGB}{158,150,204}
\definecolor{darkslategray66}{RGB}{66,66,66}
\definecolor{dimgray85}{RGB}{85,85,85}
\definecolor{gainsboro229}{RGB}{229,229,229}
\definecolor{gray119}{RGB}{119,119,119}
\definecolor{indianred2049072}{RGB}{204,90,72}
\definecolor{steelblue69133171}{RGB}{69,133,171}
\definecolor{yellowgreen13817181}{RGB}{138,171,81}

\begin{axis}[
axis background/.style={fill=gainsboro229},
axis line style={white},
tick align=outside,
x grid style={white},
xmajorticks=false,
xmin=-0.5, xmax=5.5,
xtick style={color=dimgray85},
xtick={0,1,2,3,4,5},
xticklabels={GrpA-1,GrpA-2,FCNN-1,FCNN-2,LieConv-1,LieConv-2},
y grid style={white},
ymajorgrids,
ymin=0, ymax=0.6,
ytick pos=left,
ytick style={color=dimgray85}
]
\draw[draw=none,fill=my_cp5_col6,very thin] (axis cs:-0.4,0) rectangle (axis cs:0.4,0.277973568281938);
\draw[draw=none,fill=my_cp5_col5,very thin] (axis cs:0.6,0) rectangle (axis cs:1.4,0.287665198237885);
\draw[draw=none,fill=my_cp5_col4,very thin] (axis cs:1.6,0) rectangle (axis cs:2.4,0.181497797356828);
\draw[draw=none,fill=my_cp5_col3,very thin] (axis cs:2.6,0) rectangle (axis cs:3.4,0.266850220264317);
\draw[draw=none,fill=my_cp5_col2,very thin] (axis cs:3.6,0) rectangle (axis cs:4.4,0.250330396475771);
\draw[draw=none,fill=my_cp5_col1,very thin] (axis cs:4.6,0) rectangle (axis cs:5.4,0.26079295154185);
\addplot [line width=1.08pt, darkslategray66]
table {%
0 0.270812224669603
0 0.284691629955947
};
\addplot [line width=1.08pt, darkslategray66]
table {%
1 0.275770925110132
1 0.298788546255507
};
\addplot [line width=1.08pt, darkslategray66]
table {%
2 0.170261563876652
2 0.191409691629956
};
\addplot [line width=1.08pt, darkslategray66]
table {%
3 0.234471365638767
3 0.301002202643172
};
\addplot [line width=1.08pt, darkslategray66]
table {%
4 0.240748898678414
4 0.259581497797357
};
\addplot [line width=1.08pt, darkslategray66]
table {%
5 0.245809471365639
5 0.274344713656388
};
\end{axis}

\end{tikzpicture}}
		\label{fig:multiData-two}
    \resizebox{\linewidth}{!}{
\begin{tikzpicture}

\def\axisdefaultwidth{9cm}
\def\axisdefaultheight{5.56cm}

\definecolor{my_cp5_col1}{RGB}{253, 231, 37}
\definecolor{my_cp5_col2}{RGB}{180, 222,44}
\definecolor{my_cp5_col3}{RGB}{94, 201, 98}
\definecolor{my_cp5_col4}{RGB}{33, 145, 140}
\definecolor{my_cp5_col5}{RGB}{59, 82, 139}
\definecolor{my_cp5_col6}{RGB}{68, 1, 84}

\definecolor{burlywood231187113}{RGB}{231,187,113}
\definecolor{darkgray158150204}{RGB}{158,150,204}
\definecolor{darkslategray66}{RGB}{66,66,66}
\definecolor{dimgray85}{RGB}{85,85,85}
\definecolor{gainsboro229}{RGB}{229,229,229}
\definecolor{gray119}{RGB}{119,119,119}
\definecolor{indianred2049072}{RGB}{204,90,72}
\definecolor{steelblue69133171}{RGB}{69,133,171}
\definecolor{yellowgreen13817181}{RGB}{138,171,81}

\begin{axis}[
axis background/.style={fill=gainsboro229},
axis line style={white},
tick align=outside,
x grid style={white},
xmajorticks=false,
xmin=-0.5, xmax=5.5,
xtick style={color=dimgray85},
xtick={0,1,2,3,4,5},
xticklabels={GrpA-1,GrpA-2,FCNN-1,FCNN-2,LieConv-1,LieConv-2},
y grid style={white},
ymajorgrids,
ymin=0, ymax=0.6,
ytick pos=left,
ytick style={color=dimgray85}
]
\draw[draw=none,fill=my_cp5_col6,very thin] (axis cs:-0.4,0) rectangle (axis cs:0.4,0.325859030837004);
\draw[draw=none,fill=my_cp5_col5,very thin] (axis cs:0.6,0) rectangle (axis cs:1.4,0.329801762114537);
\draw[draw=none,fill=my_cp5_col4,very thin] (axis cs:1.6,0) rectangle (axis cs:2.4,0.137301762114537);
\draw[draw=none,fill=my_cp5_col3,very thin] (axis cs:2.6,0) rectangle (axis cs:3.4,0.261310572687225);
\draw[draw=none,fill=my_cp5_col2,very thin] (axis cs:3.6,0) rectangle (axis cs:4.4,0.274779735682819);
\draw[draw=none,fill=my_cp5_col1,very thin] (axis cs:4.6,0) rectangle (axis cs:5.4,0.276277533039648);
\addplot [line width=1.08pt, darkslategray66]
table {%
0 0.320131883259912
0 0.331708700440529
};
\addplot [line width=1.08pt, darkslategray66]
table {%
1 0.318534140969163
1 0.339627477973568
};
\addplot [line width=1.08pt, darkslategray66]
table {%
2 0.106981002202643
2 0.166674284140969
};
\addplot [line width=1.08pt, darkslategray66]
table {%
3 0.219585077092511
3 0.298526156387665
};
\addplot [line width=1.08pt, darkslategray66]
table {%
4 0.263897026431718
4 0.28557654185022
};
\addplot [line width=1.08pt, darkslategray66]
table {%
5 0.262309196035242
5 0.287092511013216
};
\end{axis}

\end{tikzpicture}}

		\caption{Gaussian}
	\end{subfigure}
	\begin{subfigure}{.24\textwidth}
		\centering
		    \resizebox{\linewidth}{!}{
\begin{tikzpicture}

\def\axisdefaultwidth{9cm}
\def\axisdefaultheight{5.56cm}

\definecolor{my_cp5_col1}{RGB}{253, 231, 37}
\definecolor{my_cp5_col2}{RGB}{180, 222,44}
\definecolor{my_cp5_col3}{RGB}{94, 201, 98}
\definecolor{my_cp5_col4}{RGB}{33, 145, 140}
\definecolor{my_cp5_col5}{RGB}{59, 82, 139}
\definecolor{my_cp5_col6}{RGB}{68, 1, 84}

\definecolor{burlywood231187113}{RGB}{231,187,113}
\definecolor{darkgray158150204}{RGB}{158,150,204}
\definecolor{darkslategray66}{RGB}{66,66,66}
\definecolor{dimgray85}{RGB}{85,85,85}
\definecolor{gainsboro229}{RGB}{229,229,229}
\definecolor{gray119}{RGB}{119,119,119}
\definecolor{indianred2049072}{RGB}{204,90,72}
\definecolor{steelblue69133171}{RGB}{69,133,171}
\definecolor{yellowgreen13817181}{RGB}{138,171,81}

\begin{axis}[
axis background/.style={fill=gainsboro229},
axis line style={white},
tick align=outside,
x grid style={white},
xmajorticks=false,
xmin=-0.5, xmax=5.5,
xtick style={color=dimgray85},
xtick={0,1,2,3,4,5},
xticklabels={GrpA-1,GrpA-2,FCNN-1,FCNN-2,LieConv-1,LieConv-2},
y grid style={white},
ymajorgrids,
ymin=0, ymax=0.6,
ytick pos=left,
ytick style={color=dimgray85}
]
\draw[draw=none,fill=my_cp5_col6,very thin] (axis cs:-0.4,0) rectangle (axis cs:0.4,0.337334801762115);
\draw[draw=none,fill=my_cp5_col5,very thin] (axis cs:0.6,0) rectangle (axis cs:1.4,0.364647577092511);
\draw[draw=none,fill=my_cp5_col4,very thin] (axis cs:1.6,0) rectangle (axis cs:2.4,0.16784140969163);
\draw[draw=none,fill=my_cp5_col3,very thin] (axis cs:2.6,0) rectangle (axis cs:3.4,0.299449339207048);
\draw[draw=none,fill=my_cp5_col2,very thin] (axis cs:3.6,0) rectangle (axis cs:4.4,0.316299559471366);
\draw[draw=none,fill=my_cp5_col1,very thin] (axis cs:4.6,0) rectangle (axis cs:5.4,0.329185022026432);
\addplot [line width=1.08pt, darkslategray66]
table {%
0 0.310132158590308
0 0.361247246696035
};
\addplot [line width=1.08pt, darkslategray66]
table {%
1 0.341949339207048
1 0.385691079295154
};
\addplot [line width=1.08pt, darkslategray66]
table {%
2 0.144810022026432
2 0.19108204845815
};
\addplot [line width=1.08pt, darkslategray66]
table {%
3 0.226756607929515
3 0.362026431718062
};
\addplot [line width=1.08pt, darkslategray66]
table {%
4 0.297026431718062
4 0.33557268722467
};
\addplot [line width=1.08pt, darkslategray66]
table {%
5 0.303741740088106
5 0.353309471365639
};
\end{axis}

\end{tikzpicture}}
		\label{fig:multiData-two}
    \resizebox{\linewidth}{!}{
\begin{tikzpicture}

\def\axisdefaultwidth{9cm}
\def\axisdefaultheight{5.56cm}

\definecolor{my_cp5_col1}{RGB}{253, 231, 37}
\definecolor{my_cp5_col2}{RGB}{180, 222,44}
\definecolor{my_cp5_col3}{RGB}{94, 201, 98}
\definecolor{my_cp5_col4}{RGB}{33, 145, 140}
\definecolor{my_cp5_col5}{RGB}{59, 82, 139}
\definecolor{my_cp5_col6}{RGB}{68, 1, 84}
\definecolor{burlywood231187113}{RGB}{231,187,113}
\definecolor{darkgray158150204}{RGB}{158,150,204}
\definecolor{darkslategray66}{RGB}{66,66,66}
\definecolor{dimgray85}{RGB}{85,85,85}
\definecolor{gainsboro229}{RGB}{229,229,229}
\definecolor{gray119}{RGB}{119,119,119}
\definecolor{indianred2049072}{RGB}{204,90,72}
\definecolor{steelblue69133171}{RGB}{69,133,171}
\definecolor{yellowgreen13817181}{RGB}{138,171,81}

\begin{axis}[
axis background/.style={fill=gainsboro229},
axis line style={white},
tick align=outside,
x grid style={white},
xmajorticks=false,
xmin=-0.5, xmax=5.5,
xtick style={color=dimgray85},
xtick={0,1,2,3,4,5},
xticklabels={GrpA-1,GrpA-2,FCNN-1,FCNN-2,LieConv-1,LieConv-2},
y grid style={white},
ymajorgrids,
ymin=0, ymax=0.6,
ytick pos=left,
ytick style={color=dimgray85}
]
\draw[draw=none,fill=my_cp5_col6,very thin] (axis cs:-0.4,0) rectangle (axis cs:0.4,0.313281938325991);
\draw[draw=none,fill=my_cp5_col5,very thin] (axis cs:0.6,0) rectangle (axis cs:1.4,0.360462555066079);
\draw[draw=none,fill=my_cp5_col4,very thin] (axis cs:1.6,0) rectangle (axis cs:2.4,0.0699559471365639);
\draw[draw=none,fill=my_cp5_col3,very thin] (axis cs:2.6,0) rectangle (axis cs:3.4,0.22715859030837);
\draw[draw=none,fill=my_cp5_col2,very thin] (axis cs:3.6,0) rectangle (axis cs:4.4,0.239383259911894);
\draw[draw=none,fill=my_cp5_col1,very thin] (axis cs:4.6,0) rectangle (axis cs:5.4,0.250649779735683);
\addplot [line width=1.08pt, darkslategray66]
table {%
0 0.286109306167401
0 0.339104074889868
};
\addplot [line width=1.08pt, darkslategray66]
table {%
1 0.345886839207048
1 0.376762389867841
};
\addplot [line width=1.08pt, darkslategray66]
table {%
2 0.0583590308370044
2 0.0839647577092511
};
\addplot [line width=1.08pt, darkslategray66]
table {%
3 0.173785792951542
3 0.280291024229075
};
\addplot [line width=1.08pt, darkslategray66]
table {%
4 0.231208425110132
4 0.24848045154185
};
\addplot [line width=1.08pt, darkslategray66]
table {%
5 0.241177588105727
5 0.261234856828194
};
\end{axis}

\end{tikzpicture}}
		\caption{Grid}
		\label{fig:multiData-two}
	\end{subfigure}
		\begin{subfigure}{.24\textwidth}
		\centering
    \resizebox{\linewidth}{!}{
\begin{tikzpicture}

\def\axisdefaultwidth{9cm}
\def\axisdefaultheight{5.56cm}

\definecolor{my_cp5_col1}{RGB}{253, 231, 37}
\definecolor{my_cp5_col2}{RGB}{180, 222,44}
\definecolor{my_cp5_col3}{RGB}{94, 201, 98}
\definecolor{my_cp5_col4}{RGB}{33, 145, 140}
\definecolor{my_cp5_col5}{RGB}{59, 82, 139}
\definecolor{my_cp5_col6}{RGB}{68, 1, 84}

\definecolor{burlywood231187113}{RGB}{231,187,113}
\definecolor{darkgray158150204}{RGB}{158,150,204}
\definecolor{darkslategray66}{RGB}{66,66,66}
\definecolor{dimgray85}{RGB}{85,85,85}
\definecolor{gainsboro229}{RGB}{229,229,229}
\definecolor{gray119}{RGB}{119,119,119}
\definecolor{indianred2049072}{RGB}{204,90,72}
\definecolor{steelblue69133171}{RGB}{69,133,171}
\definecolor{yellowgreen13817181}{RGB}{138,171,81}

\begin{axis}[
axis background/.style={fill=gainsboro229},
axis line style={white},
tick align=outside,
x grid style={white},
xmajorticks=false,
xmin=-0.5, xmax=5.5,
xtick style={color=dimgray85},
xtick={0,1,2,3,4,5},
xticklabels={GrpA-1,GrpA-2,FCNN-1,FCNN-2,LieConv-1,LieConv-2},
y grid style={white},
ymajorgrids,
ymin=0, ymax=0.6,
ytick pos=left,
ytick style={color=dimgray85}
]
\draw[draw=none,fill=my_cp5_col6,very thin] (axis cs:-0.4,0) rectangle (axis cs:0.4,0.335352422907489);
\draw[draw=none,fill=my_cp5_col5,very thin] (axis cs:0.6,0) rectangle (axis cs:1.4,0.3715859030837);
\draw[draw=none,fill=my_cp5_col4,very thin] (axis cs:1.6,0) rectangle (axis cs:2.4,0.141740088105727);
\draw[draw=none,fill=my_cp5_col3,very thin] (axis cs:2.6,0) rectangle (axis cs:3.4,0.269603524229075);
\draw[draw=none,fill=my_cp5_col2,very thin] (axis cs:3.6,0) rectangle (axis cs:4.4,0.26795154185022);
\draw[draw=none,fill=my_cp5_col1,very thin] (axis cs:4.6,0) rectangle (axis cs:5.4,0.293392070484581);
\addplot [line width=1.08pt, darkslategray66]
table {%
0 0.320154185022026
0 0.350881057268722
};
\addplot [line width=1.08pt, darkslategray66]
table {%
1 0.349666850220264
1 0.388215859030837
};
\addplot [line width=1.08pt, darkslategray66]
table {%
2 0.122020925110132
2 0.161569383259912
};
\addplot [line width=1.08pt, darkslategray66]
table {%
3 0.208689427312775
3 0.320154185022026
};
\addplot [line width=1.08pt, darkslategray66]
table {%
4 0.25880781938326
4 0.277422907488987
};
\addplot [line width=1.08pt, darkslategray66]
table {%
5 0.278191079295154
5 0.306720814977974
};
\end{axis}

\end{tikzpicture}}

		\label{fig:multiData-two}
				    \resizebox{\linewidth}{!}{
\begin{tikzpicture}

\def\axisdefaultwidth{9cm}
\def\axisdefaultheight{5.56cm}

\definecolor{my_cp5_col1}{RGB}{253, 231, 37}
\definecolor{my_cp5_col2}{RGB}{180, 222,44}
\definecolor{my_cp5_col3}{RGB}{94, 201, 98}
\definecolor{my_cp5_col4}{RGB}{33, 145, 140}
\definecolor{my_cp5_col5}{RGB}{59, 82, 139}
\definecolor{my_cp5_col6}{RGB}{68, 1, 84}

\definecolor{burlywood231187113}{RGB}{231,187,113}
\definecolor{darkgray158150204}{RGB}{158,150,204}
\definecolor{darkslategray66}{RGB}{66,66,66}
\definecolor{dimgray85}{RGB}{85,85,85}
\definecolor{gainsboro229}{RGB}{229,229,229}
\definecolor{gray119}{RGB}{119,119,119}
\definecolor{indianred2049072}{RGB}{204,90,72}
\definecolor{steelblue69133171}{RGB}{69,133,171}
\definecolor{yellowgreen13817181}{RGB}{138,171,81}

\begin{axis}[
axis background/.style={fill=gainsboro229},
axis line style={white},
tick align=outside,
x grid style={white},
xmajorticks=false,
xmin=-0.5, xmax=5.5,
xtick style={color=dimgray85},
xtick={0,1,2,3,4,5},
xticklabels={GrpA-1,GrpA-2,FCNN-1,FCNN-2,LieConv-1,LieConv-2},
y grid style={white},
ymajorgrids,
ymin=0, ymax=0.6,
ytick pos=left,
ytick style={color=dimgray85}
]
\draw[draw=none,fill=my_cp5_col6,very thin] (axis cs:-0.4,0) rectangle (axis cs:0.4,0.360484581497797);
\draw[draw=none,fill=my_cp5_col5,very thin] (axis cs:0.6,0) rectangle (axis cs:1.4,0.394702643171806);
\draw[draw=none,fill=my_cp5_col4,very thin] (axis cs:1.6,0) rectangle (axis cs:2.4,0.0773788546255507);
\draw[draw=none,fill=my_cp5_col3,very thin] (axis cs:2.6,0) rectangle (axis cs:3.4,0.232522026431718);
\draw[draw=none,fill=my_cp5_col2,very thin] (axis cs:3.6,0) rectangle (axis cs:4.4,0.226773127753304);
\draw[draw=none,fill=my_cp5_col1,very thin] (axis cs:4.6,0) rectangle (axis cs:5.4,0.235759911894273);
\addplot [line width=1.08pt, darkslategray66]
table {%
0 0.349878579295154
0 0.371135737885463
};
\addplot [line width=1.08pt, darkslategray66]
table {%
1 0.383513215859031
1 0.406051762114537
};
\addplot [line width=1.08pt, darkslategray66]
table {%
2 0.0619162995594714
2 0.0928573788546256
};
\addplot [line width=1.08pt, darkslategray66]
table {%
3 0.18523127753304
3 0.265760187224669
};
\addplot [line width=1.08pt, darkslategray66]
table {%
4 0.219525605726872
4 0.234431718061674
};
\addplot [line width=1.08pt, darkslategray66]
table {%
5 0.224645925110132
5 0.246402808370044
};
\end{axis}

\end{tikzpicture}}

		\caption{Uniform}
		\label{fig:multiData-two}
	\end{subfigure}
	\caption{Test accuracy on ModelNet10 \cite{wu20153d}: Signals are defined on Sphere, Gaussian, Grid, and Uniform grids with few samples (Top row: $|\widehat{\ccalX}| = 125$) and many samples (Bottom row: $|\widehat{\ccalX}| = 1000$). Each filter is evaluated with one and two layers. }
	\label{fig_modelnet10_numsim_small}
\end{figure*}



\subsection{ModelPoint10 \cite{wu20153d} Dataset}

The ModelNet10 \cite{wu20153d} dataset contains 4,899 CAD models for 10 categories. The dataset is split into 3,991 samples for training and 908 for testing. After normalizing the dataset, we rotate the signals so they are no longer aligned. In particular, similar to the Knot dataset generation, signals are rotated at random by selecting a group action $g\in \mathsf{SO}(3)$. We project each point cloud onto the Grid, Uniform, Sphere, and Gaussian sampling schemes described in \eqref{equ_point_cloud_proj} with $a_1 = a_2 - a_3 = -1$, $b_1= b_2 = b_3 = 1$, and $R = 1$.


\subsection{RotatedMNIST \cite{larochelle2007empirical} Dataset}


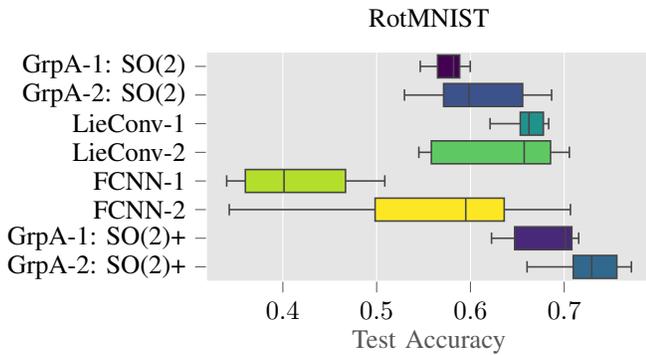
\begin{figure}
    \centering
\begin{tikzpicture}

\def\axisdefaultwidth{7.5cm}
\def\axisdefaultheight{4.63cm}

\definecolor{my_cp5_col1}{RGB}{253, 231, 37}
\definecolor{my_cp5_col2}{RGB}{180, 222,44}
\definecolor{my_cp5_col3}{RGB}{94, 201, 98}
\definecolor{my_cp5_col4}{RGB}{33, 145, 140}
\definecolor{my_cp5_col5}{RGB}{59, 82, 139}
\definecolor{my_cp5_col6}{RGB}{68, 1, 84}
\definecolor{my_cp5_col7}{RGB}{49, 104, 142}
\definecolor{my_cp5_col8}{RGB}{72, 40, 120}

\definecolor{cadetblue74171164}{RGB}{74,171,164}
\definecolor{darksalmon233149163}{RGB}{233,149,163}
\definecolor{darkslategray70}{RGB}{70,70,70}
\definecolor{dimgray85}{RGB}{85,85,85}
\definecolor{gainsboro229}{RGB}{229,229,229}
\definecolor{lightsteelblue182167234}{RGB}{182,167,234}
\definecolor{mediumseagreen72175114}{RGB}{72,175,114}
\definecolor{mediumturquoise82172205}{RGB}{82,172,205}
\definecolor{peru20215175}{RGB}{202,151,75}
\definecolor{plum231140215}{RGB}{231,140,215}
\definecolor{yellowgreen15116269}{RGB}{151,162,69}

\begin{axis}[
axis background/.style={fill=gainsboro229},
axis line style={white},
tick align=outside,
tick pos=left,
title={RotMNIST},
x grid style={white},
xlabel=\textcolor{dimgray85}{Test Accuracy},
xmajorgrids,
xmin=0.318525, xmax=0.793175,
xtick style={color=dimgray85},
y dir=reverse,
y grid style={white},
ymin=-0.5, ymax=7.5,
ytick style={color=dimgray85},
ytick={0,1,2,3,4,5,6,7},
yticklabels={
  GrpA-1: SO(2),
  GrpA-2: SO(2),
  LieConv-1,
  LieConv-2,
  FCNN-1,
  FCNN-2,
  GrpA-1: SO(2)+,
  GrpA-2: SO(2)+
}
]
\path [draw=darkslategray70, fill=my_cp5_col6, semithick]
(axis cs:0.564935,-0.4)
--(axis cs:0.564935,0.4)
--(axis cs:0.588515,0.4)
--(axis cs:0.588515,-0.4)
--(axis cs:0.564935,-0.4)
--cycle;
\path [draw=darkslategray70, fill=my_cp5_col5, semithick]
(axis cs:0.571395,0.6)
--(axis cs:0.571395,1.4)
--(axis cs:0.655665,1.4)
--(axis cs:0.655665,0.6)
--(axis cs:0.571395,0.6)
--cycle;
\path [draw=darkslategray70, fill=my_cp5_col4, semithick]
(axis cs:0.65313,1.6)
--(axis cs:0.65313,2.4)
--(axis cs:0.67776,2.4)
--(axis cs:0.67776,1.6)
--(axis cs:0.65313,1.6)
--cycle;
\path [draw=darkslategray70, fill=my_cp5_col3, semithick]
(axis cs:0.558335,2.6)
--(axis cs:0.558335,3.4)
--(axis cs:0.685385,3.4)
--(axis cs:0.685385,2.6)
--(axis cs:0.558335,2.6)
--cycle;
\path [draw=darkslategray70, fill=my_cp5_col2, semithick]
(axis cs:0.36003,3.6)
--(axis cs:0.36003,4.4)
--(axis cs:0.46681,4.4)
--(axis cs:0.46681,3.6)
--(axis cs:0.36003,3.6)
--cycle;
\path [draw=darkslategray70, fill=my_cp5_col1, semithick]
(axis cs:0.49835,4.6)
--(axis cs:0.49835,5.4)
--(axis cs:0.636005,5.4)
--(axis cs:0.636005,4.6)
--(axis cs:0.49835,4.6)
--cycle;
\path [draw=darkslategray70, fill=my_cp5_col8, semithick]
(axis cs:0.64712,5.6)
--(axis cs:0.64712,6.4)
--(axis cs:0.70809,6.4)
--(axis cs:0.70809,5.6)
--(axis cs:0.64712,5.6)
--cycle;
\path [draw=darkslategray70, fill=my_cp5_col7, semithick]
(axis cs:0.70958,6.6)
--(axis cs:0.70958,7.4)
--(axis cs:0.75608,7.4)
--(axis cs:0.75608,6.6)
--(axis cs:0.70958,6.6)
--cycle;
\addplot [semithick, darkslategray70]
table {%
0.564935 0
0.54652 0
};
\addplot [semithick, darkslategray70]
table {%
0.588515 0
0.5998 0
};
\addplot [semithick, darkslategray70]
table {%
0.54652 -0.2
0.54652 0.2
};
\addplot [semithick, darkslategray70]
table {%
0.5998 -0.2
0.5998 0.2
};
\addplot [semithick, darkslategray70]
table {%
0.571395 1
0.52966 1
};
\addplot [semithick, darkslategray70]
table {%
0.655665 1
0.68654 1
};
\addplot [semithick, darkslategray70]
table {%
0.52966 0.8
0.52966 1.2
};
\addplot [semithick, darkslategray70]
table {%
0.68654 0.8
0.68654 1.2
};
\addplot [semithick, darkslategray70]
table {%
0.65313 2
0.62094 2
};
\addplot [semithick, darkslategray70]
table {%
0.67776 2
0.6834 2
};
\addplot [semithick, darkslategray70]
table {%
0.62094 1.8
0.62094 2.2
};
\addplot [semithick, darkslategray70]
table {%
0.6834 1.8
0.6834 2.2
};
\addplot [semithick, darkslategray70]
table {%
0.558335 3
0.545 3
};
\addplot [semithick, darkslategray70]
table {%
0.685385 3
0.70548 3
};
\addplot [semithick, darkslategray70]
table {%
0.545 2.8
0.545 3.2
};
\addplot [semithick, darkslategray70]
table {%
0.70548 2.8
0.70548 3.2
};
\addplot [semithick, darkslategray70]
table {%
0.36003 4
0.3401 4
};
\addplot [semithick, darkslategray70]
table {%
0.46681 4
0.5087 4
};
\addplot [semithick, darkslategray70]
table {%
0.3401 3.8
0.3401 4.2
};
\addplot [semithick, darkslategray70]
table {%
0.5087 3.8
0.5087 4.2
};
\addplot [semithick, darkslategray70]
table {%
0.49835 5
0.34284 5
};
\addplot [semithick, darkslategray70]
table {%
0.636005 5
0.70664 5
};
\addplot [semithick, darkslategray70]
table {%
0.34284 4.8
0.34284 5.2
};
\addplot [semithick, darkslategray70]
table {%
0.70664 4.8
0.70664 5.2
};
\addplot [semithick, darkslategray70]
table {%
0.64712 6
0.6225 6
};
\addplot [semithick, darkslategray70]
table {%
0.70809 6
0.71542 6
};
\addplot [semithick, darkslategray70]
table {%
0.6225 5.8
0.6225 6.2
};
\addplot [semithick, darkslategray70]
table {%
0.71542 5.8
0.71542 6.2
};
\addplot [semithick, darkslategray70]
table {%
0.70958 7
0.66038 7
};
\addplot [semithick, darkslategray70]
table {%
0.75608 7
0.7716 7
};
\addplot [semithick, darkslategray70]
table {%
0.66038 6.8
0.66038 7.2
};
\addplot [semithick, darkslategray70]
table {%
0.7716 6.8
0.7716 7.2
};
\addplot [semithick, darkslategray70]
table {%
0.582 -0.4
0.582 0.4
};
\addplot [semithick, darkslategray70]
table {%
0.59851 0.6
0.59851 1.4
};
\addplot [semithick, darkslategray70]
table {%
0.66241 1.6
0.66241 2.4
};
\addplot [semithick, darkslategray70]
table {%
0.65733 2.6
0.65733 3.4
};
\addplot [semithick, darkslategray70]
table {%
0.40126 3.6
0.40126 4.4
};
\addplot [semithick, darkslategray70]
table {%
0.59502 4.6
0.59502 5.4
};
\addplot [semithick, darkslategray70]
table {%
0.70102 5.6
0.70102 6.4
};
\addplot [semithick, darkslategray70]
table {%
0.7293 6.6
0.7293 7.4
};
\end{axis}

\end{tikzpicture}
    \caption{Test Accuracy on RotatedMNIST \cite{larochelle2007empirical}: $\mathsf{SO}(2)+$ denotes the augmentation of our filter with the expansion group which mirrors the locality heuristic employed by \cite{finzi2020generalizing}.}
    \label{fig:rotMNIST}
\end{figure}


The RotatedMNIST\cite{larochelle2007empirical} dataset contains 12,000 randomly rotated MNIST digits with rotations sampled uniformly at random from $\left[0, 2\pi\right]$. The dataset is split into 10,000 for training and 2,000 for testing.


\subsection{Implementation details} \label{sec_implementation_details}

We train each model for 100 iterations with a batchsize of 16. We train the models with cross-entropy loss and an ADAM optimizer with no weight decay. For each dataset, we run each simulation ten times with different random seeds. For all simulations, we set the learning rate equal to $10^{-3}$. 

We evaluate the performance of our model -  which we denote by GrpA for Group Algebra - with the classification accuracy on the test data and compare it to two baselines. The first is a naive fully connected neural network (FCNN) with one hidden layer of length 50. The second is the Lie group convolution (LieConv) described in \cite{finzi2020generalizing}, where we remove the residual net architecture and evaluate the convolution layer itself. To remove the effect of model architecture (which is not considered in this work), we evaluate the performance of one layer (filter) and a simple two-layer (NN) (see Figure \ref{fig:GNNbasicfig}) architecture across all methods to showcase the performance of the proposed group convolution. We emphasize this to explain the difference between our accuracy with the state-of-the-art leader-boards. We use swish non-linearities\cite{ramachandran2017searching}, as was done in \cite{finzi2020generalizing}.


\subsection{Stability}
In this section, we describe a numerical experiment to validate and visualize the stability results described in Section \ref{sec:multigraphs}. In particular, we want to verify that the learned filters satisfy \eqref{sec:multigraphs}. In other words, we want to show that a deformation in the operator is proportional to the size of the perturbation.

Numerically, for each $\hat{g}$, we let the additive perturbation matrix $\bbQ_{0,\hat{g}}$ to be a diagonal matrix where the diagonal is generated by i.i.d. samples whose absolute value is no greater than $\varepsilon$. For simplicity, we let the multiplicative matrix $\bbQ_{1,\hat{g}}$ be equal to the identity. The model is then trained as described in Section \ref{sec_implementation_details} with the unperturbed weights. During evaluation, we use the learned coefficients $\widehat{\bba}$ with and without the perturbed transformation matrices $\widehat{\bbT}_{\hat{g}}$ and $\widehat{\bbT}_{\hat{g}} + \bbQ_{0,\hat{g}}$. We evaluate the difference between the predictions of the model on the test dataset with the $\ell_2$-norm. 

\begin{figure}
    \centering
\begin{tikzpicture}

\def\axisdefaultwidth{8.5cm}
\def\axisdefaultheight{5.25cm}

\definecolor{my_cp_col1}{RGB}{253, 231, 37}
\definecolor{my_cp_col2}{RGB}{94, 201, 98}
\definecolor{my_cp_col3}{RGB}{33, 145, 140}
\definecolor{my_cp_col4}{RGB}{59, 82, 139}
\definecolor{my_cp_col5}{RGB}{68, 1, 84}

 \definecolor{darkgray158150204}{RGB}{158,150,204}
 \definecolor{darkslategray71}{RGB}{71,71,71}
 \definecolor{dimgray85}{RGB}{85,85,85}
 \definecolor{gainsboro229}{RGB}{229,229,229}
 \definecolor{gray119}{RGB}{119,119,119}

\begin{axis}[
axis background/.style={fill=gainsboro229},
axis line style={white},
log basis y={10},
tick align=outside,
tick pos=left,
x grid style={white},
xlabel=\textcolor{dimgray85}{$\epsilon$},
xmin=-0.5, xmax=3.5,
xtick style={color=dimgray85},
xtick={0,1,2,3},
xticklabels={0.0001,0.001,0.01,0.1},
y grid style={white},
ylabel=\textcolor{dimgray85}{$\displaystyle\Vert \boldsymbol{\Phi} - \widehat{\boldsymbol{\Phi}}\Vert_2$},
ymajorgrids,
ymin=1.0492022810478e-05, ymax=0.204191480772931,
ymode=log,
ytick style={color=dimgray85}
]
\path [draw=darkslategray71, fill=my_cp_col1, semithick]
(axis cs:-0.4,2.80972343400204e-05)
--(axis cs:0.4,2.80972343400204e-05)
--(axis cs:0.4,0.00011592602333221)
--(axis cs:-0.4,0.00011592602333221)
--(axis cs:-0.4,2.80972343400204e-05)
--cycle;
\path [draw=darkslategray71, fill=my_cp_col2, semithick]
(axis cs:0.6,0.000832990881499015)
--(axis cs:1.4,0.000832990881499015)
--(axis cs:1.4,0.000995070240795132)
--(axis cs:0.6,0.000995070240795132)
--(axis cs:0.6,0.000832990881499015)
--cycle;
\path [draw=darkslategray71, fill=my_cp_col3, semithick]
(axis cs:1.6,0.00915008470727624)
--(axis cs:2.4,0.00915008470727624)
--(axis cs:2.4,0.0174365942969233)
--(axis cs:1.6,0.0174365942969233)
--(axis cs:1.6,0.00915008470727624)
--cycle;
\path [draw=darkslategray71, fill=my_cp_col4, semithick]
(axis cs:2.6,0.0638297013800146)
--(axis cs:3.4,0.0638297013800146)
--(axis cs:3.4,0.0888474182937662)
--(axis cs:2.6,0.0888474182937662)
--(axis cs:2.6,0.0638297013800146)
--cycle;
\addplot [semithick, darkslategray71]
table {%
0 2.80972343400204e-05
0 1.6436974458924e-05
};
\addplot [semithick, darkslategray71]
table {%
0 0.00011592602333221
0 0.000126187025043404
};
\addplot [semithick, darkslategray71]
table {%
-0.2 1.6436974458924e-05
0.2 1.6436974458924e-05
};
\addplot [semithick, darkslategray71]
table {%
-0.2 0.000126187025043404
0.2 0.000126187025043404
};
\addplot [semithick, darkslategray71]
table {%
1 0.000832990881499015
1 0.000832990881499015
};
\addplot [semithick, darkslategray71]
table {%
1 0.000995070240795132
1 0.000995070240795132
};
\addplot [semithick, darkslategray71]
table {%
0.8 0.000832990881499015
1.2 0.000832990881499015
};
\addplot [semithick, darkslategray71]
table {%
0.8 0.000995070240795132
1.2 0.000995070240795132
};
\addplot [black, mark=diamond*, mark size=2.5, mark options={solid,fill=darkslategray71}, only marks]
table {%
1 0.00012288988067422
1 0.0019836098779344
};
\addplot [semithick, darkslategray71]
table {%
2 0.00915008470727624
2 0.00206422857917563
};
\addplot [semithick, darkslategray71]
table {%
2 0.0174365942969233
2 0.0181646451458484
};
\addplot [semithick, darkslategray71]
table {%
1.8 0.00206422857917563
2.2 0.00206422857917563
};
\addplot [semithick, darkslategray71]
table {%
1.8 0.0181646451458484
2.2 0.0181646451458484
};
\addplot [semithick, darkslategray71]
table {%
3 0.0638297013800146
3 0.0550036066754042
};
\addplot [semithick, darkslategray71]
table {%
3 0.0888474182937662
3 0.0888474182937662
};
\addplot [semithick, darkslategray71]
table {%
2.8 0.0550036066754042
3.2 0.0550036066754042
};
\addplot [semithick, darkslategray71]
table {%
2.8 0.0888474182937662
3.2 0.0888474182937662
};
\addplot [black, mark=diamond*, mark size=2.5, mark options={solid,fill=darkslategray71}, only marks]
table {%
3 0.130339173996327
};
\addplot [semithick, darkslategray71]
table {%
-0.4 3.65120010580238e-05
0.4 3.65120010580238e-05
};
\addplot [semithick, darkslategray71]
table {%
0.6 0.000979825673783452
1.4 0.000979825673783452
};
\addplot [semithick, darkslategray71]
table {%
1.6 0.0120925452387801
2.4 0.0120925452387801
};
\addplot [semithick, darkslategray71]
table {%
2.6 0.0749102842302174
3.4 0.0749102842302174
};
\end{axis}

\end{tikzpicture}
    \caption{Stability results on the Knot dataset with Sphere sampling over ten random seeds. Filters are first trained without perturbations to obtain the coefficients. Then they are evaluated with ($\widehat{\boldsymbol{\Phi}}:= p\left( \bbQ\left(\widehat{\bbT}\right) \right)$) and without ($\boldsymbol{\Phi}:= p\left( \bbQ\left(\bbT\right) \right)$) the perturbations.  We consider an additive perturbation model with diagonal matrix whose elements are generated randomly and are no greater in absolute value than $\varepsilon$. }
    \label{fig:stability}
\end{figure}
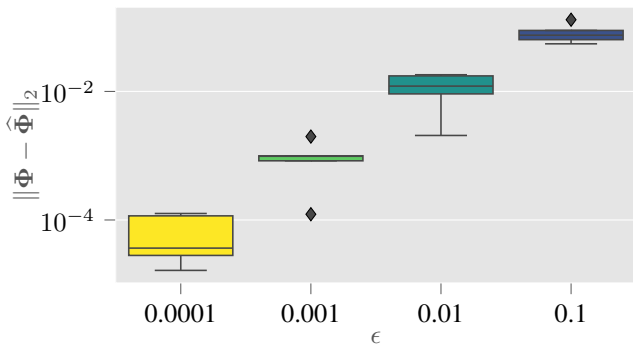

Figure \ref{fig:stability} shows the evaluation for GrpA-1 (i.e. one layer filter of our method) when training on the Knot dataset with the Sphere sampling scheme. All hyperparameters are the same as described in Section \ref{sec_implementation_details}, and the box plots describe the behavior over ten different random seeds. These results corroborate the expected results as defined in Definition \ref{def_operator_stab} because the size of the error $\Vert \boldsymbol{\Phi} - \widehat{\boldsymbol{\Phi}}\Vert_2$ is proportional to the size of the perturbation $\varepsilon$.


\subsection{Discussion}

First, let us consider the binary knot classification task. The top and bottom rows of Figure \ref{fig_knot_numsim_small} show the accuracy on test data when the size of the sampled set $|X|$ is 125 and 1000 respectively. As expected, the sampling scheme plays an important role on the performance of all of the models. The Gaussian sampling has the worst performance over all test samples, while Grid and Sphere sampling tend to have higher accuracies. Unsurprisingly, the two-layer network improves on the one-layer filter network. The difference between the models is small when the sample size is low (125); however, it becomes strikingly apparent in the large sample regime (1000). The performance of GrpA outperforms the FCNNs on all four sampling schemes; however, LieConv does outperform GrpA more consistently.

Next, we look at the ModelPoint10 classification task as shown in figure \ref{fig_modelnet10_numsim_small}. Similar to the binary knot classification problem, the sampling scheme plays a large role in the performance of the model. Again, we see that the Gaussian sampling is lower in general with Sphere and Grid being preferred. In this example, not only does GrpA outperform the baselines in the high sample regime, but it also does so with fewer samples. 

Finally, the RotatedMNIST results are shown in Figure \ref{fig:rotMNIST}. Unlike the experiments in $\mathsf{SO}(3)$, where GrpA generally improved upon LieConv, the experiments in $\mathsf{SO}(2)$ show the opposite. Recall that LieConv of \cite{finzi2020generalizing} uses a locality heuristic in order to account for the locality of the filter. By similarly augmenting $\mathsf{SO(2)}$ of GrpA with the expansion group (which we denote by $\mathsf{SO}(2)+$), we see the expected improvement of GrpA over LieConv.

There is an important caveat to mention for LieConv in the numerical results for the high sampling scheme (bottom row). The implementation of the group convolution requires sampling from the neighborhood in the lifted coordinates. In high dimensions, this resulted in the model running out of memory as the sampling is done at each call of the model. To make the experiments tractable, we had to reduce the size of the locality (i.e. fewer neighbors to sample) which is a change from the default hyperparameters otherwise used. Specifically, we reduced the fraction of the input used for the local neighborhood from $1/3$ to $1/32$. This further points to the advantage of our approach to do all of the sampling offline as in Algorithm \ref{alg:Transformation}.

To summarize, the numerical simulations reveal the following main insights. First, the sampling of the signal plays a large role in the overall performance of \emph{all} models. Second, GrpA not only outperforms LieConv and FCNN in most settings but also enables group convolutions in large dimensional spaces through the use of sparse transformation matrices.




\section{Conclusion} \label{sec:discussion}

In this work, we introduced a method to implement group convolutions through the lens of ASP. In particular, we showed that the Lie group algebra homomorphism can be leveraged to develop a convolution of which the conventional group convolution is a particular case. To realize the filter, we proposed two separate sampling methods. The first was the sampling of the filter, and the second is the sampling of the signal. We showed that the sampled filter uniquely corresponds with a continuous filter with finite bandwidth, and we characterized a relationship between the proposed sampling scheme and the bandwidth of the filter. We then showed that the sampling of the signal, which is generally an inherent sampling of the problem, results in algebraically justified interpolation to realize the filter with sparse matrices. Finally, we draw connections to multigraph signal processing, where we inherit stability concerning perturbations. The numerical experiments corroborated our findings on signals with $\mathsf{SO}(2)$ and $\mathsf{SO}(3)$ symmetries.

Notice that the error bound in Theorem~\ref{thm_imageinterpolator} describes the relationship between a filter and its reconstructed version in $L^{2}(G)\subset L^{1}(G)$ from a set of samples on $G$, and such error estimate does not involve the interpolation map $\tau$ that is included then the filters are implemented in $\ccalB(\ccalH)$. Additionally, note that the effect of $\tau$ on the filters implemented in $\ccalB(\ccalH)$ can be included in the perturbation model in equation~\eqref{eq_perturb_model}. Therefore, the results and constants in Corollary~\ref{corollary_stability} provide a measure of the stability of the filters concerning the properties of $\tau$. In this context, the construction of the filters holds for an arbitrary $\tau$, and the stability results point to the fact that the ability of the filers to capture the symmetries in $G$ on $\widehat{\ccalH}$ is affected by $\tau$ in a proportional way to the ability of $\tau$ to map elements from $\ccalH$ to $\overline{\ccalH}$. In this regard, there are some avenues for future work. To begin, the effect of different interpolation schemes could be considered. In this work, we evaluated using trilinear and Barycentric interpolation methods; however, using better interpolation algorithms tailored to the sampling scheme (either structured or unstructured) should increase the performance of the filter. Moreover, the filter could be improved by implementing a pruning algorithm on the diffusion tree (see Fig \ref{fig:diffusion_tree}). Pruning the tree has shown to be effective in multigraph signal processing, as it removes elements that are close together allowing the tree to grow. Finally, pooling algorithms should also be considered. Given the graph-like structure of the signal, graph pooling algorithms may be desirable. Note that in most group convolutional literature, a global maxpooling step is implemented, which results in a major loss of information. Focusing on local geometry preserving pooling should improve the quality of the filter.

\bibliography{bibliography}
\bibliographystyle{unsrt}

\ifCLASSOPTIONcaptionsoff
  \newpage
\fi


 \appendices



\section{Proofs}


\subsection{Proof of Theorem \ref{corr:add_degree}} \label{sec:proof_thm_add_degree}

\begin{proof}
    It suffices to show that $\widehat{G}_{\delta,N}^{k} \subseteq \hat G_{\delta,N}^{k'}$. By Definition~\ref{def_Ghat_monomials}, it follows that all of the monomials included up until order $k$ are included in order $k'>k$. The case for $N'$ can be shown similarly.
\end{proof}


\subsection{Proof of Theorem \ref{corr:decrease_resolution}} \label{sec:proof_them_decrease_resolution}

\begin{proof}
    It suffices to show that $\widehat{G}_{\delta, N}\subset~\widehat{G}_{\delta', N'}$. Let $x\in \hat G_{\delta, N}$. By definition\eqref{def_Ghat}, there exist $n\in [-N, N], \mathfrak{X} \in \mathfrak{G}$ such that $x = e^{\delta n \mathfrak{X}}$. Let $n' =  hn$ and $\delta' = \delta / h$. Then $e^{\delta'n'\mathfrak{X}} = e^{hn\delta\mathfrak{X}/h} \in \hat G_{\delta', N'}$, which completes the proof.
\end{proof}


\subsection{Proof of Theorem \ref{thm:equivariance_LR}} \label{sec:proof_equivariance_LR}
\begin{proof}
Taking into account that
\begin{equation*}
\rho\left( 
           \boldsymbol{a}
       \right)
       \bbT_{h}
       f
       =
        \int_{G}
           \boldsymbol{a}(g)\bbT_{g}\bbT_{h}
           f
           d\mu(g)
       =    
         \int_{G}
           \boldsymbol{a}(g)\bbT_{gh}
           f
           d\mu(g)
           .
\end{equation*}
Then, making $u=gh$ and taking into account that $\mu$ is right invariant we have
\begin{equation*}
         \int_{G}
           \boldsymbol{a}(g)\bbT_{gh}
           f
           d\mu(g)
       =
        \int_{G}
           \boldsymbol{a}\left( uh^{-1} \right)\bbT_{u}
           f
           d\mu(u)
       =    
\rho\left(
         R_{h^{-1}}\boldsymbol{a}
    \right)
    f
           .
\end{equation*}

Now, we take into account that
\begin{equation*}
\bbT_{h}\rho\left( 
           \boldsymbol{a}
       \right)
       f
       =
        \int_{G}
           \boldsymbol{a}(g)\bbT_{h}\bbT_{g}
           f
           d\mu(g)
       =    
         \int_{G}
           \boldsymbol{a}(g)\bbT_{hg}
           f
           d\mu(g)
           .
\end{equation*}
Then, making $u=hg$ and that $\mu$ is left invariant we have
\begin{equation*}
\int_{G}
        \boldsymbol{a}(g)\bbT_{hg}
        f
        d\mu(g)
       =
        \int_{G}
           \boldsymbol{a}\left( h^{-1}u \right)\bbT_{u}
           f
           d\mu(u)
       =    
\rho\left(
         L_{h^{-1}}\boldsymbol{a}
    \right)
    f
           .
\end{equation*}

we complete the proof.

\end{proof}


\subsection{Proof of Theorem~\ref{thm_imageinterpolator}}
\label{sec_proof_of_thm_imageinterpolator}

Taking into account Theorem~4.1.8 in~\cite{kadison1997fundamentals} and the fact that $L^{1}(G)$ is a Banach $\ast$-algebra that can be completed to a $C^{\ast}$-algebra (see~\cite{kenneth1996c}, page 182) it follows that
\begin{equation}
\left\Vert 
      \rho\left( 
             \boldsymbol{a}
          \right)
      -    
      \rho\left( \mathtt{I}_{\hat{g}}^{(\ell)}
                \left(
                            \boldsymbol{a}
                         \right)   
                \right)
\right\Vert
        \leq
\left\Vert 
             \boldsymbol{a}
      -    
          \mathtt{I}_{\hat{g}}^{(\ell)}
                \left(
                            \boldsymbol{a}
                         \right)   
\right\Vert
        .
\end{equation}
 Then, taking into account Theorem~\ref{thm:pesenson} and~\eqref{eq_interpolator_filter} the proof is completed.

\newpage


\section{Summary of Symbols used}


\begin{tabular}{cp{0.6\textwidth}}
  $G$ & Lie group \\
  $\boldsymbol{\Delta}$ & Laplace Beltrami operator on $G$\\
  $L^{1}(G)$ & Set of absolutely integrable functions on $G$ \\
  $L^{2}(G)$ & Set of square integrable functions on $G$ \\
  $E_{\omega}(\boldsymbol{D})$ & Set of $\omega$-bandlimited filters in $L^{2}(G)$\\
  $B(g,r)$ & Open ball on $G$ with center $g$ and radius $r$ \\
  $r_{max}$ & Maximum geodesic distance on $G$\\
  $Y_{\widehat{G}}(r)$ & Union of balls $B(g,r)$ with $g\in\widehat{G}$ \\
  $\omega(r)$ & Maximum bandwidth that satisfies~\eqref{equ:thm1_ineqality}\\
  $\overline{\ccalH}$ & Space of piece-wise constant functions \\
  $\widehat{G}$ & Discrete subset of $G$\\
  $\widehat{G}_{\delta, N}$ & Discrete subset of $G$ from Definition~\ref{def_Ghat}\\
  $ \widehat{G}^k_{\delta, N} $ & Discrete subset of $G$ from Definition~\ref{def_Ghat_monomials}\\
  $ L^{1}\left( \widehat{G}^k_{\delta, N}\right) $ & The set of functions on $\widehat{G}^k_{\delta, N}$ \\
  $\widehat{\boldsymbol{a}}$ & Discrete representation of $\boldsymbol{a}\in L^{1}(G)$ \\ 
  $\mathfrak{g}$ & Lie algebra associated with $G$\\
  $\mathfrak{G}$ & Basis of $\mathfrak{g}$ \\
  $\ccalX$       & A subset of $\mbR^n$ \\
  $T_{g}$ & Action of $g\in G$ \\
  $\ccalH$ & Hilbert space of functions from $G$ to $\mbC$ \\
  $\boldsymbol{f}$ & Element in $\ccalH$ \\
  $\ccalB(\ccalH)$ & Set of bounded operators from $\ccalH$ to $\ccalH$\\
  $\rho$ & Homomorphism from $L^{1}(G)$ to $\ccalB(\ccalH)$ \\
  $\bbT_{g}$ & Operator in $\ccalB(\ccalH)$ induced by $T_{g}$ \\
  $\overline{\boldsymbol{f}(x)}$ & Complex conjugate of $\boldsymbol{f}(x)$\\
  $\widehat{\ccalX}$ & Discrete subset of $\ccalX$\\
  $\widehat{\ccalH}$ & Space of discrete functions on $\widehat{\ccalX}$\\
  $\widehat{\bbT}_{g}$ & Operator acting on $\widehat{\ccalH}$ induced by $\bbT_g$ \\
  $\ccalC_i$ & Voronoi cells from \eqref{equ:voronoi_cells}\\
  $\theta$ & Isomorphism between $\ccalH$ and $\overline{\ccalH}$\\
  $\mathsf{N}$ & set of indices of KNN from Alg \ref{alg:BarycentricInterpolator}\\
  $\mathsf{W}$ & set of weights of KNN from Alg \ref{alg:BarycentricInterpolator}\\
  $\bbQ$ & perturbation model from \eqref{eq_perturb_model}\\
  
\end{tabular}\\



\section{Preliminaries for Lie Algebraic Signal Processing}
\label{sec_background}

In this section, we discuss the basics of Lie groups, Lie Algebras, and Lie Group Algebras. We start with the following basic definition.


\begin{definition}[Lie group~\cite{hall_liealg}]\label{def:LieGroup}	
A \textbf{Lie group} is a smooth manifold $G$ which is also a group, such that the group product and the inverse map are smooth.
\end{definition}


Manifolds by definition look locally that subsets of $\mbR^n$, i.e. they can be parametrized in $\mbR^n$. Then, from Definition~\ref{def:LieGroup} we can see that a Lie group is roughly speaking a group that can be parametrized by a set of variables taking values on $\mbR$. 

It is worth pointing out that Definition~\ref{def:LieGroup} can be made even more tangible when considering matrix Lie groups~\cite{hall_liealg,stillwell2008naive} as subsets of the set of all invertible matrices of size $n\times n$, i.e. $\mathsf{GL}(n;\mbC)$. To see this we state the following definition.


\begin{definition}[matrix Lie group~\cite{hall_liealg}]\label{def_matrix_LieGroup}	
	A \textbf{matrix Lie group} is a subgroup $G$ of $\mathsf{GL}(n;\mbC)$ with the following property: If $\bbA_m$ is any sequence of matrices in $G$, and $\bbA_m$ converges to some matrix $\bbA$, then either $\bbA$ is in $G$ or $\bbA$ is not invertible.
\end{definition}


The notion of convergence in Definition~\ref{def_matrix_LieGroup} can be conceived associating $M_{n\times n}$ -- the set of matrices of size $n\times n$ -- with $\mbC^{n^2}$ and considering $G$ as a subset of this set. In particular, we say that the sequence $\{ \bbA_m \}_m$ of matrices converge to $\bbA$ if $\{ \bbA_m (i,j) \}_m$ converges to $\bbA (i,j)$ for each $(i,j)$. This is the sequence of entries of $\{ \bbA_m \}_m$ at the $(i,j)$ position converge to the $(i,j)$th entry of $\bbA$. Additionally, notice that Definition~\ref{def_matrix_LieGroup} can be rephrased saying that $G$ is a matrix Lie group when it is a closed subgroup of $\mathsf{GL}(n;\mbC)$.

In what follows we introduce some classical examples of Lie groups.


\begin{example}
(\cite{hall_liealg})\normalfont $G = \mathsf{SO}(2,\mbR)$,
The set of orthogonal matrices, $\bbA \bbA^{\mathsf{T}} = \bbA^{\mathsf{T}} \bbA = \bbI$, of size $n\times n$ whose determinant is equal to 1. The group product is given by the ordinary product of matrices. Geometrically, elements of $\mathsf{SO}(2,\mbR)$ are rotations.
\end{example}


%
%
%


Other important examples of Lie groups include:

\begin{itemize}
    \item The unitary group $\mathsf{U}(n)$: The set of unitary matrices of size $n\times n$. Unitary matrix: $\bbA \bbA^\ast = \bbA^\ast \bbA = \bbI$.
	\item The special unitary group $\mathsf{SU}(n)$: The set of unitary matrices of size $n\times n$ whose determinant is equal to 1.
	\item The orthogonal group $\mathsf{O}(n)$: The set of orthogonal matrices of size $n\times n$. Orthogonal matrix: $\bbA \bbA^{\mathsf{T}} = \bbA^{\mathsf{T}} \bbA = \bbI$.
	\item The special linear group  $\mathsf{SU}(2,\mbR)$: The set of all $2\times 2$ matrices whose determinant is equal to one. 
\end{itemize}

Although Lie groups are naturally non linear since they are manifolds, there is a linear structure associated to them called the \textit{Lie algebra} whose formal definition we introduce next.


\begin{definition}[Lie Algebra~\cite{hall_liealg}]\label{def_LieAlgebra} A \textbf{Lie Alegbra} is a finite-dimensional real or complex vector space $\mathfrak{g}$, together with a mapping $\left[\cdot, \cdot\right]$ that is bilinear, skew symmetric, and satisfies the Jacobi identity ($[X,[Y,Z]] + [Y, [Z, X]] + [Z, [X,Y]] = 0$).
\end{definition}


Each matrix Lie group $G$ has an associated Lie algebra $\mathfrak{g}$, and every element in $G$ can be obtained from $\mathfrak{g}$ via the exponential map $e^{t(\cdot)}: \mathfrak{g}\to G$, where $t\in\mbR$. Then, for every $X\in\mathfrak{g}$ it follows that $e^{tX}\in G$ for any $t\in\mbR$. The Lie algebra $\mathfrak{g}$ is the tangent space at the identity to $G$, and the exponential map is onto and one-to-one in the neighborhood of the identity. For many Lie groups of practical importance, their associated Lie algebras are determined by a finite number of generators -- in the sense of vector spaces. This simplifies the characterization of the Lie group. We will leverage this fact in Section~\ref{sec_grpasm}.

Given a Lie group $G$ there is a space, $L^{1}(G)$, of continuous and integrable functions defined on $G$ that allows us to characterize properties of $G$ itself using the Haar measure~\cite{folland2016course,deitmar2014principles,hewitt1994abstract}. To define properly $L^{1}(G)$ we recall that a left Haar measure on a group $G$ is a nonzero Radon measure $\mu$ on $G$ such that $\mu(xE)=\mu(E)$ for any Borel set $E\subset G$ and any group element $x\in G$. As pointed out in~\cite{folland2016course} (Theorem~2.10), every locally compact group has a left Haar measure. Let us denote by $C_{c}(G)$ the set of compactly supported continuous functions on $G$ and $L_{y}\boldsymbol{a}(x) = \boldsymbol{a}(y^{-1}x)$ for $\boldsymbol{a}\in C_{c}(G)$. If $\mu$ is a left Haar measure, then for any $\boldsymbol{a}\in C_{c}(G)$ it follows that

\begin{equation}
 \int_G   L_{y}\boldsymbol{a} d\mu 
 =
 \int_G \boldsymbol{a} d\mu
 .
\end{equation}

As indicated in~\cite{folland2016course} for any Lie group it is possible to derive a Haar measure using tools from differential and Riemmanian geometry. With a clear notion of integration for functions on $G$, we define the Lie Group Algebra as 


\begin{definition}[Lie Group Algebra]\label{def_LieGroupAlgebra} Given a group $G$ and Haar measure $\mu$,  the \textbf{Lie group algebra}, denoted $L^1(G)$, is defined by
\begin{equation}
 L^1 (G)
       :=
       \left\lbrace 
               \boldsymbol{a}\in C_{c}(G) 
       \left\vert
                \int_G \vert \boldsymbol{a}\vert d\mu
                <\infty
       \right.            
       \right\rbrace
       .
\end{equation}
\end{definition}


%
%
%

For reasons that become clear in Section~\ref{subsec_asp_BanachA}, $L^{1}(G)$ is referred to as the \textit{group algebra} associated to $G$.

It is possible to formulate a notion of integration considering a right Haar measure, that at the same time can be uniquely related to a left Haar measure. If $\mu$ is a left Haar measure on the group $G$ and $x\in G$ we can define another measure $\mu_{x}(E) = \mu (Ex)$. As shown in~\cite{folland2016course} there exists a unique function $\Delta (x)$ -- independent of $\mu$ -- such that $\mu_{x} = \Delta(x)\mu$. The term $\Delta: G \rightarrow (0,\infty)$ is known as the modular function of $G$. As shown in~\cite{folland2016course} we have that

\begin{equation}
\int_G R_{y}\boldsymbol{a}d\mu
       = 
       \Delta(y^{-1})\int_G \boldsymbol{a}d\mu
,
\end{equation}
where $R_{y}\boldsymbol{a}(x) = \boldsymbol{a}(xy)$. Additionally, it can be shown that if $z\in G$ is a fixed element we have $d\mu (xz) = \Delta(z)d\mu (x)$. When $\Delta (x) = 1$ for all $x\in G$ we say that $G$ is unimodular. In this scenario we have that left and right Haar measures are equal. As pointed out in~\cite{hewitt1994abstract,folland2016course} discrete groups, finite groups, compact groups, connected semi-simple Lie groups, and nilpotent connected Lie groups are all unimodular, i.e $\Delta(x)=1$ for all $x\in G$. We emphasize here that the modular function $\Delta$ also captures properties of $G$. We will leverage the properties of $\Delta$ in Section~\ref{subsec_asp_BanachA}.



\section{Connections to conventional CNNs}\label{rem:GrpA_as_CNN}\normalfont

For readers who are more familiar with conventional CNNs for image processing, we provide an example to show the equivalence between a discrete convolutional filter and the proposed filter \eqref{equ:lieGAH_discrete}.  Consider the $3\times 3$ CNN kernel $\kappa$
\begin{equation} \label{equ_CNN}\kappa = \begin{bmatrix}
    a_{x^{-1}y} & a_{y_1} & a_{xy} \\
    a_{x^{-1}} & a_0 & a_x \\
    a_{x^{-1}y^{-1}} & a_{y^{-1}} & a_{xy^{-1}}
\end{bmatrix},
\end{equation}
which is convolved with an image $I \in \mbR^{M\times N}$ by $I' = \kappa *I$.
Let $\bbT_x$ and $\bbT_y$ be the induced transformations which translate the signal (2D-image) one pixel to the right and one pixel up respectively. The corresponding proposed filter in the form of \eqref{equ:lieGAH_discrete} takes the form
$$\hat \rho(\hat \bba) I= \left(a_0 \bbT_0 + a_x \bbT_x + a_y \bbT_y + a_{xy}\bbT_{xy} +\dots\right)I,$$
where the sum is taken over all induced transformations and coefficients in $\kappa$, and $\bbT_{xy} \equiv\bbT_x \circ \bbT_y$. Indeed, learning the coefficients in the proposed filter is equivalent to finding the kernel in \eqref{equ_CNN}.

\end{document}